\documentclass[rmp,10pt,byrevtex,titlepage,secnumarabic,nofootinbib,a4paper,longbibliography,preprintnumbers]{revtex4-1}
\usepackage[colorlinks=true,urlcolor=blue,linkcolor=blue,citecolor=blue]{hyperref}

\usepackage{graphicx}
\usepackage{amsmath}
\usepackage{amssymb}
\usepackage{float}
\usepackage{siunitx}
\usepackage[utf8]{inputenc}
\usepackage[english]{babel}

\usepackage{xcolor}
\usepackage{colortbl}
\usepackage{mathtools}
\usepackage{natbib}
\usepackage{lineno}
\usepackage{array}

\AtBeginDocument{\usepackage{booktabs}}
\renewcommand{\thesection}{\arabic{section}}
\renewcommand{\thefigure}{\arabic{figure}}
\renewcommand{\thetable}{\arabic{table}}

\definecolor{BLUE2}{HTML}{084594}
\begin{document}
\author{José María Martín-Olalla}
\email{olalla@us.es}
\homepage{https://twitter.com/MartinOlalla_JM}
\affiliation{Universidad de Sevilla. Facultad de Física. Departamento de Física de la Materia Condensada. ES41012 Sevilla. Spain}
\title{The long term impact of Daylight Saving Time regulations in daily life at several circles of latitude}
\preprint{Published version: \emph{Scientific Reports} 2019 \textbf{9} 54990 \url{http://www.nature.com/articles/s41598-019-54990-6}}
\date{November 22, 2019}
\begin{abstract}
  This is a post-peer-review, pre-copyedit version of an article published in \emph{Scientific Reports}. The final authenticated version is available online at: \url{http://dx.doi.org/10.1038/s41598-019-54990-6}

  We analyze large scale ($N\sim\num{10000}$) time use surveys in United States, Spain, Italy, France and Great Britain to ascertain seasonal variations in the sleep/wake cycle and the labor cycle after daylight saving time regulations have stood up for at least forty years. That is, not the usual search for the impact of the biannual transitions, but a search for how industrialized societies have answered to DST regulations at different circles of latitude.

  Results show that the labor cycle is equally distributed through seasons if measured in local time. It is an everyday experience which is a major outcome of DST. The sleep/wake cycle displays disturbances punctuated by solar events: sunrise, sunset and noon. In week-ends, under free preferences, sleep onset delays in summer, opposing to the regulation and following the delay in sunset time, while sleep offset advances, despite clock time already advanced in the spring transition. This advance still follows the advance in sunrise times. The best explanation for these findings is that human cycles are not misaligned by the size and direction of DST regulations, which explains the success of that practice.

  The sleep/wake cycle in Great Britain and France exhibit fewer statistically significant excursions than the sleep/wake cycle in Spain, Italy and United States, despite light and dark seasonal deviations are larger. That could be indicating that the preference for a seasonal regulation of time decreases with increasing latitude above $\ang{47}$.

  The preferences for a seasonal regulation of clocks and for the choice of permanent summer time or permanent winter time are sketched from a previous report on human activity. 
\end{abstract}

\maketitle
\tableofcontents

\begin{widetext}
    This is a post-peer-review, pre-copyedit version of an article published in \emph{Scientific Rerpots}. The final authenticated version is available online at: \url{http://dx.doi.org/10.1038/s41598-019-54990-6}
  \end{widetext}

\section{Introduction}
\label{sec:introduccion}

Legislative bodies in America and Europe are currently pondering the continuity of Daylight Saving Time regulations ---DST---, also known as Summer Time Arrangements. They consist of a forward change of clocks in spring, which is reversed in autumn. They have become a part of everyday life in some industrial, urban societies since its inception one hundred years ago, amidst World War One, or since the 1970s, amidst the energy crisis.

DST regulations are conveniently described as a time zone shift ---for instance from UTC+1 to UTC+2 in Central Europe at the spring transition--- however, if nothing else change, they are simply a seasonal regulation in the phase of human cycles, all equal to a seasonal regulation of opening times. As an example in 1810, well before the standardization of time,  the Spanish National Assembly started sessions at \textsc{10am} from October to April but at \textsc{9am} from May to September\cite{Luxan1810}.

 Either way, the seasonal regulation of human activity in industrialized, urban societies synced to clocks meets the biannual transitions that characterizes DST. Physiologists and medical doctors think of these transitions as a disruption  to our circadian system\cite{Kantermann2007,Roenneberg2016}. A series of studies has tried to unveil their impact in the stocks markets\cite{Kamstra2000}, in suicide rates\cite{Berk2008}, in the rates of myocardial infarction\cite{Manfredini2018}, in accident rates \cite{Robb2018}, in accident and emergency visits and return visits\cite{Ferrazzi2018} or in the rate of autopsies\cite{Lindenberger2019}. Generally speaking hazards happen to increase immediately after the spring transition. Physiologists then advocate for discontinuing DST\cite{MeiraeCruz2019,Roenneberg2019,Watson2019} and setting winter time permanently or, in practical terms, delaying summer activity by one hour. However, and almost inadvertently, they also acknowledge the convenience of the practice, at least for some people\cite{Roenneberg2019a}.

The ``benefits'' of DST have been usually summarized in extended outdoor leisure activities and marginal savings in energy consumption. Some economists, decision-makers push to extend these benefits to winter, by discontinuing DST in the ``permanent summer time'' mode\cite{Calandrillo2008,Coate2004,Hill2010}. If nothing else changes this choice would advance human activity by one hour in winter pushing for start times ahead of sunrise in winter. The natural abhorrence for getting activated in the dark plays against this choice. When physiologists advocate for permanent winter time they are preventing against this kind of stress on the circadian system\cite{Roenneberg2019,Watson2019}. To be socially accepted an early start of human activity require tangible outcomes which usually come in the form of increasing scores of end times ahead of winter sunset.

The seasonal regulation of human activity ---and also DST regulations--- questions when human activity should start in view of the seasonal spread of sunrise times, sunset times, daytime and noon insolation. That is, upon which conditions individuals and societies synced to clock time would prefer seasonally varying opening times instead of year round opening times, even though annoying biannual transitions would be needed in the former. The answer to this question  may depend on a number of issues that includes chronotype ---preference for morningness versus eveningness--- or economic sector ---outdoor versus indoor activities---. From the point of view of the physical science the answer is dominated by latitude, a role that is virtually omitted in DST discussions\cite{Martin-Olalla2019a}. For instance DST regulations have never been a choice in the Tropics, because there seasonal variations are small. Likewise, the preference for advancing human activity relative to winter sunrise is more frequent at high latitude just because there sunrise delays the most and sunset advances the most in winter. Generally speaking every circle of latitude exhibits a distinct pattern of seasonal light and dark cycle and human activity adapts to that accordingly\cite{Monsivais2017a,Martin-Olalla2018,Martin-Olalla2019b}.

The role of latitude is of the utmost importance in the European Union, where population and decision-makers spread from $\ang{35}$ latitude ---Malta, Crete and Cyprus--- to $\ang{70}$ latitude ---Finnish Lapland--- exhibiting vividly different natural conditions\cite{EPRS2017}. However, and  based on internal market efficiency, the European Commission has been pushing for synchronous DST arrangements, preventing member states from opting-out. The Commission considered the impact of six possible scenarios in which a few members have asynchronous arrangements\cite{Kearney2014}. However no scenario simulated that the preference for DST regulations could stratify with latitude. In sharp contrast this preference does stratify with latitude in Australia, Brazil and Chile. 

Time zone and time zone regulation only set arbitrary references of time. They have stood up unchanged for almost one hundred years in Great Britain, Ireland, Portugal, significant cities of United States; and for some fifty/forty years elsewhere in Europe and America. These settings allow people to make decisions based on clock time which are rationally linked to the cycle of light and dark.\cite{Martin-Olalla2018}.

This manuscript is aimed to inspect how these regulations impacted in human life through sensitive decisions. Time use surveys ---large scale studies ($N\sim\num{.e4}$) which try to ascertain how people share a standard day\cite{chenu2006}--- are the appropriate tool for that. This paper will analyze the case of five industrialized countries (United States, Spain, Italy, France and United Kingdom), where seasonal time arrangements have occurred in the forty years prior to the date of the survey. The two most basic daily cycles will be studied: the sleep/wake cycle ---regulated by our physiology--- and the labor cycle ---strongly related to our social life---. The basic goal is to assess seasonal deviations in these cycles including their statistical significance and a sketch of the impact of latitude. It must be emphasized that this manuscript will focus on the long term ---seasonal--- effects of the regulation of time, not on effects around the transitions.

No counterfactual will be considered here. The obvious choice is a similar analysis of the time use surveys in independent countries where DST has not been applied regularly. The Republic of Korea and Japan are the best candidates.

\section{Methods}
\label{sec:methods}

\subsection{The data sets}
\label{sec:data-sets}

Microdata from time use surveys in  Spain\cite{estus-2010e} and United States\cite{ustus-2012} are freely available at the internet. Microdata from surveys in Italy\cite{ittus-2010e}, France\cite{frtus-2010e}, Great Britain\cite{uktus-2003} could be obtained after petition to public authorities in 2014.

Every respondent of a time use survey filled a diary which consisted of $N_0=\num{144}$ time slots ---each one representing ten minutes--- or indexes. It is possible to discriminate the date of the diary at least to the level of a trimester and the day of the week. In this work the year will be partitioned into two semesters or seasons: summer (April to September, trimesters two and three) and winter (October to March, trimesters one and four). Notice that October is counted in the winter season although regulations at the time of the surveys extend summer time until end of October (Europe) or until start of November (United States). It is also assumed that seasonal partition does not introduce a bias in the distribution of chronotypes and economic sectors.

\begin{table*}
  \centering
\footnotesize\sf
\setlength{\tabcolsep}{4pt}
	\begin{tabular}{llccccrrrcr}
	\toprule
	Time Use Survey&&Time&Latitude&Rise/set&Inso efficiency&\multicolumn{3}{c}{Sample size}&\multicolumn{1}{c}{Ratio}&\\
	&&offset&$\phi$&spread&$\cos\theta_w/\cos\theta_s$&\multicolumn{1}{c}{$N_t$}&\multicolumn{1}{c}{$N_w$}&\multicolumn{1}{c}{$N_s$}&\multicolumn{1}{c}{$N_s/N_w$}&\multicolumn{1}{c}{$\sqrt{N_h}$}\\
	\midrule
	\textbf{Great Britain}\cite{uktus-2003}&&\SI{6}{\minute}&\ang{52.3}&04h39m&$\SI{25}{\percent}/\SI{88}{\percent}$&&&&&\\
	\multicolumn{2}{l}{1. Mon-Fri (employees)}&&&&&\num{4081}&\num{1745}&\num{2336}&\num{1.34}&\num{44.7}\\
	\multicolumn{2}{l}{2. Mon-Fri (non-employees) age $\geq20$}&&&&&\num{3768}&\num{1481}&\num{2287}&\num{1.54}&\num{42.4}\\
	\multicolumn{2}{l}{3. Sat-Sun (employees)}&&&&&\num{1378}&\num{599}&\num{779}&\num{1.30}&\num{26.0}\\
	\multicolumn{2}{l}{4. Sat-Sun (non-employees), age $\geq20$}&&&&&\num{6473}&\num{2611}&\num{3862}&\num{1.48}&\num{55.8}\\
	\multicolumn{10}{c}{ }\\
	\textbf{France}\cite{frtus-2010e}&&\SI{50}{\minute}&\ang{47.8}&03h37m&$\SI{32}{\percent}/\SI{91}{\percent}$&&&&&\\
	\multicolumn{2}{l}{1. Mon-Fri (employees)}&&&&&\num{6243}&\num{2872}&\num{3371}&\num{1.17}&\num{55.7}\\
	\multicolumn{2}{l}{2. Mon-Fri (non-employees) age $\geq20$}&&&&&\num{6095}&\num{2622}&\num{3473}&\num{1.32}&\num{54.7}\\
	\multicolumn{2}{l}{3. Sat-Sun (employees)}&&&&&\num{1890}&\num{877}&\num{1013}&\num{1.16}&\num{30.7}\\
	\multicolumn{2}{l}{4. Sat-Sun (non-employees), age $\geq20$}&&&&&\num{8544}&\num{3729}&\num{4815}&\num{1.29}&\num{64.8}\\
	\multicolumn{10}{c}{ }\\
	\textbf{Italy}\cite{ittus-2010e}&&\SI{11}{\minute}&\ang{43.6}&03h11m&$\SI{39}{\percent}/\SI{94}{\percent}$&&&&&\\
	\multicolumn{2}{l}{1. Mon-Fri (employees)}&&&&&\num{5576}&\num{2784}&\num{2792}&\num{1.00}&\num{52.8}\\
	\multicolumn{2}{l}{2. Mon-Fri (non-employees) age $\geq20$}&&&&&\num{5390}&\num{2719}&\num{2671}&\num{0.98}&\num{51.9}\\
	\multicolumn{2}{l}{3. Sat-Sun (employees)}&&&&&\num{3596}&\num{1805}&\num{1791}&\num{0.99}&\num{42.4}\\
	\multicolumn{2}{l}{4. Sat-Sun (non-employees), age $\geq20$}&&&&&\num{15707}&\num{7698}&\num{8009}&\num{1.04}&\num{88.6}\\
	\multicolumn{10}{c}{ }\\
	\textbf{Spain}\cite{estus-2010e}&&\SI{72}{\minute}&\ang{40.4}&02h49m&$\SI{44}{\percent}/\SI{96}{\percent}$&&&&&\\
	\multicolumn{2}{l}{1. Mon-Fri (employees)}&&&&&\num{4271}&\num{2053}&\num{2218}&\num{1.08}&\num{46.2}\\
	\multicolumn{2}{l}{2. Mon-Fri (non-employees) age $\geq20$}&&&&&\num{4889}&\num{2364}&\num{2525}&\num{1.07}&\num{49.4}\\
	\multicolumn{2}{l}{3. Sat-Sun (employees)}&&&&&\num{971}&\num{505}&\num{466}&\num{0.92}&\num{22.0}\\
	\multicolumn{2}{l}{4. Sat-Sun (non-employees), age $\geq20$}&&&&&\num{4965}&\num{2431}&\num{2534}&\num{1.04}&\num{49.8}\\
	\multicolumn{10}{c}{ }\\
	\textbf{United States}\cite{ustus-2012}&&\SI{8}{\minute}&\ang{38.5}&02h36m&$\SI{47}{\percent}/\SI{97}{\percent}$&&&&&\\
	\multicolumn{2}{l}{1. Mon-Fri (employees)}&&&&&\num{36498}&\num{18629}&\num{17869}&\num{0.96}&\num{135.1}\\
	\multicolumn{2}{l}{2. Mon-Fri (non-employees) age $\geq20$}&&&&&\num{25589}&\num{13014}&\num{12575}&\num{0.97}&\num{113.1}\\
	\multicolumn{2}{l}{3. Sat-Sun (employees)}&&&&&\num{14775}&\num{7631}&\num{7144}&\num{0.94}&\num{85.9}\\
	\multicolumn{2}{l}{4. Sat-Sun (non-employees), age $\geq20$}&&&&&\num{47877}&\num{24408}&\num{23469}&\num{0.96}&\num{154.7}\\
	\multicolumn{10}{c}{ }\\
	\bottomrule
	\end{tabular}
  \caption{List of participant countries with geophysical information: time offset, latitude $\phi$, the spread of sunrise/sunset times and the efficiency of insolation at noon in winter $\cos\theta_w$ and summer $\cos\theta_s$. Notice that DST regulations apparently increase the spread of sunset times by one hour, decrease the spread of sunrise times by one hour. Also time offset increases by one hour after DST onset. The table also lists the sample size for every group of respondents. $N_t$ is the whole number of respondents in group, $N_s$ stands for the number of summer respondents, $N_w$ is the number of winter respondents,  $N_s/N_w$ is the ratio and $\sqrt{N_h}$ is the square root of the harmonic mean of $N_s$ and $N_w$.}
  \label{tab:geo}
\end{table*}

Diaries will be grouped in four items: (1) week day (Monday to Friday) diaries that report some working activity; (2) week day diaries that do not report any working activity; (3) week-end (Saturday and Sunday) diaries that report some working activity; and (4) week-end diaries that do not report any working activity.

Respondents in groups 1 and 3 will be labeled as ``employees'' hereafter. Groups 2 and 4 will only consider diaries filled by respondents aged at least twenty years; they will be labeled as ``non-employees''.

Group 1 encloses working respondents in a week day. They are least prone to free preferences and most prone to tight social timing, including the use of alarm clocks for getting activated. Group 4 is the most populated in every survey. It encloses non-working respondents in a week-end. It does not distinguish between those who worked during week days and those who not. Either case, respondents in this group are most prone to free preferences ---for instance rising up at their will---. Contrastingly Group 3 is the least populated since the labor activity in week-ends is comparatively small. Table~\ref{tab:geo} lists sample sizes of every survey and group analyzed.

In the cases of Spain, France and United States only contiguous regions will be analyzed. Also Arizona respondents will not enter in the study since this state does not observe DST regulations.

 Eventually the seasonal index will be permuted, re-sampled or re-shuffled and two partitions of the permuted samples will be compared to the unshuffled, seasonal, original partition to set the statistical significance of seasonal deviations.
\subsection{The mathematical framework}
\label{sec:math-fram}

The sleep/wake cycle ---being awake or sleeping--- and the labor cycle ---being working or not working--- will be studied. For either cycle two kind of magnitudes will be characterized: (1) the daily rhythm $R(i)$ ---the shares of respondents doing a prescribed activity on a given index--- and (2) the experimental (sample) cumulative distribution function $P(i)$ and sample average value $E$ of a set of five stochastic variables which characterizes each cycle for every individual. Figure~\ref{fig:acti} in Supplementary Material sketches this set which is composed of:
\begin{itemize}
\item The daily duration of the activity $d_1$: the daily working time and the daily sleep or wake time.
\item The center of gravity $t_2$: the moment when half the daily duration has been consumed and half remains.
\item The first occurrence of the activity $t_1$ or onset time.
\item The last occurrence of the activity $t_3$ or offset time.
\item The distance from onset to offset $d_2$.
\end{itemize}
For these quantities the cycle begins at 4am, around the point of daily minimum human activity. If the cycle is unimodal ---just one period of activity, with no breaks--- the center of gravity matches to the midpoint from onset to offset. 

The domain of any of these stochastic variables consists of the $N_0$ indexes that fill one cycle (day). Therefore they all are discrete variables.

Seasonal differences will be obtained by subtracting results from winter respondents to results form summer respondents. The statistical significance of the seasonal deviations $R_s(i)-R_w(i)$ will be determined by the Welch's $t$-test for unequal variances\cite{Welch1947}. The statistical significance of $P_s-P_w$ and $E_s-E_w$ will be determined by analyzing random permutations of the seasonal index. The size of the permutation test will be $M=\num{.e7}$, much smaller than the whole number of permutations $\sim10^N$ and large enough for the purpose. The $p$-value will be the fraction of random permutations which yields larger deviations than the tested seasonal value, thus questioning whether the observed difference is seasonal or random. Significance will be set at the standard level $\alpha=\SI{5}{\percent}$. This test will be one-tailed to assess first-order stochastic ordering.

It should be noticed that for $R$, $t_1$, $t_2$ and $t_3$ the seasonal deviation will compare $t$ to $t-\SI{1}{\hour}$ if $t$ is given as a universal time and DST regulations apply. 

For the purpose of comparing samples with different sizes $R_s-R_w$ will be scaled by variance and sample size to get the Welch's statistics of the deviations and $P_s-P_w$ will be scaled by sample size to get the Kolmogorov-Smirnov statistics of the deviations. In Supplementary material the mathematical framework is thoroughly described (see Section~\ref{sec:mathframe}).

\subsection{The geophysical framework}
\label{sec:geophys-fram-1}

Summer time arrangements are related to seasons, which arises from the obliquity $\varepsilon=\ang{23.5}$ of Earth's rotation axis relative to the revolution axis. Without obliquity there would be no seasons. The Sun would climb up to the same zenith angle day after day. Sunrises and sunsets will happen with a recurrence of $T=\SI{24}{\hour}$ (Earth's rotation period) due East (dawn) and due West (dusk) year round.

\begin{figure*}[tb]
  \centering
  \includegraphics[width=\textwidth]{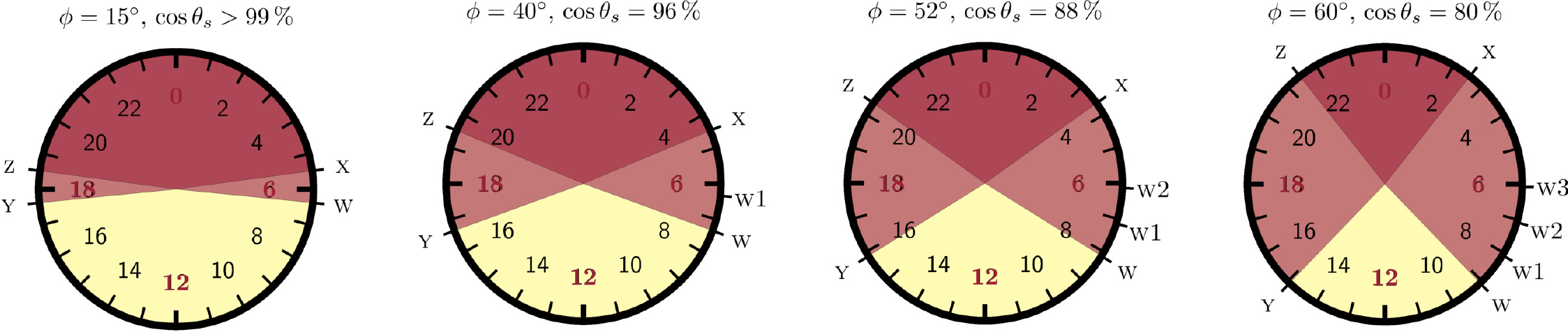}
  \caption{The seasonal cycle of light and dark depicted in four clocks with \textsc{24-h} analog dial. Clocks are located at circles of latitude $\phi=\ang{15}, \ang{40}, \ang{52}, \ang{60}$. Every clock locates morning hours on the right half, from \textsc{0} (midnight) to \textsc{12} (noon); afternoon hours (\textsc{12-24}), on the left half. The equinoctial night is located on the top half, from \textsc{18} (sunset) to \textsc{6} (sunrise) and the equinoctial daytime, on the bottom half. The red ink displays permanent night, the yellow ink displays permanent daytime, and the pink ink displays the region where light and dark alternate seasonally. Point \textsc{z} denotes summer sunset; \textsc{x}, summer sunrise; \textsc{y}, winter sunset; \textsc{w}, winter sunrise; and the number in \textsc{w}$n$ annotates hours before winter sunrise. Notice that \textsc{w} and \textsc{z} are separated by some \SI{12}{\hour} irrespective of latitude. Clock faces display mean solar time and local time at time meridians. Time offset ---see Table~\ref{tab:geo}--- must be added to find local time values or, alternatively, clock faces must be turned counterclockwise one degree per every four minutes of time offset.}
  \label{fig:clocks}
\end{figure*}

With non-zero obliquity every of these magnitudes change from date to date in a quantity prescribed solely by latitude. Table~\ref{tab:geo} lists time offset, latitude $\phi$, the spread of sunrise/sunset times from summer to winter, and the cosine of the solar zenith angle at noon in winter $\theta_w$ and summer $\theta_s$. Instead of the bare angle $\theta$, its  cosine is a more sensitive magnitude for understanding human behavior since insolation is proportional to it. Figure~\ref{fig:clocks} summarizes the spread of sunrise and sunset times at four representative circles of latitude helped by clocks with \textsc{24-h} analog dial.

In Supplementary material (Section~\ref{sec:geophys-fram}) the seasonal variations are thoroughly described.

Notice that as per DST regulations, time offset increases by one whole hour in spring, and sets back to the value listed in Table~\ref{tab:geo} in autumn; the spread of sunrise apparently decreases by one whole hour as measured local time, while the spread of sunset apparently increases by one whole hour.

\section{Results}
\label{sec:results}

\begin{figure*}[!t]
  \centering
  \includegraphics[height=.85\textheight]{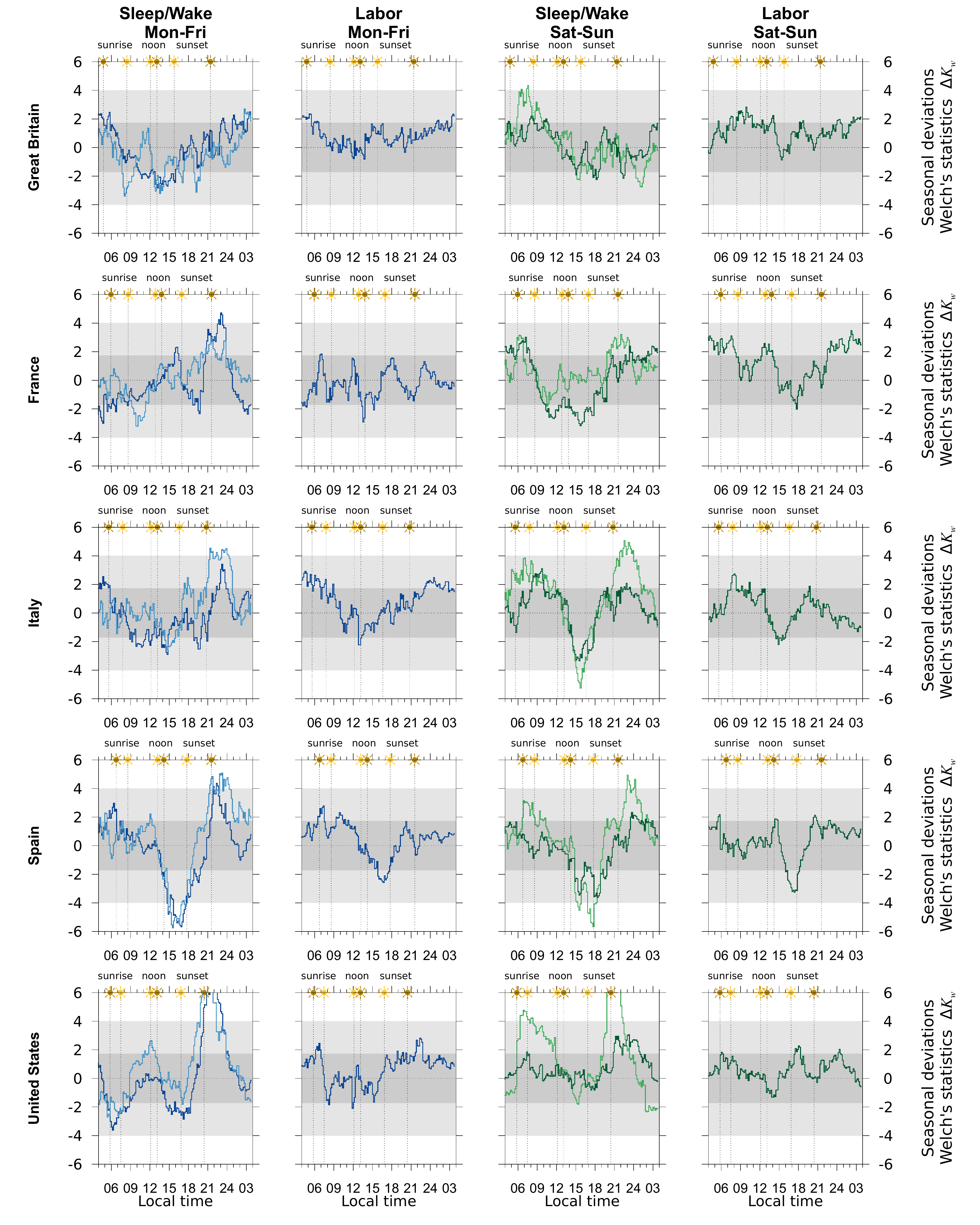}
  \caption{Normalized seasonal deviations $\Delta K_w=\sqrt{\mathcal{N}}\cdot(R_s-R_w)/s_{\text{rms}}\sqrt{2}$. Blueish inks display week-day statistics (groups 1 and 2). Greenish inks display week-end statistics (groups 3 and 4). Darker lines apply to employees (groups 1 and 3); lighter lines to non-employees (groups 2 and 4). The darker horizontal gray band highlight the region where the Welch's $t$-test sustains the null hypothesis $H_0:R_s-R_w=0$ at the standard level of significance $\alpha=\SI{5}{\percent}$. The lighter horizontal gray bands highlight the region $10^{-7}\leq p<\alpha$. For $\Delta K_w=6$ the $p$-value equals to $10^{-8.69}$. Vertical lines show solar ephemerides (sunrise, noon and sunset) in winter and summer as measured in local time. Darker sun applies to summer.}
  \label{fig:devrhythm}
\end{figure*}

Normalized seasonal differences of daily rhythms are shown in Figure~\ref{fig:devrhythm}. As Supplementary material seasonal daily rhythms from which seasonal differences were computed are shown in Figure~\ref{fig:rhythmlv}. There, seasonal differences are usually hard to visualize. Sleep/wake daily rhythm looks like a window function showing the overwhelming preference for monophasic sleep ---one continued period of sleep, followed by one continued period of wake per day--- and a great deal of coordination in the transitions, with the wake tied to sunrise and sleep determined by homeostasis\cite{Martin-Olalla2019b}. The only deviation from this behavior occurs in Italy and Spain; it is due to \emph{siesta}, an afternoon, short take of sleep. Labor daily rhythms are notably more complex.

Seasonal deviations in thirty daily rhythms are shown in Figure~\ref{fig:devrhythm}. They are arranged in $5\times 4$ panels. On every row a time use survey is presented; on every column a daily rhythm: sleep/wake in a week-day (groups 1 and 2), labor in a week-day (group 1) and then the same for a week-end. Darker lines refer to employees, lighter lines to non-employees. Solar ephemerides ---sunrise, noon and sunset--- are noted by dotted vertical lines and range from summer to winter values.

Darker horizontal gray bands highlight the region where Welch's $t$-test would sustain the null hypothesis $H_0:R_s(i)-R_w(i)=0$ at the standard level of significance. Lighter bands highlight the region where $\num{.e-7}<p<\alpha$. As Supplementary material Figure~\ref{fig:prhythm} shows the corresponding values $p(i)$ of these seasonal deviations, and Table~\ref{tab:laborWelchRandom} reports the occurrence of significant deviations $p(i)<\alpha$ in morning and afternoon hours.

\begin{table*}[t!]
  \centering
\footnotesize\sf
\setlength{\tabcolsep}{5pt}
	\begin{tabular}{lccccc}
	\toprule
	$\Delta K_w$ breakdown&\multicolumn{1}{c}{\begin{tabular}[t]{@{}c@{}}Group 1\\Mon-Fri\\employees\end{tabular}}&\multicolumn{1}{c}{\begin{tabular}[t]{@{}c@{}}Group 2\\Mon-Fri\\non-employees\end{tabular}}&\multicolumn{1}{c}{\begin{tabular}[t]{@{}c@{}}Group 3\\Sat-Sun\\employees\end{tabular}}&\multicolumn{1}{c}{\begin{tabular}[t]{@{}c@{}}Group 4\\Sat-Sun\\non-employees\end{tabular}}&All groups\\
	\multicolumn{1}{r}{$p<\alpha$}&\textsc{am/pm}&\textsc{am/pm}&\textsc{am/pm}&\textsc{am/pm}&total\\
	\midrule
	\multicolumn{6}{l}{\textbf{Sleep/wake}}\\
	Great Britain&\textcolor{black}{$\num{17}/\num{20}$}&\textcolor{black}{$\num{16}/\num{16}$}&\textcolor{black}{$\num{4}/\num{4}$}&\textcolor{black}{$\num{31}/\num{3}$}&\textcolor{black}{$\num{111}$}\\
	France&\textcolor{black}{$\num{14}/\num{21}$}&\textcolor{black}{$\num{13}/\num{11}$}&\textcolor{black}{$\num{31}/\num{24}$}&\textcolor{black}{$\num{7}/\num{21}$}&\textcolor{black}{$\num{142}$}\\
	Italy&\textcolor{black}{$\num{16}/\num{17}$}&\textcolor{black}{$\num{7}/\num{26}$}&\textcolor{black}{$\num{11}/\num{19}$}&\textcolor{black}{$\num{43}/\num{32}$}&\textcolor{black}{$\num{171}$}\\
	Spain&\textcolor{black}{$\num{6}/\num{44}$}&\textcolor{black}{$\num{13}/\num{48}$}&\textcolor{black}{$\num{0}/\num{26}$}&\textcolor{black}{$\num{18}/\num{39}$}&\textcolor{black}{$\num{194}$}\\
	United States&\textcolor{black}{$\num{21}/\num{44}$}&\textcolor{black}{$\num{12}/\num{34}$}&\textcolor{black}{$\num{7}/\num{15}$}&\textcolor{black}{$\num{45}/\num{24}$}&\textcolor{black}{$\num{202}$}\\
	\midrule
	\multicolumn{6}{l}{\textbf{Labor}}\\
	Great Britain&\textcolor{black}{$\num{13}/\num{0}$}&&\textcolor{black}{$\num{22}/\num{2}$}&&\textcolor{black}{$\num{37}$}\\
	France&\textcolor{black}{$\num{0}/\num{3}$}&&\textcolor{black}{$\num{48}/\num{16}$}&&\textcolor{black}{$\num{67}$}\\
	Italy&\textcolor{black}{$\num{31}/\num{4}$}&&\textcolor{black}{$\num{6}/\num{1}$}&&\textcolor{black}{$\num{42}$}\\
	Spain&\textcolor{black}{$\num{10}/\num{9}$}&&\textcolor{black}{$\num{2}/\num{13}$}&&\textcolor{black}{$\num{34}$}\\
	United States&\textcolor{black}{$\num{4}/\num{7}$}&&\textcolor{black}{$\num{1}/\num{4}$}&&\textcolor{black}{$\num{16}$}\\
	\midrule
	\end{tabular}
  \caption{Breakdown of statistically significant deviations in $\Delta K_w$ after a Welch's $t$-test of unequal variance was performed. Significance is taken at the standard level $\alpha=\SI{5}{\percent}$. Every group of respondents splits occurrences in morning/afternoon (\textsc{am/pm}) indexes. The remaining indexes up to a total of $\num{72}/\num{72}$ sustained the null hypothesis $H_0: R_s(i)-R_w(i)=0$.  The right-most column list the whole number of excursions $p<\alpha$, from a total of $4N_0=\num{576}$ (sleep/wake) and $2N_0=\num{288}$ (labor) analysis. In Spain \textsc{am/pm} divide was set at \textsc{1pm} local time, see time offset in Table~\ref{tab:geo}.}
  \label{tab:laborWelchRandom}
\end{table*}

\begin{figure*}[t!]
  \centering
  \includegraphics[height=0.75\textheight]{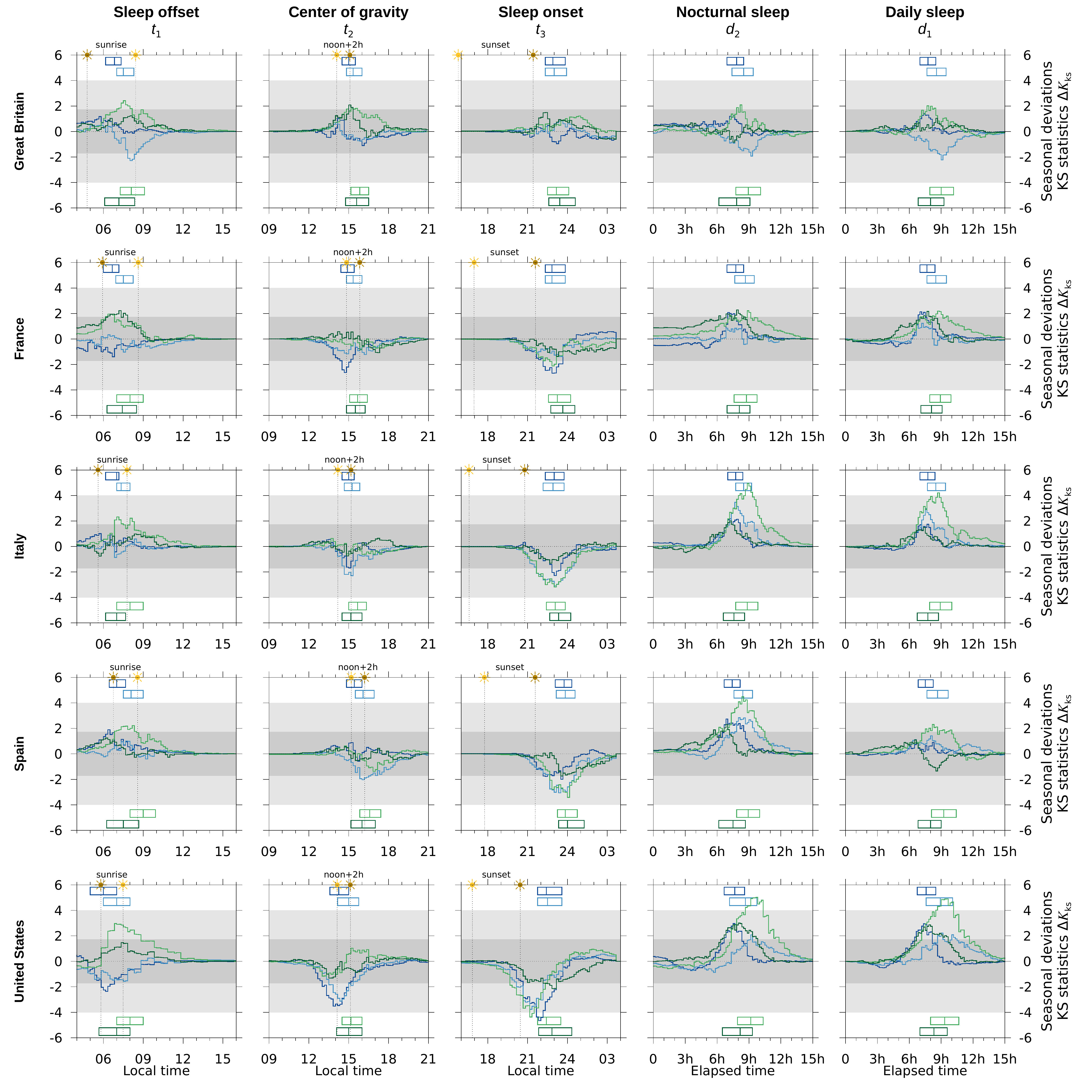}
  \caption{Normalized seasonal deviations $\Delta K_{ks}=\sqrt{N_h}\cdot(P_s-P_w)$ for the stochastic variables related to the sleep/wake cycle. Along a row a time use survey is shown, along a column one variable is shown. Blueish lines show week-days respondents. Greenish lines show week-end respondents. Darker lines show employees; lighter lines, non-employees. Boxes locate first, second and third quartiles of the distributions. For sided tests on continuous samples, which is never the case here, the darker horizontal gray band highlights the region where $p<\alpha$; the lighter band, the region $10^{-7}<p<\alpha$; and for $\Delta K_{ks}=6$ the $p$-value equals $10^{-15.7}$. In the left most three columns vertical dotted lines locate solar ephemerides ---sunrise, noon, sunset--- in winter and summer. Darkest sun applies to summer.}
  \label{fig:devsueno}
\end{figure*}

\begin{figure*}[t!]
  \centering
  \includegraphics[height=.75\textheight]{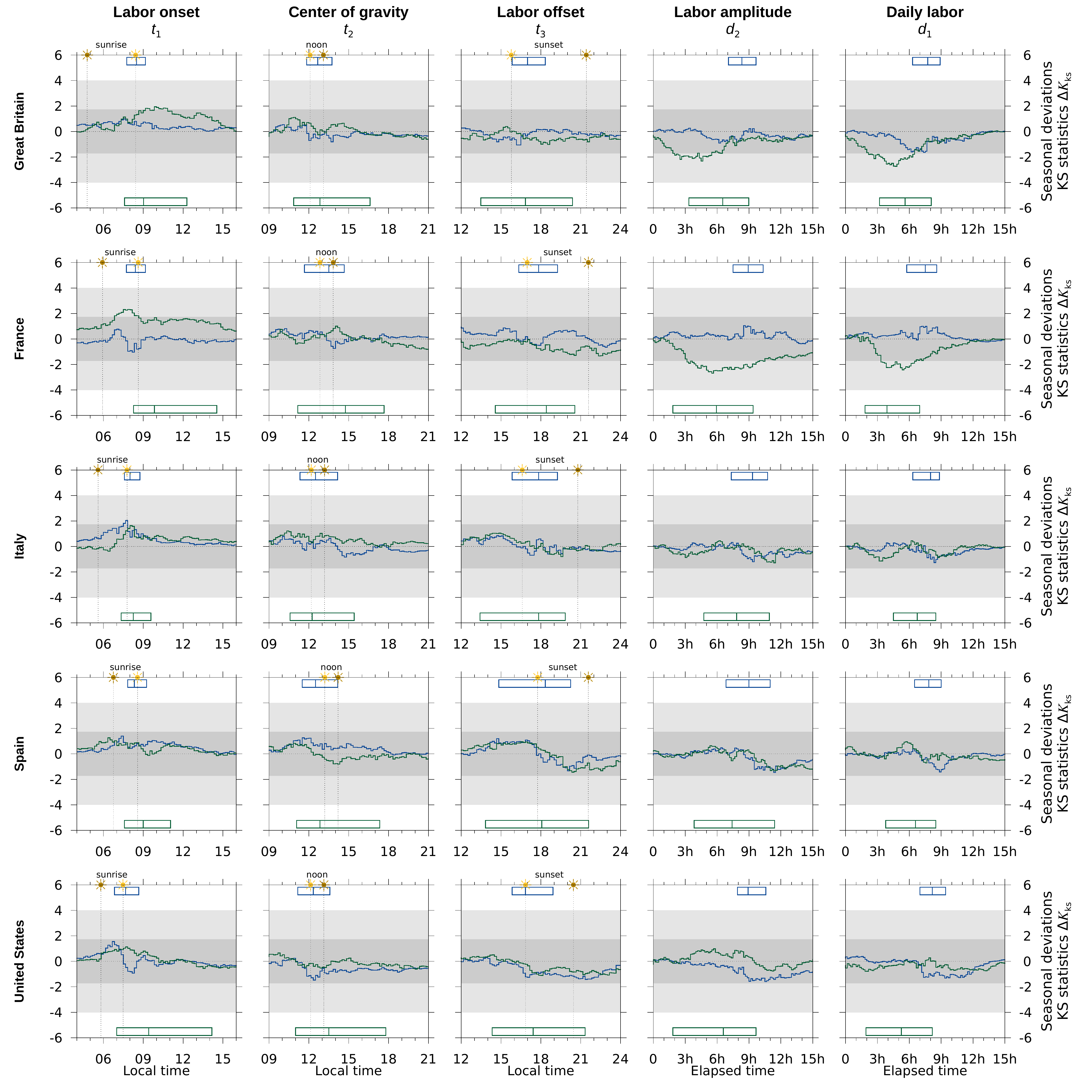}
  \caption{Same as Figure~\ref{fig:devsueno} but for the stochastic variables related to the labor cycle.}
  \label{fig:devlabor}
\end{figure*}

Likewise, Figure~\ref{fig:devsueno} (sleep/wake) and Figure~\ref{fig:devlabor} (labor) show the normalized seasonal deviations  of the stochastic variables obtained from each cycle. As Supplementary material  Figure~\ref{fig:pwakeemp} (sleep/wake cycle) and Figure~\ref{fig:plabor} (labor cycle) show the seasonal sample cumulative distribution functions from which seasonal differences were computed; in most of the circumstances seasonal differences are hard to visualize. 

Figure~\ref{fig:devsueno} displays one-hundred seasonal deviations and Figure~\ref{fig:devlabor}, fifty. They are arranged in $5\times 5$ panels each showing results from four groups (sleep/wake) and two groups (labor). Vertically, panels display the results of one stochastic variable: first three columns display time marks, last two columns durations; along a row panels display the results of one time use survey. On every panel the horizontal axis displays the variable ---either a local time or an elapsed time--- and  the vertical axis displays the Kolmogorov-Smirnov distance, see Eq.~(S8).

As in Figure~\ref{fig:devrhythm} blueish lines refer to week-day groups; greenish lines to week-end groups; darker lines refer to employees; lighter lines to non-employees. Finally, boxes locate the first, median and third quartile of the distributions; and for time-marks ---three right most columns--- the panel also displays solar ephemerides: sunrise, sunset and noon in winter and summer (vertical dotted lines).

Unlike in Figure~\ref{fig:devrhythm}, the direction of deviations in Figure~\ref{fig:devsueno} and Figure~\ref{fig:devlabor} is easy to address: positive excursions ($P_s>P_w$) indicate that the summer distribution advances relative to the winter distribution, earlier times for time-marks, shorter durations for elapsed times. Negative excursions ($P_s<P_w$) indicate the opposite. One single peak usually indicates a shift in the mean of the distributions.

The darker gray band highlights the region where the null hypothesis $H_0: P_s-P_w=0$ would sustain at the standard level of significance if the distribution were continuous, which was never the case here. The lighter gray strip highlight the region where  $\num{.e-7}<p<\alpha$ still in the continuous case.

 Table~\ref{tab:sleepKScounts} lists the occurrences of the four possible outcomes of the one-tailed permutation test: $P_s=P_w$, $P_s\neq P_w$, $P_s>P_w$ and $P_s<P_w$. A score is obtained  after assigning $+1$ to  $P_s>P_w$, $-1$ to $P_s<P_w$ and $0$ otherwise. As Supplementary material Table~\ref{tab:sleepKS} (sleep/wake) and Table~\ref{tab:laborKS} (labor) list the results of the permutation test for every of the \num{100} (sleep/wake) and \num{50} (labor) analyses. 

\begin{table*}[tb]
  \centering
\footnotesize\sf
\setlength{\tabcolsep}{3pt}
	\begin{tabular}{lccccrccccrccccrccccrccccr}
	\toprule
	 Counts&\multicolumn{5}{c}{\begin{tabular}[t]{@{}c@{}}Group 1\\Mon-Fri\\employees\end{tabular}}&\multicolumn{5}{c}{\begin{tabular}[t]{@{}c@{}}Group 2\\Mon-Fri\\non-employees\end{tabular}}&\multicolumn{5}{c}{\begin{tabular}[t]{@{}c@{}}Group 3\\Sat-Sun\\non-employees\end{tabular}}&\multicolumn{5}{c}{\begin{tabular}[t]{@{}c@{}}Group 4\\Sat-Sun\\employees\end{tabular}}&\multicolumn{5}{c}{All Groups}\\
	&\multicolumn{4}{c}{$P_s\leftrightarrow P_w$}&\multicolumn{1}{c}{Score}&\multicolumn{4}{c}{$P_s\leftrightarrow P_w$}&\multicolumn{1}{c}{Score}&\multicolumn{4}{c}{$P_s\leftrightarrow P_w$}&\multicolumn{1}{c}{Score}&\multicolumn{4}{c}{$P_s\leftrightarrow P_w$}&\multicolumn{1}{c}{Score}&\multicolumn{4}{c}{$P_s\leftrightarrow P_w$}&\multicolumn{1}{c}{Score}\\
&$<$&$=$&$\neq$&$>$&&$<$&$=$&$\neq$&$>$&&$<$&$=$&$\neq$&$>$&&$<$&$=$&$\neq$&$>$&&$<$&$=$&$\neq$&$>$&\\
	\midrule
	Sleep offset&\num{2}&\num{2}&\num{0}&\num{1}&$\num[retain-explicit-plus=true]{-0.2}$&\num{1}&\num{2}&\num{1}&\num{1}&$\num[retain-explicit-plus=true]{+0.0}$&\num{0}&\num{3}&\num{0}&\num{2}&$\num[retain-explicit-plus=true]{+0.4}$&\num{0}&\num{0}&\num{0}&\num{5}&$\num[retain-explicit-plus=true]{+1.0}$&\num{3}&\num{7}&\num{1}&\num{9}&$\num[retain-explicit-plus=true]{+0.3}$\\
	Center of gravity&\num{3}&\num{2}&\num{0}&\num{0}&$\num[retain-explicit-plus=true]{-0.6}$&\num{3}&\num{2}&\num{0}&\num{0}&$\num[retain-explicit-plus=true]{-0.6}$&\num{0}&\num{4}&\num{0}&\num{1}&$\num[retain-explicit-plus=true]{+0.2}$&\num{0}&\num{4}&\num{0}&\num{1}&$\num[retain-explicit-plus=true]{+0.2}$&\num{6}&\num{12}&\num{0}&\num{2}&$\num[retain-explicit-plus=true]{-0.2}$\\
	Sleep onset&\num{4}&\num{1}&\num{0}&\num{0}&$\num[retain-explicit-plus=true]{-0.8}$&\num{4}&\num{1}&\num{0}&\num{0}&$\num[retain-explicit-plus=true]{-0.8}$&\num{2}&\num{3}&\num{0}&\num{0}&$\num[retain-explicit-plus=true]{-0.4}$&\num{4}&\num{1}&\num{0}&\num{0}&$\num[retain-explicit-plus=true]{-0.8}$&\num{14}&\num{6}&\num{0}&\num{0}&$\num[retain-explicit-plus=true]{-0.7}$\\
	Onset to offset&\num{0}&\num{1}&\num{0}&\num{4}&$\num[retain-explicit-plus=true]{+0.8}$&\num{1}&\num{0}&\num{1}&\num{3}&$\num[retain-explicit-plus=true]{+0.4}$&\num{0}&\num{1}&\num{0}&\num{4}&$\num[retain-explicit-plus=true]{+0.8}$&\num{0}&\num{0}&\num{0}&\num{5}&$\num[retain-explicit-plus=true]{+1.0}$&\num{1}&\num{2}&\num{1}&\num{16}&$\num[retain-explicit-plus=true]{+0.8}$\\
	Sleep time&\num{0}&\num{2}&\num{0}&\num{3}&$\num[retain-explicit-plus=true]{+0.6}$&\num{1}&\num{2}&\num{0}&\num{2}&$\num[retain-explicit-plus=true]{+0.2}$&\num{0}&\num{3}&\num{0}&\num{2}&$\num[retain-explicit-plus=true]{+0.4}$&\num{0}&\num{0}&\num{0}&\num{5}&$\num[retain-explicit-plus=true]{+1.0}$&\num{1}&\num{7}&\num{0}&\num{12}&$\num[retain-explicit-plus=true]{+0.6}$\\
	\midrule
	 All combined&\num{9}&\num{8}&\num{0}&\num{8}&$\num[retain-explicit-plus=true]{-0.0}$&\num{10}&\num{7}&\num{2}&\num{6}&$\num[retain-explicit-plus=true]{-0.2}$&\num{2}&\num{14}&\num{0}&\num{9}&$\num[retain-explicit-plus=true]{+0.3}$&\num{4}&\num{5}&\num{0}&\num{16}&$\num[retain-explicit-plus=true]{+0.5}$&\num{25}&\num{34}&\num{2}&\num{39}&$\num[retain-explicit-plus=true]{+0.1}$\\
	\midrule
\\
	Labor onset&\num{0}&\num{4}&\num{0}&\num{1}&$\num[retain-explicit-plus=true]{+0.2}$&&&&&&\num{0}&\num{3}&\num{0}&\num{2}&$\num[retain-explicit-plus=true]{+0.4}$&&&&&&\num{0}&\num{7}&\num{0}&\num{3}&$\num[retain-explicit-plus=true]{+0.3}$\\
	Center of gravity&\num{0}&\num{5}&\num{0}&\num{0}&$\num[retain-explicit-plus=true]{+0.0}$&&&&&&\num{0}&\num{5}&\num{0}&\num{0}&$\num[retain-explicit-plus=true]{+0.0}$&&&&&&\num{0}&\num{10}&\num{0}&\num{0}&$\num[retain-explicit-plus=true]{+0.0}$\\
	Labor offset&\num{0}&\num{5}&\num{0}&\num{0}&$\num[retain-explicit-plus=true]{+0.0}$&&&&&&\num{0}&\num{5}&\num{0}&\num{0}&$\num[retain-explicit-plus=true]{+0.0}$&&&&&&\num{0}&\num{10}&\num{0}&\num{0}&$\num[retain-explicit-plus=true]{+0.0}$\\
	Onset to offset&\num{0}&\num{5}&\num{0}&\num{0}&$\num[retain-explicit-plus=true]{+0.0}$&&&&&&\num{2}&\num{3}&\num{0}&\num{0}&$\num[retain-explicit-plus=true]{-0.4}$&&&&&&\num{2}&\num{8}&\num{0}&\num{0}&$\num[retain-explicit-plus=true]{-0.2}$\\
	Labor time&\num{0}&\num{5}&\num{0}&\num{0}&$\num[retain-explicit-plus=true]{+0.0}$&&&&&&\num{2}&\num{3}&\num{0}&\num{0}&$\num[retain-explicit-plus=true]{-0.4}$&&&&&&\num{2}&\num{8}&\num{0}&\num{0}&$\num[retain-explicit-plus=true]{-0.2}$\\
	\midrule
	 All combined&\num{0}&\num{24}&\num{0}&\num{1}&$\num[retain-explicit-plus=true]{+0.0}$&&&&&&\num{4}&\num{19}&\num{0}&\num{2}&$\num[retain-explicit-plus=true]{-0.1}$&&&&&&\num{4}&\num{43}&\num{0}&\num{3}&$\num[retain-explicit-plus=true]{-0.0}$\\
	\midrule
\\
	 Survey&\multicolumn{5}{c}{Sleep/Wake}&\multicolumn{5}{c}{}&\multicolumn{5}{c}{Labor}&\multicolumn{5}{c}{}&\multicolumn{5}{c}{}\\
&$<$&$=$&$\neq$&$>$&Score&&&&&&$<$&$=$&$\neq$&$>$&Score&&&&&&&&&&\\
	\midrule
	 Great Britain&\num{3}&\num{12}&\num{0}&\num{5}&$\num[retain-explicit-plus=true]{+0.1}$&&&&&&\num{2}&\num{7}&\num{0}&\num{1}&$\num[retain-explicit-plus=true]{-0.1}$&&&&&&&&&&\\
	 France&\num{5}&\num{5}&\num{2}&\num{8}&$\num[retain-explicit-plus=true]{+0.1}$&&&&&&\num{2}&\num{7}&\num{0}&\num{1}&$\num[retain-explicit-plus=true]{-0.1}$&&&&&&&&&&\\
	 Italy&\num{5}&\num{7}&\num{0}&\num{8}&$\num[retain-explicit-plus=true]{+0.1}$&&&&&&\num{0}&\num{9}&\num{0}&\num{1}&$\num[retain-explicit-plus=true]{+0.1}$&&&&&&&&&&\\
	 Spain&\num{5}&\num{6}&\num{0}&\num{9}&$\num[retain-explicit-plus=true]{+0.2}$&&&&&&\num{0}&\num{10}&\num{0}&\num{0}&$\num[retain-explicit-plus=true]{+0.0}$&&&&&&&&&&\\
	 United States&\num{7}&\num{4}&\num{0}&\num{9}&$\num[retain-explicit-plus=true]{+0.1}$&&&&&&\num{0}&\num{10}&\num{0}&\num{0}&$\num[retain-explicit-plus=true]{+0.0}$&&&&&&&&&&\\
\midrule
	 All combined&\num{25}&\num{34}&\num{2}&\num{39}&$\num[retain-explicit-plus=true]{+0.1}$&&&&&&\num{4}&\num{43}&\num{0}&\num{3}&$\num[retain-explicit-plus=true]{-0.0}$&&&&&&&&&&\\
	\bottomrule
	\end{tabular}

  \caption{Occurrences of hypotheses $P_s>P_w$, $P_s\neq P_w$, $P_s=P_w$ and  $P_s<P_w$ after a permutation test of size $M=\num{.e7}$ in the seasonal indexes was performed. Detailed results are shown as Supplementary material in Table~\ref{tab:sleepKS} and Table~\ref{tab:laborKS}. The score is obtained after assigning $+1,0,0,-1$ to each outcome and scaling by the whole number of tests.}
  \label{tab:sleepKScounts}
\end{table*}

The occurrence of outcomes for the test on the average values $E_s-E_w$ is not different from results in Table~\ref{tab:sleepKScounts} save for the fact that inequalities are shifted: $P_s>P_w$ leads to $E_s<E_w$ and conversely. As Supplementary material  Table~\ref{tab:vigil} (sleep/wake cycle) and Table~\ref{tab:laborWelch} (labor cycle) lists the results for every of the analyses. Probability differences $P_s(i)-P_w(i)$ ---vertical differences in Figure~\ref{fig:pwakeemp}--- can be transposed into quantile or ``horizontal'' differences. As Supplementary material and for the sake of completeness Table~\ref{tab:sleepQQ} (sleep/wake) and Table~\ref{tab:laborQQ} (labor) list quartile differences.

\section{Discussion}
\label{sec:discussion}

Figure~\ref{fig:devrhythm} highlights the stability of the labor cycle in week-days ---column 2--- and week-ends ---column 4---. Seasonal deviations seldom climb up to $\num{2}$ and  the null  hypothesis $H_0: R_s(i)-R_w(i)=0$ is seldom rejected at the standard level of significance, see Table~\ref{tab:laborWelchRandom}. Labor cycle in a week-end (group 3) is slightly more prone to seasonal alterations, although it must be taken into account that week-end labor cycle differs significantly from week-day labor cycle, see Figure~\ref{fig:rhythmlv} and that sample sizes are smaller in group 3 by a factor ranging from $\num{2}$ to $\num{8}$, see Table~\ref{tab:geo}.

In the same way Figure~\ref{fig:devlabor} displays modest seasonal deviations in variables related to the labor cycle, which do not break the null hypothesis $H_0: P_s-P_w=0$ (see Table~\ref{tab:sleepKScounts}) at the standard level of significance as deduced from a permutation test of size $M=\num{.e7}$. Also it is seldom the case that the null hypothesis $H_0:E_s-E_w=0$  (see Table~\ref{tab:laborWelch}) is rejected. Therefore labor cycle is equally distributed through seasons in environments under DST regulations.

The stability of the labor cycle is probably the main silent goal and the main outcome of summer time regulations. Indeed time regulations have always been aimed towards social life, of which working hours is one of its most significant examples.

It is a question to understand the seasonal deviations in the labor cycle of  countries with the same degree of social development, latitude and not exposed to seasonal clock changing. The obvious choices would be the Republic of Korea and Japan. Irrespective of that the softest conclusion of the preceding results is that the labor cycle tolerates a seasonal shift of its phase equal to one hour in the direction prescribed by the regulations (advance in spring, delay in autumn). It is out of the scope of this manuscript to correlate this practice with economic outcomes like the Gross Domestic Product.

On its way the sleep/wake cycle is more significantly altered and it does so in different ways. Figure~\ref{fig:devrhythm} ---columns 1 and 3--- suffices to understand this idea: $\Delta K_w$ climbs up to $\num{4}$ or $\num{6}$, sometime towards the positive side, sometime towards the negative side, and excursions $p(i)<\alpha$ are more frequent, see Table~\ref{tab:laborWelchRandom}. It is also worthy to mention that excursions weigh on the afternoon, which makes sense taking into account the direction of DST regulations, which pushes for stabilizing morning times only. Finally Table~\ref{tab:sleepKScounts} lists occurrences of stochastic dominance in the sleep/wake cycle and in the labor cycle, and provides a score: differences among both cycles are self evident, with the sleep/wake cycle heavily exhibiting stochastic dominance and labor cycle reluctant to that.

Virtually every sleep/wake cycle in Figure~\ref{fig:devrhythm} is signaling a large excursion by the time of summer sunset, when the permanent night begins. Since, by then, the daily rhythm is decreasing as more people is coming to bed, and the excursion goes towards the positive side, a delay in $R_s$ is observed or $R_s$ fades out later as a result of later sunsets.

In the morning, smaller but still positive excursions are observed in some groups, notably in the week-end. They occur in the range when light and dark seasonally alternate. Since, by then, the daily rhythm is uprising, as more people is getting up, the positive excursion is now signaling an advance in $R_s$, which soars earlier.

Figure~\ref{fig:devsueno} shows these results in the form of probability distributions. Here sleep onset has generally a negative excursion, which leads to stochastic dominance $P_s<P_w$ or $P_s$ delays, while sleep offset display the opposite dominance $P_s>P_w$, or $P_s$ advances. Table~\ref{tab:sleepKScounts} provides a summary for this observation: sleep offset gets a positive (advanced) score; sleep onset, a negative negative (delayed) score. 

As for durations are concerned, sleep time and nocturnal sleep (onset to offset time) more than often decreases as a result of relatively delayed sleep onset and slightly advanced sleep offset. Both elapsed times get positive (decrease) scores in Table~\ref{tab:sleepKScounts}. Likewise the center of gravity tends to delay only in a few cases.

No matter how statistically significant any of these deviations can be, they have little significance in average values as seen in Table~\ref{tab:vigil}: $E_s-E_w$ lie in the range of few minutes, much smaller than the size of clock changing and the spread of sunrise time and sunset time. In the same way, quartile differences are most frequently equal or less than one time slot or ten minutes, see Table~\ref{tab:sleepQQ}.

Notwithstanding all this the fact that sleep offset and sleep onset seasonal deviations are opposite must be further discussed in relation to the impact of DST. If sleep onset and sleep offset had got the same type of stochastic dominance  ---be that an advance or a delay--- then it could be argued that the size of the seasonal change is too small ---if both tended to advanced, amplifying the advance of clock time--- or too large ---if both tended to delay, fighting against the advance of clock time.  Either way societies would be expressing a global preference for pushing or pulling the sleep/wake cycle. With stochastic dominance leaning in opposite directions such a preference can not be identified and DST can not be easily challenged. Indeed they are showing that sunrise and sunset still impact human behaviour, even after having altered clocks, and that the sleep/wake cycle is not misaligned by the regulations.

Since DST advances clocks only in spring, delays them only in autumn, thus pushing for stabilizing sunrise times as measured local time, morning results must be further discussed. As for the labor cycle is concerned the point to note is that DST does not prevent employers and employees from delaying their timing after the spring transition or from advancing their timing after the autumn transition. However results show that this is seldom the case. Also it is seldom the case that they need a further advance  in spring, or a further delay in winter. In view of that all, data show seasonal regulations of time operate timely, easing the preference for earlier times in summer and the preference for later times in winter. It also does so with a high degree of coordination.

As for the sleep/wake cycle is concerned the group most prone to free preferences ---non-employees in a week-end, group 4--- exhibits the largest shift of sleep offset in summer with $P_s>P_w$ in the five surveys (see Table~\ref{tab:sleepKScounts} and Table~\ref{tab:sleepKS} in Supplementary material). Historically winter time and winter activity is taken as the normal, since transitions were first set on spring. Upon this view, results suggest that  virtually no respondent in the surveys wants to delay summer morning duties, whereas a number of people pushes for an advance in summer, even though local time was already advanced by time regulations. This is compatible with the spontaneous preference for permanent summer time in polls\cite{Commission2018}, since the primary outcome of winter permanent time would be a delay in summer human activity if nothing else changes.

With only five surveys, the impact of latitude in seasonal behavior can only be sketched with some caution. The most apparent result is reported in Table~\ref{tab:laborWelchRandom}: the occurrence of significant deviations ($p(i)<\alpha$) in the sleep/wake cycle increases with decreasing latitude. In Figure~\ref{fig:devrhythm} and Figure~\ref{fig:devsueno} this is noted by the stability of sleep/wake cycles in Great Britain or France compared to those elsewhere, even though the spread of sunrise and sunset times is larger the higher the latitude. These results are in accordance with the impact of latitude in periods of low calling activity in mobile phone data of a large number ($N\sim\num{.e6}$) of individuals for short range of latitudes\cite{Monsivais2017a} ($\phi=\{\ang{37},\ang{40},\ang{42.5}\}$, see  Figure~2 in the Reference). There the lowest latitude showed the strongest seasonal variations in the periods.

The rationale is that human activity is less able to track large spreads of sunrise and sunset times. Therefore, even if the clock is regulated seasonally, societies at lower latitudes societies are willing to finely tune the sleep/wake cycle and solar activity, whereas at higher latitude they could be finding cues of synchronization by clock time only. As an example, a great deal of excursions $p(i)<\alpha$ in Spain and Italy come in the afternoon (see Figure~\ref{fig:devrhythm}) as a result of the seasonal changes in \emph{siesta}, a short take of sleep in the afternoon. \emph{Siesta} is a transfer of sleep time from morning to afternoon, much preferred in summer due to the high noon insolation, see Table~\ref{tab:geo}. It could be argued \emph{siesta} is favored by DST since it brings sleep offset closer to summer sunrise. But with the word ``siesta'' meaning ``six'' ---the ancient hour for noon--- a more sensitive description is that DST plus year round timing is the modern way in which this ancient seasonal behavior is achieved. A dip in human activity around noon is also found in the Tropics\cite{Siegmund1998,Yetish2015,Smit2019}, where insolation is similar.

In the same sense, an without overemphasizing the importance of historical records, a few of them put forward differences in latitude. For instance Willet's pamphlet\cite{Willet1907} (1907) advocating for DST shows British human activity delayed in summer by 1907 perhaps after it got synced to clock time.  On the contrary Cádiz Cortes, Spanish first National Assembly, opening and closing times shows in 1810 a seasonal regulation\cite{Luxan1810}  ---\textsc{10am} to \textsc{2pm} from October to April, but \textsc{9am} to \textsc{1pm} from May to September---, in every leg equivalent to modern DST, from which it is one hundred years ahead. Even Benjamin Franklin observations\cite{Franklin1784,Franklin2005} after moving (1784) from Pennsylvania ($\phi=\ang{40}$) to Paris ($\phi=\ang{49}$) may be showing the shocking changes of human behavior due to latitude.

Much more recently in 2015 Chile, which spread from $\phi\sim\ang{20}$ to $\phi\sim\ang{55}$,  tried to discontinue DST, switching to permanent summer time. The new layout was socially abhorred and only lasted for one winter. However, since 2017, the Magallanes Province ($\phi\sim\ang{53}$) sustains permanent summer time.

We are now in a position to address the relevant issue to which DST regulation is related: when should human activity start in view of the seasonal spread of sunrise and sunset times? To answer this question the clocks with \textsc{24-h} analog dial shown in Figure~\ref{fig:clocks} come handy. As for human activity is concerned data extracted from time use surveys and reported in Ref.~\cite{Martin-Olalla2018} ---see Figure~2 in this reference--- will help to sketch real scenarios. 

Sunrise has always been a lodestar for the start of human activity, even after the advent of efficient artificial light. Our physiology makes us prone to be activated by natural light and abhor getting activated too early in the morning darkness\cite{Borbely2016}.  In the Tropics sunrise and sunset times are pretty stable and insolation is high ---see the left most panel in Figure~\ref{fig:clocks}, which displays the case for $\phi=\ang{15}$---. Human activity in pre-industrial societies starts by the sunrise ---around \textsc{6am}--- despite photoperiod is twelve hours, as a way of preventing noon insolation.

As latitude increases winter sunrise \textsc{w}, which marks the starting point of permanent daytime, increasingly lags. At $\phi=\ang{40}$ it delays more than one hour with respect to the normal \textsc{6am}. Winter sunset \textsc{y} ---the end of permanent daytime--- comes $\SI{9.5}{\hour}$ later so that a great deal of human activity can be developed in daytime, irrespective of season. Not surprisingly \textsc{w} is the choice for human start of activity in United States and Spain. A paradigm runs as follows: school children can come to school after sunrise and can go before sunset year round.

In summer, the starting point \textsc{w} would delay  some $\SI{3}{\hour}$ from sunrise  ---now occurring at \textsc{x}---. Human activity finds relief in advancing activity, pushing towards the normal \textsc{6am} and keeping sunrise as a lodestar. This is met in modern times by DST regulations, which bring the start point to \textsc{w1} in summer if nothing else changes. A clear, ancient outcome for this shift is linked to latitude: it prevents exposition to noon insolation and overheating. The paradigm would be: there is no  need to subject school children to arrive to school some three hours after sunrise if insolation efficiency is going to climb up to $\SI{96}{\percent}$. A modern outcome is the trade of morning daytime leisure for afternoon daytime leisure. However this outcome is only related to the direction of the regulation and it is found elsewhere, irrespective of latitude. 

In this scenario permanent summer time policy pushes to make \textsc{w1} the starting point of activity year round. This is not always welcome at $\phi=\ang{40}$ since there is no need to transfer human activity into dawn in winter ---or  there is no need to subject school children to morning darkness if daytime lasts for more than $\SI{9.5}{\hour}$---. Indeed permanent summer time failed in United States and Portugal (1970s), and recently in Chile. To some extend it also failed in Spain, where permanent summer time in 1945 was met with a delay in social timing which brought the normal starting point back to \textsc{w}.

The seasonal arrangement \textsc{w}$\leftrightarrow$\textsc{w1} is also found at $\phi=\ang{52}$ in Great Britain and Ireland, despite \textsc{w} delays and \textsc{y} advances so that human activity increasingly occurs in the darkness  ---now, school children start being subjected to darkness, either at dawn or at dusk---. Some individuals may find relief in trading activity at dusk for activity at dawn. One way of so doing is a permanent summer time setting which advocates making \textsc{w1} the normal. That was the case in Saskatchewan (Canada, 1960) and Magallanes Province(Chile, 2017). It is a bet for non-seasonal behavior, without biannual transitions. It must be noted though that this choice failed in Great Britain and Ireland in the 1970s, due to morning darkness.

Nevertheless \textsc{w1} can become the normal starting point of activity just by advancing social timing, finding relief in earlier end times. That is the case in Germany and Poland. Still then, it is possible that DST regulations apply in summer. That would bring the start point in summer to \textsc{w2}. Long after the seasonal regulation of clocks occurred people may push for discontinuing DST. A permanent summer time would now  make \textsc{w2} the normal starting point. This is two hours before winter sunrise and it may come too early so that the choice may not sustain for a long time. As an example Russia switched to permanent summer time in 2011 and to permanent winter time in 2014.

At $\phi=\ang{60}$ latitude the spread of sunrise time is even larger. Winter sunrise (\textsc{w}) delays so much that it can hardly be the normal for the start of activity. Indeed Finland, Norway and Lithuania find start times come close to point \textsc{w2}, some two hours before winter sunrise, struggling to meet the normal \textsc{6am} and to find end times closer to a much advanced winter sunset \textsc{y}. DST regulations still advances human activity  in summer and brings the starting point to  \textsc{w3}. Finally, a push for permanent summer time in this scenario could also bring the start of the activity in winter too much advanced relative to winter sunrise, even for the standards at $\phi=\ang{60}$.

 Irrespective of latitude once human activity gets advanced \emph{in winter} relative to sunrise, summer sunset time puts pressure against seasonal clock changing: as the start of human activity shifts from \textsc{w} to \textsc{w1}, to \textsc{w2} and to \textsc{w3}, it comes close to the normal \textsc{6am} but also to summer sunset (point \textsc{z}, on the other side of the clock), which separates slightly more than twelve hours from \textsc{w}, irrespective of latitude. Indeed, with the start of activity at \textsc{w3} as in $\phi=\ang{60}$, summer sunset \textsc{z} is only some nine hours earlier. That struggles against human preference for going to bed in darkness.  However this setting shows good similarity with previous records from the Age of Enlightenment\cite{VanEgmond2019,Martin-Olalla2019d} at $\phi=\ang{60}$.

Figure~\ref{fig:clocks} and the preceding discussion sketch the intriguing way by which  historically artificial light has impacted human activity at dawn, after having easily impacted human activity at night. The current wave of DST discussion is just showing that a preference for a non-seasonal behavior may be becoming increasingly popular, specifically at high enough latitude, where sunrise ceases to be a lodestar and noon insolation is less intense. Helped by artificial light the starting point of activity can be appropriately set ahead of winter sunrise; thereafter people would just witness the formidable changes in ambient light conditions that happen from winter to summer. The shocking paradox is that both societies in the Tropics and societies at high latitudes could be more willing to share a preference for non-seasonal behavior, albeit for opposite reasons: too little and too large solar seasonal variations.

\section{Conclusion}
\label{sec:conclusion}

DST turned out to be the way by which seasonal variability was introduced into non-seasonal mechanical clocks, thus  pushing for ---but not forcing to--- a seasonal behavior in urban, industrialized societies. After decades of practice, it has successfully addressed a three-fold problem: the seasonal variations in the light and dark cycle, the inability of mechanical clocks to track this variability and the appetence for a regular way of life.

In so doing DST has promoted a highly coordinated seasonal adaptation that has matched to its layout: one-hour advance of human activity in summer relative to the winter activity. This is noted in time use surveys by the equal distribution of the labor cycle through seasons as measured local time. The sleep/wake cycle is more prone to seasonal deviations, specially at lower latitudes. Nonetheless these deviations are still linked to solar activity ---the advance of sunrise and the delay of sunset, and the changes in the efficiency of the insolation at noon--- which reveals that human activity is not misaligned by DST regulations.

Latitude weighs the scores of people pushing for non-seasonal behaviour.  A soft divide can be guessed around a latitude equal to $2\varepsilon=\ang{47}$ where the seasonal spread of sunrise/sunset time is a bit smaller than $\SI{4}{\hour}$ and the efficiency of the insolation at noon ($\cos\theta_s=\SI{92}{\percent}$) meets with the value at the Equator. Time use surveys reveal more frequent seasonal deviation below this line, which indicate a coupling with solar activity and a preference for seasonal regulation. Above the line, data reports less frequent seasonal deviations, which suggests a decoupling between social and solar activity and, ultimately, a cue that seasonal regulation of time could be more easily abandoned as latitude increases.

In view of all that the European Commission should be open to consider thoroughly the impact of latitude in summertime arrangements.

\section*{Acknowledgments}

 The author expresses his gratitude to the institutions which granted access to time use survey microdata. 

 The author wishes to thank Prof. Pablo Maynar Blanco and Prof. María Isabel García de Soria Lucena from Universidad de Sevilla for kind discussion on two-state systems.

 The author also wishes to thank Prof. Jorge Mira Pérez from Universidade de Santiago de Compostela for his support.

 The author thanks the Spanish think-tank Politikon \href{http://www.politikon.es}{http://www.politikon.es} for helping to disseminate ideas in this manuscript.
 
 Time use survey microdata were parsed with a \texttt{language C} code compiled by \texttt{gcc 5.4} and read in \texttt{octave 4.0}. This software was used to compute daily rhythms, the distribution of stochastic variables, the Welch's $t$-test and the permutation tests. All tabular material and data displayed on figures were automatically produced from the same input files. Border lines in figure~\ref{fig:azimut} were obtained from \href{https://www.naturalearthdata.com/}{https://www.naturalearthdata.com/} and the list of populated places in \href{https://www.geonames.org/}{https://www.geonames.org/} was used to compute population weighted median values of geographical data.

 Manuscript was originally written in \LaTeXe, typed in \texttt{GNU Emacs 24.5} assisted by \texttt{AucTeX 12.1}.   Graphs were produced by \texttt{gnuplot 5.1} \texttt{pdfcairo} terminal. \texttt{Mendeley-Desktop 1.9} helped handling bibliography. All this on three different computers each running a \texttt{Xubuntu 18.04 LTS Bionic Beaver} distro and synced by \texttt{ownCloud}. \texttt{Mendeley} and \texttt{ownCloud} services were locally provided by author's institution Universidad de Sevilla. 

  This project was started on Feb 27, 2019.

 \section*{Competing interests}
\label{sec:competing-interests}

The author declares no competing interest.

\section*{Author's contributions}
\label{sec:auth-contr}
It does not apply.
\section*{Funding}

This work was not funded.
\section*{Research Ethics}

It does not apply.
\section*{Animal Ethics}

It does not apply.

\section*{Permissions to carry out fieldwork}

\label{sec:data-availability}
Time Use Surveys are official surveys carried out by public Institutions. In US it is the Bureau of Labor Statistics and the United States Census Bureau. In UK it is the Office of National Statistics. They are a key tool for sociologists and economists to understand how we share time in a standard day. Any of them is a traceable, well-identified set of data.

Spanish and American Time Use Survey microdata were freely available in the internet at the time of writing this manuscript. A link to these surveys is provided in their corresponding references. 

French, British, and Italian Time Use Survey microdata were obtained after a formal request was addressed to the institutions. The microdata were then sent to the author by email, a hard-copy or as a downloadable link. In 2015, these institutions definitely were happy to send their microdata to any researcher who may need them.

\appendix
\setcounter{figure}{0}
\setcounter{table}{0}
\setcounter{subsection}{0}
\setcounter{section}{19}
\renewcommand{\thefigure}{\thesection.\arabic{figure}}
\renewcommand{\thetable}{\thesection.\arabic{table}}
\renewcommand{\thesubsection}{\thesection.\arabic{subsection}}

\section*{Supplementary material}
\label{sec:additional-figures}

\subsection{Mathematical framework}
\label{sec:mathframe}

Every respondent of a time use survey filled a diary which consisted of $N_0=\num{144}$ time slots ---each one representing ten minutes--- or indexes. For every index, respondents indicated which activity was being performed, where and with whom. The sleep/wake cycle and the labour cycle ---two of the most basic, universal human activities--- will be studied after identifying in every diary activity codes corresponding to sleeping or working.

An activity will be represented by a state function $A$ which can take only two values $a_1=1$, if the respondent is doing the activity at a given index, and $a_0=0$ if the respondent is not doing the activity. Therefore we deal with at two-state system.

The average value of $A$ is called the daily rhythm $R$ as it counts the shares of population doing the activity as a function of time. For a survey of size $N$ if $m$ individuals are doing the activity and $N-m$ are not doing the activity then:
\begin{equation}
  \label{eq:7}
 R(i)=\frac{m(i)\times a_1+(N-m(i))\times a_0}{N}=\frac{1}{N}m(i).
\end{equation}

Daily rhythms can be computed for winter $R_w$ and summer $R_s$ and then seasonal differences are given by $\Delta R (i)=R_s(i)-R_w(i)$. It is very important of understand that this seasonal difference are taken at constant $i$, that is at a given local time. Taking into account summer time regulations seasonal differences are computed for universal times differing one whole hour: $R_s(t)-R_w(t+\SI{1}{\hour})$.

If human activity were not seasonal then $\Delta R$ should fade to zero for large enough values of sample size. But with finite samples $\Delta R$ is experimentally non-zero and a test statistic is needed to sustain or reject the null hypothesis $H_0:R_s(i)-R_w(i)=0$, tested on every index $i$. The appropriate test in this case is the Welch's unequal variance $t$-test since $R$ is an average value and its variance can be obtained easily as:
\begin{equation}
  \label{eq:8}
  s(i)^2=R(i)\cdot(1-R(i)).
\end{equation}

The Welch's statistics  normalises the raw difference of average values $\Delta R$ by sample size and sample variance with the following formula:
\begin{equation}
  \label{eq:9}
  \Delta K_w(i)=\sqrt{\mathcal{N}(i)}\cdot\frac{\Delta R(i)}{s_{\text{rms}}(i)\sqrt{2}}
\end{equation}
where $s_{\text{rms}}(i)$ is the root mean square of the variances and $\mathcal{N}(i)$ is the harmonic average of sample sizes weighted by the variances:
\begin{equation}
  \label{eq:11}
  \mathcal{N}(i)=\frac{s_s^2(i)+s_w^2(i)}{\dfrac{s_s^2(i)}{N_s}+\dfrac{s_w^2(i)}{N_w}}
\end{equation}
As a result $\Delta K_w$ can compare differences from surveys and groups differing in sample size. Finally evaluating the cumulative distribution function of the $t$-Student at $\Delta K_w(i)$ a $p$-value is obtained for the null hypothesis $H_0: R_s(i)-R_w(i)=0$. Notice that $p(i)$. The number of degrees of freedom for this computus is  determined by the Welch-Satterthwaite equation:
\begin{equation}
  \label{eq:10}
  \nu(i)=\frac{\left(N_ws_w^2(i)+N_ss_s^2(i)\right)^2}{N_w^2s_w^4(i)+N_s^2s_s^4(i)}\mathfrak{N}(i)
\end{equation}
where $\mathfrak{N}(i)$ is the harmonic average of the sample sizes weighted by $Ns^2$.

An alternative description of an activity can be obtained from five stochastic variables that result from analysing the state of an activity during one cycle for every individual. A paradigmatic state function $A$ is sketched in  Figure~\ref{fig:acti}. The shaded area is the duration of the activity computed just by adding up all contributions to the activity in one diary:

\begin{figure*}
  \centering
\includegraphics[width=0.75\textwidth]{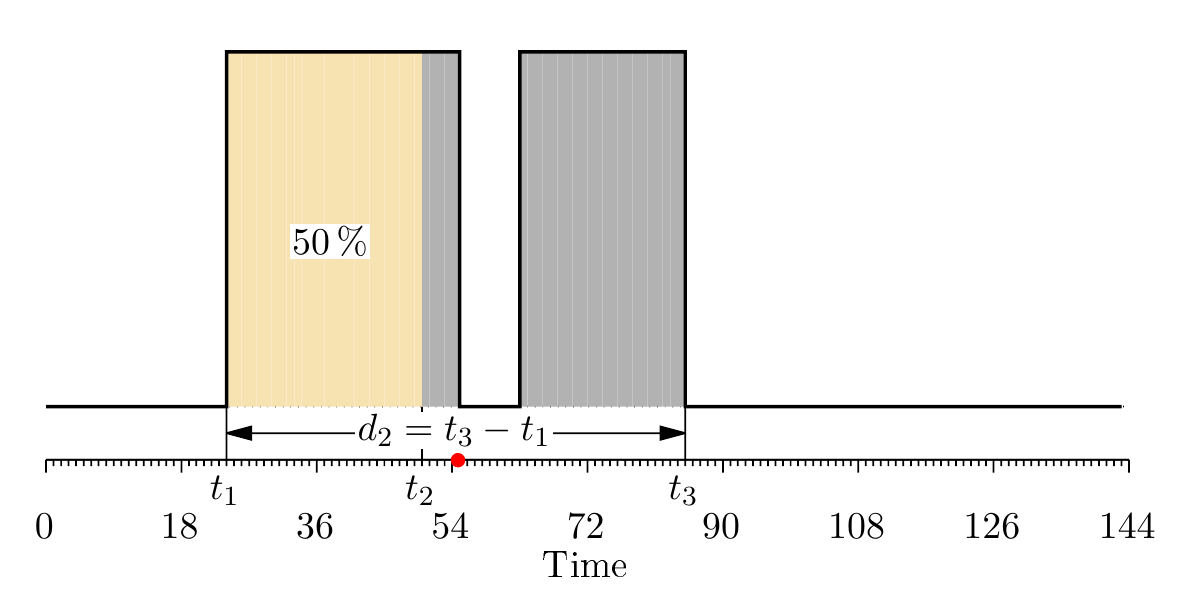}  
  \caption{A sketch of the state function for some activity and some individual. The function is activated (upper bound) or deactivated (lower bound). The shaded area computes the duration of the activity $d_1$: the first half is noted in a lighter ink; the second half, in a darker ink.  Onset time $t_1$ is the first occurrence of the activity. Offset time $t_3$ is the last occurrence. The amplitude $d_2$ is the time distance from onset to offset. The center of gravity $t_2$ is the moment when half the duration has been consumed. The red circle indicates the midpoint from $t_1$ and $t_3$. If the activity were unimodal ---no breaks--- the midpoint and the center of gravity would match to each other. It is not the case in this sketch.}
  \label{fig:acti}
\end{figure*}

\begin{equation}
  \label{eq:3}
d_1=\sum_{i=1}^{N_0}A(i).  
\end{equation}
From this quantity the center of gravity $t_2$ can be determined as the moment when half duration has been consumed and half remains to be consumed.

The onset time $t_1$ can be defined as the first occurrence of the activity and the offset time $t_3$ as the last occurrence. The amplitude $d_2$ is the time distance from offset to onset.

It must be noted that \emph{duration} is unequivocally computed. So is the \emph{center of gravity}, up to the arbitrary starting point of the cycle. On the contrary $t_1$, $t_2$ and $d_2$ can only be identified with onset, offset, and amplitude as long as $A$ is not too fragmented and, perhaps, as long as the respondent was actually sleeping by 4am. The activity of a tiny fraction of individuals part ways from this simple scheme, which may influence results on these variables.

On the other hand three of these quantities ---offset, onset and the center of gravity, labelled with the letter $t$--- are time-marks and will always be expressed local time, thus relative to a time zone. Therefore they are sensitive to time arrangements. The remaining two quantities ---daily duration and amplitude, labelled with the letter $d$--- are elapsed times and do not depend on time zones. 

Every of these five quantities is a stochastic discrete variable with only $N_0=\num{144}$ possible outcomes. It can then be determined how many counts (respondents) $n(i)$ populate a given index $i$ and the experimental (or sample) cumulative distribution function:
\begin{equation}
  \label{eq:6}
 P(i)=\frac{1}{N}\sum_{k=1}^in(k) 
\end{equation}

These cumulative probabilities can be computed for the winter $P_w$ and summer $P_s$ partitions and the seasonal difference is just $\Delta P(i)=P_s(i)-P_w(i)$.

As for daily rhythms, a test is needed to assess the significance of seasonal deviations. However some differences must be noted. First here the null hypothesis $H_0: P_s-P_w=0$ inspects whether two experimental cumulative distributions come from the same parent, unknown distribution.  Therefore the null hypothesis is tested for one stochastic variable and it is the maximum deviation that carries the relevant information. Instead $R$ is an average value only, and the test is run for every index $i$.

For $P_s-P_w$ the Kolmogorov-Smirnov test is the appropriate choice but due to the discrete nature of the variables no $p$-values can be obtained from the test. As an alternative a permutation test will be carried out for this purpose. The seasonal index ---which links one respondent to one season--- will be permuted or reshuffled $M$ times. For every permutation two samples of size $N_s$ and $N_w$ will be extracted. The deviations of the permuted samples will be compared to the parent seasonal deviation in search of larger the deviations. The $p$-value of the test is the fraction $p=M_0/M$ of permutations with larger results than the seasonal parent sample.

Moreover, in this case tailed (one-sided) tests are meaningful since, unlike $R$, $P$ is monotonic on the index $i$. Therefore the seasonal absolute deviation $D_0=\max\{|P_s(i)-P_w(i)|\}$ and the signed deviations $D_0^+=\max\{P_s(i)-P_w(i)\}$ and $D_0^-=\min\{P_s(i)-P_w(i)\}$ will be compared with the shuffled deviations to test the null hypotheses $H_0:P_s-P_w=0$, $H_0:P_s-P_w\geq0$ and $H_0:P_s-P_w\leq0$ against the alternates $H_1:P_s-P_w\neq0$, $H_1:P_s-P_w<0$ and $H_1:P_s-P_w>0$. Four possible outcomes are possible:

\begin{description}
\item [\emph{equal} $(=)$] if $D_j\geq D_0$ at least in $\SI{5}{\percent}$ of the permutations so that the null hypothesis $H_0: P_s-P_w=0$ sustains.
  
\item [\emph{less} $(<)$] if $D^-_j<D^-_0$ in less than $\SI{5}{\percent}$ of the permutations and $D^+_j\geq D^+_0$ at least in $\SI{5}{\percent}$ of the permutations so that the null hypothesis $H_0: P_s-P_w\leq0$ sustains and the null hypothesis $H_0: P_s-P_w\geq0$ is rejected. First-order stochastic dominance results in the summer distribution delaying relative to the winter distribution for some index.

\item  [\emph{greater} $(>)$] if $D^+_j>D^+_0$ in less than $\SI{5}{\percent}$ of the permutations and $D^-_j\leq D^-_0$ at least in $\SI{5}{\percent}$ of the permutations so that the null hypothesis $H_0: P_s-P_w\geq0$ sustains and the null hypothesis $H_0:P_s-P_w\leq0$ is rejected. First-order stochastic dominance results in the summer distribution advancing relative to the winter distribution for some index.

\item [\emph{not equal} $(\neq)$]  if both $D^+_j>D^+_0$ and $D^-_j<D^-_0$ in less than $\SI{5}{\percent}$ of the permutations so that all the three null hypotheses are rejected and no stochastic dominance is indicated.
\end{description}

Permutation tests only require the knowledge of the raw seasonal difference $P_s-P_w$ but for comparing the size of deviations among surveys which differs in size, the Kolmogorov-Smirnov (KS) statistics or distance provides a normalised deviation. It is expressed as:
\begin{equation}
  \label{eq:2}
  \Delta K_{ks}(i)=\sqrt{N_h}\cdot\Delta P(i)=\sqrt{N_h}\cdot(P_s(i)-P_w(i)),
\end{equation}
where $N_h$ is the unweighted harmonic average of sample sizes:
\begin{equation}
  \label{eq:1}
  N_h=\frac{2}{\dfrac{1}{N_w}+\dfrac{1}{N_s}}.
\end{equation}

The Kolmogorov-Smirnov cumulative distribution function evaluated at  $\max|\Delta K_{ks}|/\sqrt{2}$ provides $p$-values for the null hypothesis in the continuous case. Also in that case, the $p$-value in sided tests turns to be equal to $\exp\left\{-N_h\left(D_0^{\pm}\right)^2\right\}$. In this study these $p$-values will only be used as a reference for graphical purposes.


\subsection{Geophysical framework}
\label{sec:geophys-fram}

\begin{figure*}[t!]
  \centering
  \includegraphics[width=\textwidth]{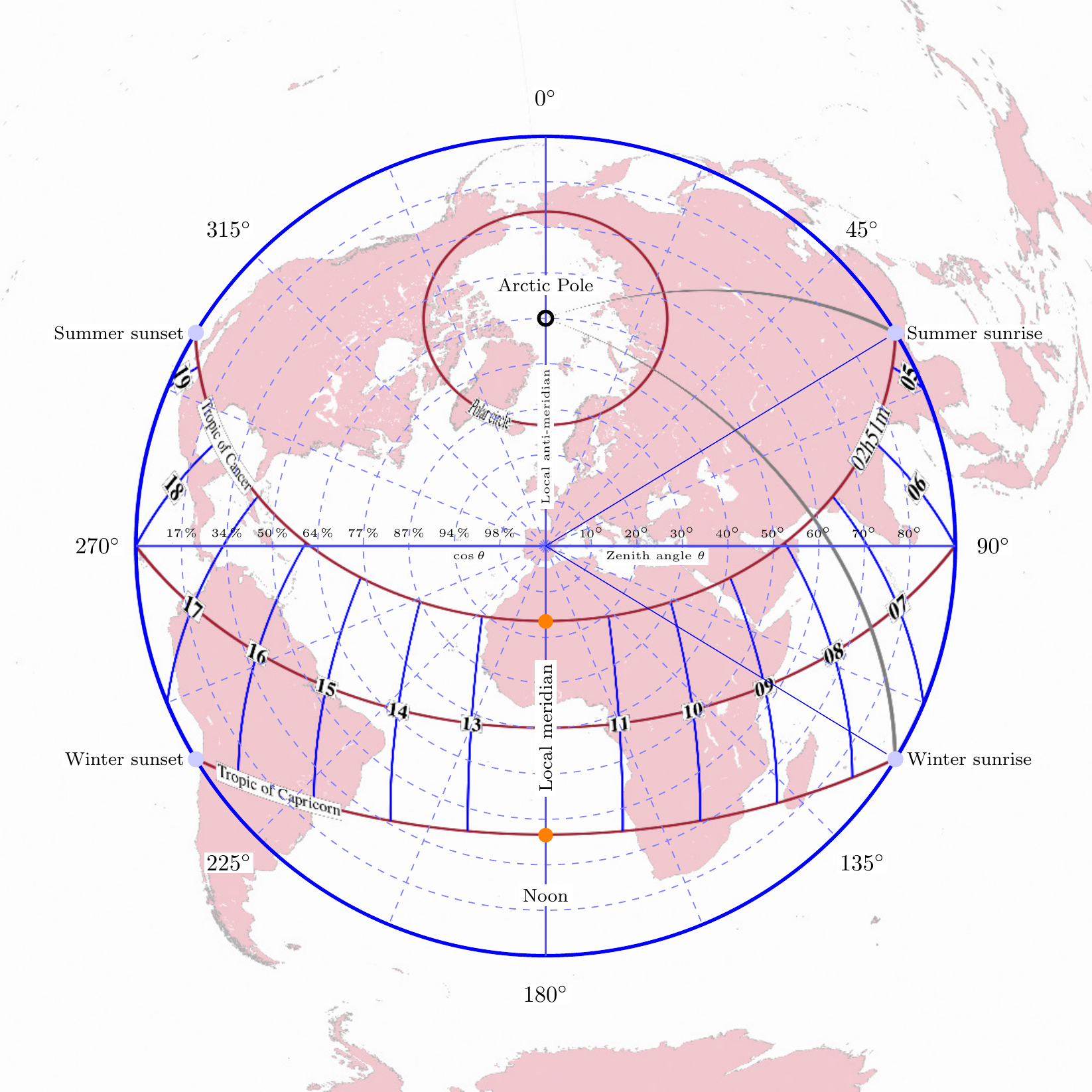}
  \caption{The Earth as seen from an observer located in the zenith of the Iberian peninsula (the center, longitude $\ang{-3.0}$, latitude $\ang{40}$). The compass rose signals azimuths, great circles leaving the location on every direction, alike meridians in the Pole. Dashed blue circles display distance to the center, converted into angle ---zenith angle--- after scaling by Earth's radius, alike parallels in the Pole. The thick blue line is the local horizon, alike the Equator for the Poles. It locates $\ang{90}$  from the center ---or $\sim\SI{10000}{\kilo\meter}$ as per the original definition of meter---. Beyond this line nothing can be viewed from the observer ---though Earth boundaries are shown in the picture--- as it lies in the opposite hemisphere of the center.  As Earth revolves around the Sun and rotates around the Pole the subsolar point ---where the Sun is overhead--- transits the Tropic of Cancer (by the June solstice), the Equator (at the equinoctes) and the Tropic of Capricorn (by the December solstice). Every day it crosses the $\ang{90}$ circle twice. Whenever this happens the Sun is up the horizon in the center and the observer sees it in the line of umbra light hemisphere of the Earth. Whenever the subsolar point transits the local meridian it is noon, the subsolar point is in its daily shortest range to the center, the solar zenith angle is the smallest and solar altitude, the highest. Six ephemerides are noted: (1) orange dots punctuate winter and summer noon, they account for the zenith or ``vertical'' seasonal variation; (2) light blue dots punctuate the subsolar point at winter and summer sunrises and sunsets, they carry the azimuth or ``horizontal'' variation, noted by the thin, solid blue lines on the center. Numbered thick solid lines display means solar time, they are great circles passing through the Poles. The gray great circles that join subsolar point at summer sunrise and at winter sunrise to the Pole punctuate the spread of sunrise times, which is annotated over the Tropic line. Notice that in summer, at the time of the winter sunrise, the center is seeing the Sun  some $\ang{60}$ from the zenith or $\ang{30}$ up the horizon. The setting is not altered if the longitude of the center is altered, save for the map of the Earth in the background. Instead, if latitude increases then the orange dots move downwards, summer sunrise/sunset opens towards the local anti-meridian and winter sunrise/sunset closes towards the local meridian. The opposite occurs if latitude decreases. Boundary lines and shapes were taken from  \href{https://www.naturalearthdata.com/}{https://www.naturalearthdata.com/}. A solar location diagram for latitude \ang{51;30;} is on display at the United Kingdom Hydrographic Office \href{http://astro.ukho.gov.uk/nao/services/ais58.pdf}{http://astro.ukho.gov.uk/nao/services/ais58.pdf}.}
  \label{fig:azimut}
\end{figure*}

Figure~\ref{fig:azimut} shows a solar location diagram for latitude \ang{40}.   The thick blue circle is the local horizon. Around this circle local azimuths or bearings are displayed. Dashed blue circles display equidistant points to the center, which can be expressed as an angular distance ---the zenith angle--- taking into account Earth's radius. Numbered thick blue lines display mean solar time.

The map of the Earth provides a context as a background image. Accidentally it is centered at longitude $-\ang{3}$. Had the diagram been plotted for latitude \ang{90}, the local horizon would have been the Equator, azimuths would have been meridians and zenith lines would have been circles of latitude.

It is daytime in the center and an observer located on its zenith sees it in the light hemisphere of the Earth whenever the subsolar point ---the point with the Sun overhead, which apparently moves as Earth rotates--- lies inside the thick blue circle. Otherwise, the observer could not the Sun and would see the center in the dark hemisphere of the Earth.

Whenever the subsolar point touches the thick blue circle the Sun is rising or setting in the center, and the observer would be seeing the line of umbra crossing the center. Solar zenith angle is invariably $\ang{90}$ and solar altitude is invariably $\ang{0}$.

The subsolar point is confined year round to the tropical range. Therefore the intersect of the local horizon with the Tropics defines the locus of sunrises and sunsets which extends ``horizontally'' over a range of azimuths. Figure~\ref{fig:azimut} shows the limiting conditions ---winter sunrise and sunset, summer sunrise and sunset--- as light blue points with labels. The spread in azimuth is the angle subtended by the center and the summer sunrise and winter sunrise points, noted by a thin blue line. The spread of sunrise/sunset times is defined by the spherical angle subtended by the pole and rise points, shown in the figure by dashed, gray arcs. This angle can be converted into time taking into account Earth's angular speed of rotation $\Omega=\SI{15}{\degree\per\hour}$ and can be projected into a clock with \textsc{24-h} analog dial as in Figure~\ref{L1-fig:clocks}. 

Noon occurs invariably when the subsolar point intersects the local meridian or the local anti-meridian. The subsolar point reaches the daily, shortest distance to the center and the lowest zenith angle. In the figure winter noon and summer noon are noted by orange dots.  In the extra-tropical range the azimuth is invariable but the zenith angle scores $\theta_s=|\phi|-\varepsilon$ in summer and $\theta_w=|\phi|+\varepsilon$ in winter. The ``vertical'' spread invariably amounts to $\theta_s-\theta_w=2\varepsilon=\ang{47}$. Notwithstanding all this, the cosine of the solar zenith angle is a more sensitive quantity to trace since it expresses the efficiency of solar insolation as a fraction of the insolation measured at the subsolar point, where the Sun is overhead, the zenith angle is zero and no shadow is cast.

As a summary of seasonal changes, the intersect of winter sunrise meridian with the Tropic of Cancer (summer solstice line) occurs at zenith angle $\sim\ang{60}$. That is at the 40th circle of latitude, by the time of winter sunrise, the Sun has already climbed up to $\ang{30}$ ---$\cos\theta_s=\SI{50}{\percent}$--- above the horizon in summer. 


If the diagram is plotted for a higher latitude, orange dots signalling solar zenith angle at noon will come closer to the horizon, the spread will still be $2\varepsilon$ but values would differ, altering also $\cos\theta$. Light blue dots will open towards the local anti-meridian (summer) and will close towards the local meridian (winter), enlarging the spread of horizontal variations. 
\clearpage

\subsection{Figures}
\label{sec:figures}

\begin{figure*}[h]
  \centering
  \includegraphics[height=0.8\textheight]{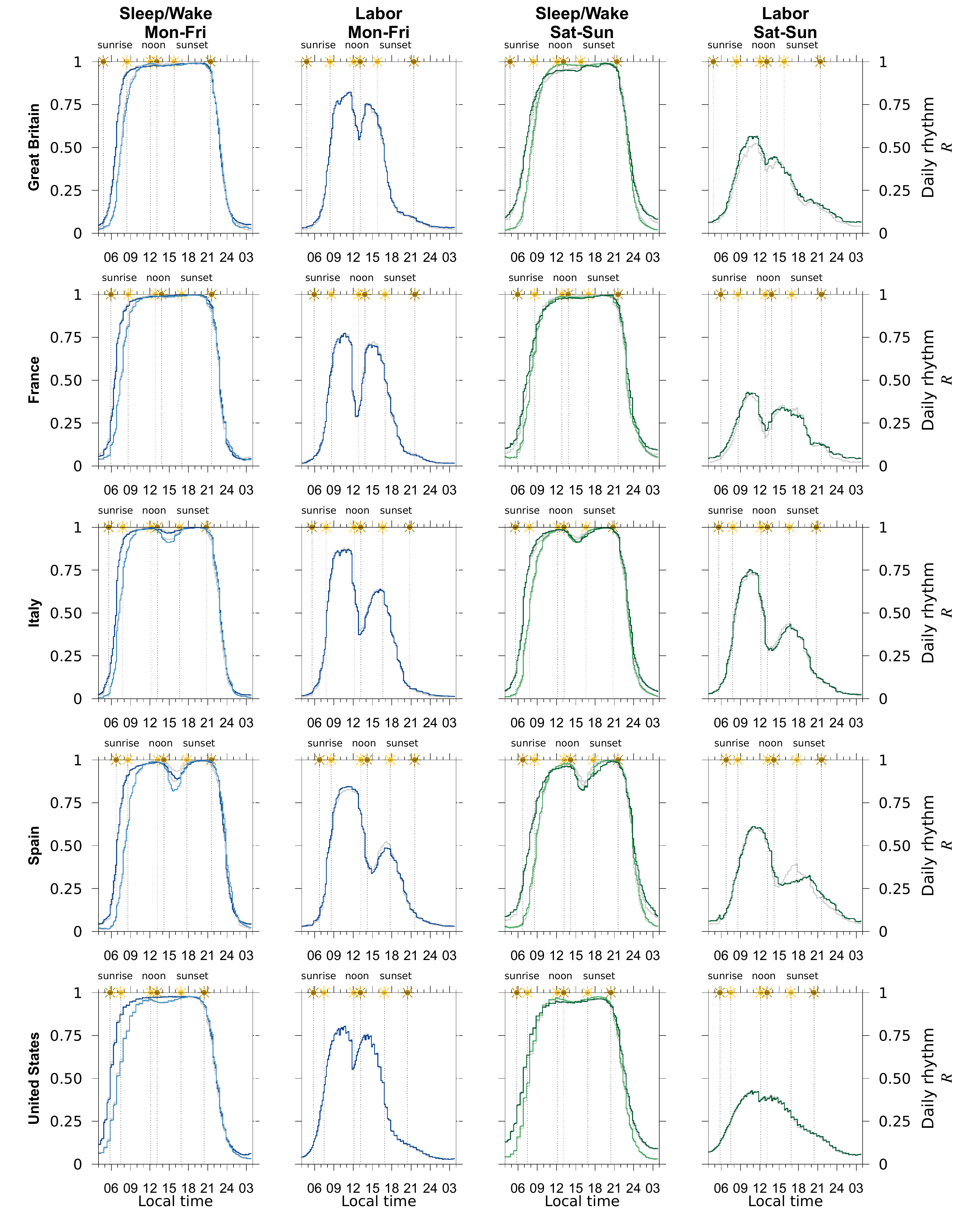}
  \caption{The sleep/wake and labour daily rhythms in winter and summer. Blueish inks display week-day rhythms (groups 1 and 2). Greenish inks display week-end rhythms (groups 3 and 4). Darker lines apply to workers (groups 1 and 3) in summer, lighter lines apply to non-workers (groups 2 and 4) in summer. Even lighter lines, hard to visualise, display winter daily rhythms. Vertical lines highlight solar events ---sunrise, noon and sunset--- obtained for the population weighted median latitude and longitude of each country.}
  \label{fig:rhythmlv}
\end{figure*}

\begin{figure*}[h]
  \centering
  \includegraphics[height=.85\textheight]{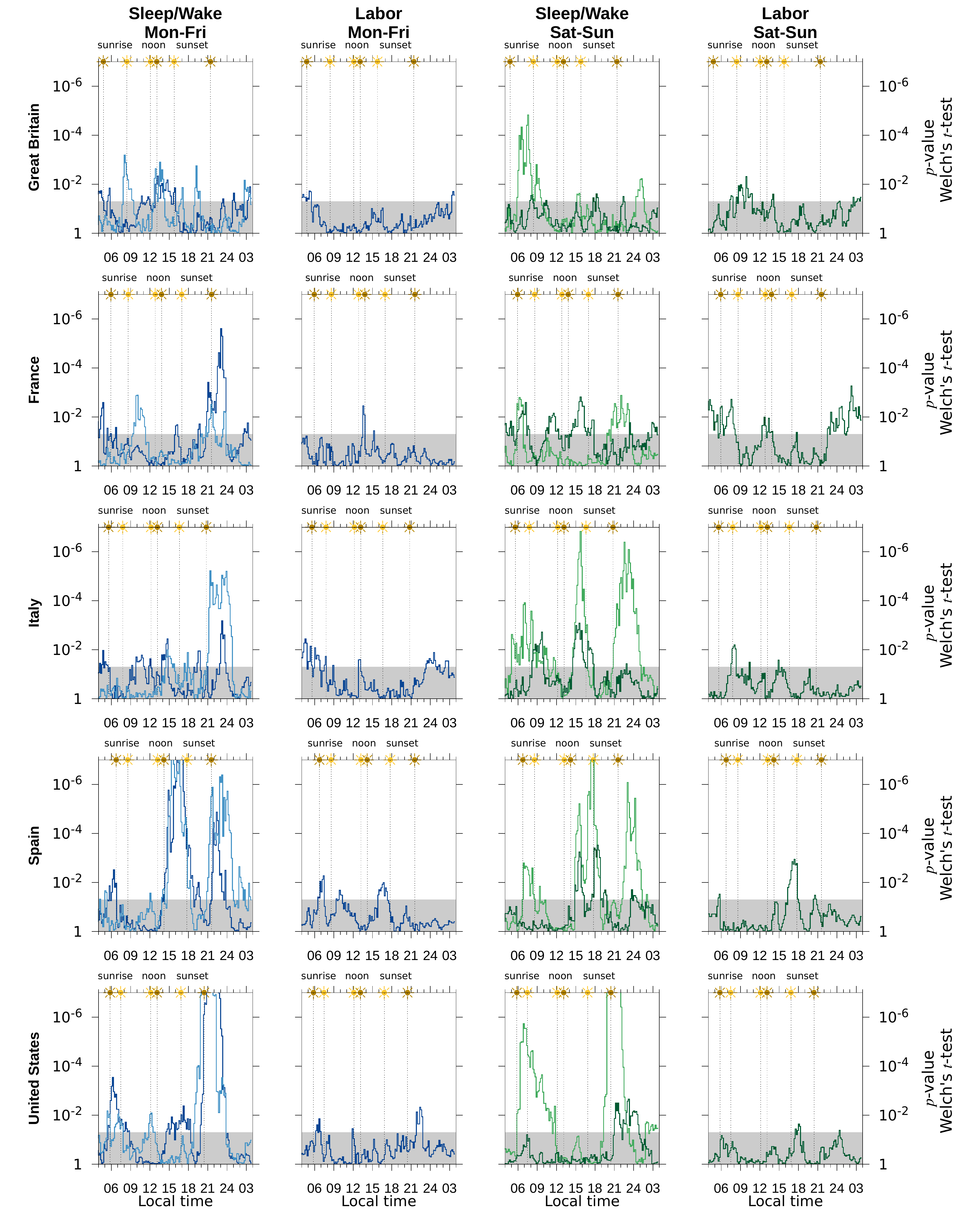}
  \caption{Probabilistic values obtained from the Welch's $t$-test on daily rhythms (see Eqs~(\ref{eq:9}) to~(\ref{eq:10})). The gray band highlights the region $p(i)>\alpha$ where the null hypothesis $H_0:R_s(i)-R_w(i)=0$ would sustain at the standard level of significance $\alpha=\num{0.05}=\num{.e-1.30}$. Vertical lines show solar ephemerides (sunrise, noon and sunset) in winter and summer as measured in local time. Darker sun applies to summer. Colour lines follow Figure~\ref{fig:rhythmlv}.}
  \label{fig:prhythm}
\end{figure*}

\begin{figure*}[h]
  \centering
  \includegraphics[height=.8\textheight]{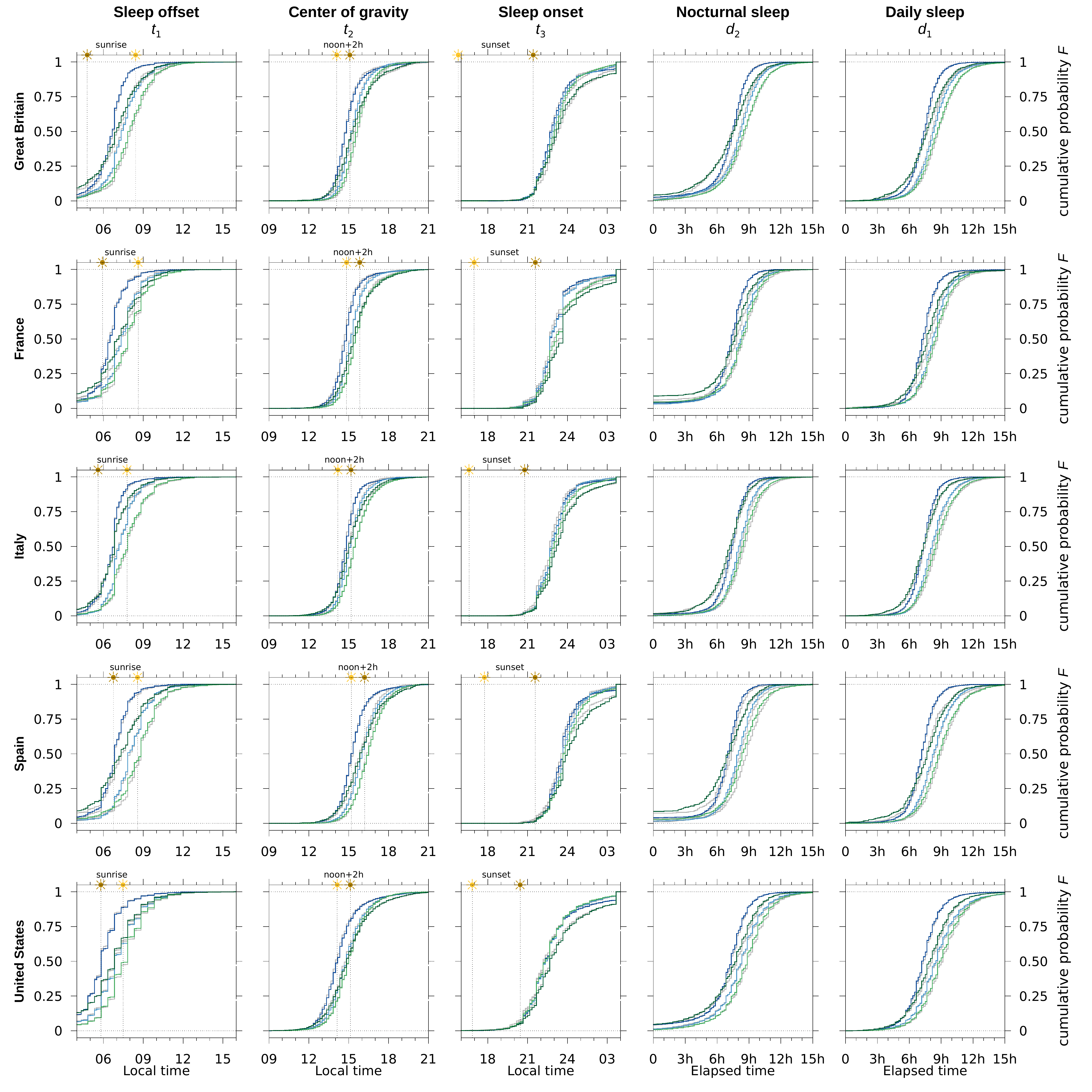}
  \caption{The sample cumulative distribution function of the sleep/wake stochastic variables (columns) for different countries (rows) and seasons: thicker lines display summer values, thinner lines (hard to visualise) display winter values. Blueish inks display groups 1 and 2 (week days) and greenish inks, groups 3 and 4 (week ends). Vertical lines display sunrise, noon and sunset times as measured local time.}
  \label{fig:pwakeemp}
\end{figure*}


\begin{figure*}[h]
  \centering
  \includegraphics[height=.8\textheight]{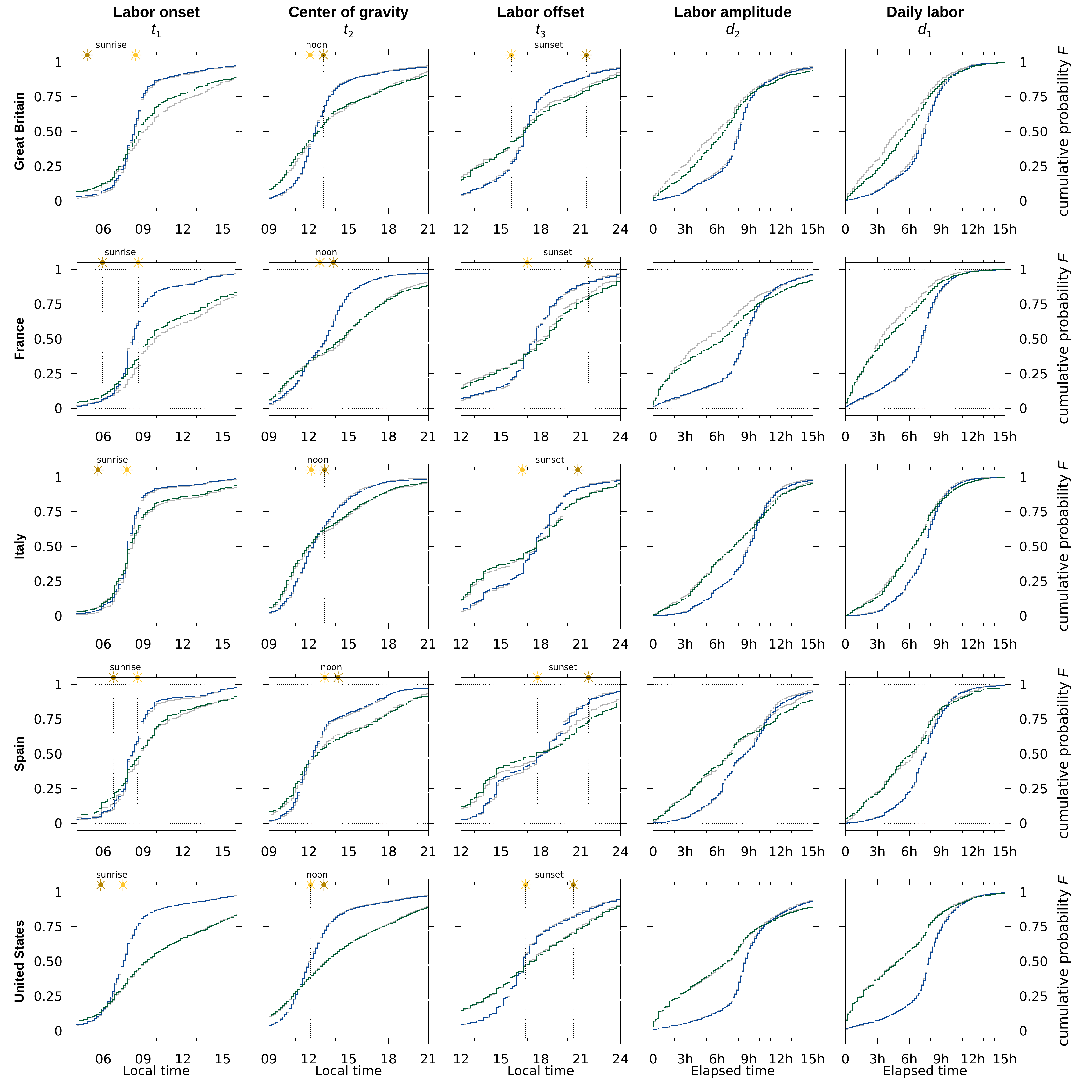}
  \caption{Same as Figure~\ref{fig:pwakeemp} but for the labour stochastic variables.}
  \label{fig:plabor}
\end{figure*}

\clearpage
\subsection{Tabular material}
\label{sec:tabular-material-1}

\begin{table*}[h]
  \centering
\footnotesize\sf
\setlength{\tabcolsep}{3pt}
	\begin{tabular}{lcccccccccccc}
	\toprule
	KS statistics&\multicolumn{3}{c}{\begin{tabular}[t]{@{}c@{}}Group 1\\Mon-Fri\\employees\end{tabular}}&\multicolumn{3}{c}{\begin{tabular}[t]{@{}c@{}}Group 2\\Mon-Fri\\non-employees\end{tabular}}&\multicolumn{3}{c}{\begin{tabular}[t]{@{}c@{}}Group 3\\Sat-Sun\\non-employees\end{tabular}}&\multicolumn{3}{c}{\begin{tabular}[t]{@{}c@{}}Group 4\\Sat-Sun\\employees\end{tabular}}\\
	&$t_m$&$P_s\leftrightarrow P_w$&$p$&$t_m$&$P_s\leftrightarrow P_w$&$p$&$t_m$&$P_s\leftrightarrow P_w$&$p$&$t_m$&$P_s\leftrightarrow P_w$&$p$\\
	\midrule
	\multicolumn{13}{l}{\textbf{Great Britain}}\\
	Sleep offset&\cellcolor{white}{05:50}&$\num{0.19}=\num{0.16}$&$10^{\num{-0.53}}$&\cellcolor{white}{08:10}&$\num{0.72}<\num{0.77}$&$10^{\num{-2.69}}$&\cellcolor{white}{08:10}&$\num{0.77}=\num{0.72}$&$10^{\num{-0.53}}$&\cellcolor{white}{07:40}&$\num{0.38}>\num{0.33}$&$10^{\num{-2.92}}$\\
	Center of gravity&\cellcolor{white}{16:10}&$\num{0.89}=\num{0.91}$&$10^{\num{-0.46}}$&\cellcolor{white}{16:00}&$\num{0.76}=\num{0.78}$&$10^{\num{-0.13}}$&\cellcolor{white}{15:10}&$\num{0.44}>\num{0.36}$&$10^{\num{-2.23}}$&\cellcolor{white}{15:40}&$\num{0.49}>\num{0.46}$&$10^{\num{-1.77}}$\\
	Sleep onset&\cellcolor{white}{23:20}&$\num{0.65}=\num{0.67}$&$10^{\num{-0.27}}$&\cellcolor{white}{23:40}&$\num{0.70}=\num{0.68}$&$10^{\num{-0.17}}$&\cellcolor{white}{22:30}&$\num{0.28}=\num{0.22}$&$10^{\num{-0.88}}$&\cellcolor{white}{01:00}&$\num{0.90}=\num{0.87}$&$10^{\num{-0.54}}$\\
	Onset to offset&\cellcolor{white}{07h50m}&$\num{0.54}=\num{0.51}$&$10^{\num{-0.55}}$&\cellcolor{white}{09h30m}&$\num{0.76}<\num{0.80}$&$10^{\num{-1.93}}$&\cellcolor{white}{07h50m}&$\num{0.51}=\num{0.54}$&$10^{\num{-0.13}}$&\cellcolor{white}{08h30m}&$\num{0.43}>\num{0.40}$&$10^{\num{-2.18}}$\\
	Sleep time&\cellcolor{white}{07h40m}&$\num{0.50}=\num{0.47}$&$10^{\num{-0.81}}$&\cellcolor{white}{09h20m}&$\num{0.70}<\num{0.76}$&$10^{\num{-2.53}}$&\cellcolor{white}{08h50m}&$\num{0.71}=\num{0.67}$&$10^{\num{-0.31}}$&\cellcolor{white}{08h00m}&$\num{0.29}>\num{0.26}$&$10^{\num{-2.05}}$\\
\\
	\multicolumn{13}{l}{\textbf{France}}\\
	Sleep offset&\cellcolor{white}{06:50}&$\num{0.57}<\num{0.59}$&$10^{\num{-3.34}}$&\cellcolor{white}{07:40}&$\num{0.55}\neq\num{0.57}$&$<10^{-S}$&\cellcolor{white}{07:20}&$\num{0.52}>\num{0.45}$&$10^{\num{-2.47}}$&\cellcolor{white}{06:30}&$\num{0.18}>\num{0.15}$&$10^{\num{-3.19}}$\\
	Center of gravity&\cellcolor{white}{14:50}&$\num{0.48}<\num{0.53}$&$10^{\num{-3.69}}$&\cellcolor{white}{16:20}&$\num{0.86}=\num{0.89}$&$10^{\num{-0.56}}$&\cellcolor{white}{17:20}&$\num{0.89}=\num{0.92}$&$10^{\num{-0.43}}$&\cellcolor{white}{16:00}&$\num{0.67}=\num{0.68}$&$10^{\num{-0.22}}$\\
	Sleep onset&\cellcolor{white}{23:00}&$\num{0.55}<\num{0.60}$&$10^{\num{-7.00}}$&\cellcolor{white}{23:30}&$\num{0.66}<\num{0.69}$&$<10^{-S}$&\cellcolor{white}{23:20}&$\num{0.44}=\num{0.49}$&$10^{\num{-0.88}}$&\cellcolor{white}{23:00}&$\num{0.45}<\num{0.48}$&$10^{\num{-3.21}}$\\
	Onset to offset&\cellcolor{white}{07h20m}&$\num{0.41}>\num{0.37}$&$10^{\num{-4.55}}$&\cellcolor{white}{07h10m}&$\num{0.22}\neq\num{0.20}$&$10^{\num{-6.15}}$&\cellcolor{white}{08h10m}&$\num{0.61}>\num{0.53}$&$10^{\num{-2.57}}$&\cellcolor{white}{09h10m}&$\num{0.68}>\num{0.64}$&$10^{\num{-3.40}}$\\
	Sleep time&\cellcolor{white}{07h20m}&$\num{0.42}>\num{0.38}$&$10^{\num{-2.36}}$&\cellcolor{white}{07h30m}&$\num{0.24}=\num{0.22}$&$10^{\num{-0.56}}$&\cellcolor{white}{08h00m}&$\num{0.54}>\num{0.47}$&$10^{\num{-2.38}}$&\cellcolor{white}{09h00m}&$\num{0.62}>\num{0.58}$&$10^{\num{-2.44}}$\\
\\
	\multicolumn{13}{l}{\textbf{Italy}}\\
	Sleep offset&\cellcolor{white}{05:50}&$\num{0.15}=\num{0.14}$&$10^{\num{-0.36}}$&\cellcolor{white}{07:00}&$\num{0.41}=\num{0.43}$&$10^{\num{-0.21}}$&\cellcolor{white}{08:30}&$\num{0.90}=\num{0.88}$&$10^{\num{-0.30}}$&\cellcolor{white}{07:10}&$\num{0.34}>\num{0.31}$&$10^{\num{-2.72}}$\\
	Center of gravity&\cellcolor{white}{15:10}&$\num{0.66}<\num{0.69}$&$10^{\num{-1.61}}$&\cellcolor{white}{15:20}&$\num{0.55}<\num{0.60}$&$10^{\num{-2.75}}$&\cellcolor{white}{15:00}&$\num{0.45}=\num{0.47}$&$10^{\num{-0.10}}$&\cellcolor{white}{16:10}&$\num{0.72}=\num{0.72}$&$10^{\num{-0.05}}$\\
	Sleep onset&\cellcolor{white}{23:10}&$\num{0.59}<\num{0.63}$&$10^{\num{-2.76}}$&\cellcolor{white}{23:10}&$\num{0.60}<\num{0.66}$&$10^{\num{-4.77}}$&\cellcolor{white}{22:50}&$\num{0.35}=\num{0.38}$&$10^{\num{-0.67}}$&\cellcolor{white}{23:10}&$\num{0.53}<\num{0.57}$&$10^{\num{-5.02}}$\\
	Onset to offset&\cellcolor{white}{07h30m}&$\num{0.45}>\num{0.41}$&$10^{\num{-2.44}}$&\cellcolor{white}{08h00m}&$\num{0.39}>\num{0.32}$&$10^{\num{-5.80}}$&\cellcolor{white}{07h30m}&$\num{0.51}>\num{0.46}$&$10^{\num{-2.05}}$&\cellcolor{white}{09h10m}&$\num{0.66}>\num{0.60}$&$<10^{-S}$\\
	Sleep time&\cellcolor{white}{08h00m}&$\num{0.67}>\num{0.64}$&$10^{\num{-1.63}}$&\cellcolor{white}{07h50m}&$\num{0.34}>\num{0.28}$&$10^{\num{-4.29}}$&\cellcolor{white}{07h40m}&$\num{0.52}=\num{0.49}$&$10^{\num{-0.88}}$&\cellcolor{white}{09h00m}&$\num{0.59}>\num{0.54}$&$<10^{-S}$\\
\\
	\multicolumn{13}{l}{\textbf{Spain}}\\
	Sleep offset&\cellcolor{white}{06:30}&$\num{0.33}>\num{0.28}$&$10^{\num{-1.94}}$&\cellcolor{white}{07:50}&$\num{0.37}>\num{0.35}$&$10^{\num{-3.13}}$&\cellcolor{white}{06:20}&$\num{0.29}>\num{0.24}$&$10^{\num{-4.68}}$&\cellcolor{white}{08:20}&$\num{0.37}>\num{0.33}$&$<10^{-S}$\\
	Center of gravity&\cellcolor{white}{15:20}&$\num{0.50}=\num{0.48}$&$10^{\num{-0.22}}$&\cellcolor{white}{16:10}&$\num{0.53}<\num{0.57}$&$10^{\num{-2.23}}$&\cellcolor{white}{14:50}&$\num{0.19}=\num{0.17}$&$10^{\num{-0.01}}$&\cellcolor{white}{17:00}&$\num{0.66}=\num{0.69}$&$10^{\num{-0.79}}$\\
	Sleep onset&\cellcolor{white}{23:40}&$\num{0.46}<\num{0.50}$&$10^{\num{-2.15}}$&\cellcolor{white}{23:30}&$\num{0.33}<\num{0.40}$&$<10^{-S}$&\cellcolor{white}{23:40}&$\num{0.31}<\num{0.39}$&$10^{\num{-6.00}}$&\cellcolor{white}{00:10}&$\num{0.59}<\num{0.66}$&$<10^{-S}$\\
	Onset to offset&\cellcolor{white}{07h10m}&$\num{0.48}>\num{0.42}$&$10^{\num{-3.72}}$&\cellcolor{white}{08h20m}&$\num{0.52}>\num{0.47}$&$<10^{-S}$&\cellcolor{white}{07h10m}&$\num{0.49}>\num{0.39}$&$<10^{-S}$&\cellcolor{white}{08h40m}&$\num{0.51}>\num{0.42}$&$<10^{-S}$\\
	Sleep time&\cellcolor{white}{08h30m}&$\num{0.84}=\num{0.82}$&$10^{\num{-0.22}}$&\cellcolor{white}{08h20m}&$\num{0.44}=\num{0.41}$&$10^{\num{-0.78}}$&\cellcolor{white}{08h50m}&$\num{0.70}=\num{0.76}$&$10^{\num{-0.68}}$&\cellcolor{white}{08h30m}&$\num{0.37}>\num{0.33}$&$10^{\num{-2.64}}$\\
\\
	\multicolumn{13}{l}{\textbf{United States}}\\
	Sleep offset&\cellcolor{white}{06:20}&$\num{0.59}<\num{0.61}$&$10^{\num{-2.97}}$&\cellcolor{white}{07:40}&$\num{0.63}=\num{0.64}$&$10^{\num{-1.10}}$&\cellcolor{white}{07:30}&$\num{0.67}=\num{0.65}$&$10^{\num{-0.94}}$&\cellcolor{white}{07:00}&$\num{0.40}>\num{0.38}$&$10^{\num{-4.49}}$\\
	Center of gravity&\cellcolor{white}{14:00}&$\num{0.42}<\num{0.44}$&$10^{\num{-6.30}}$&\cellcolor{white}{14:30}&$\num{0.38}<\num{0.40}$&$10^{\num{-4.78}}$&\cellcolor{white}{13:40}&$\num{0.18}=\num{0.19}$&$10^{\num{-0.65}}$&\cellcolor{white}{13:10}&$\num{0.05}=\num{0.06}$&$10^{\num{-0.48}}$\\
	Sleep onset&\cellcolor{white}{22:00}&$\num{0.38}<\num{0.41}$&$<10^{-S}$&\cellcolor{white}{21:20}&$\num{0.19}<\num{0.22}$&$<10^{-S}$&\cellcolor{white}{23:00}&$\num{0.55}<\num{0.57}$&$10^{\num{-2.32}}$&\cellcolor{white}{21:20}&$\num{0.20}<\num{0.23}$&$<10^{-S}$\\
	Onset to offset&\cellcolor{white}{07h50m}&$\num{0.59}>\num{0.56}$&$10^{\num{-4.28}}$&\cellcolor{white}{09h30m}&$\num{0.70}>\num{0.68}$&$10^{\num{-2.19}}$&\cellcolor{white}{08h20m}&$\num{0.59}>\num{0.55}$&$10^{\num{-4.33}}$&\cellcolor{white}{10h00m}&$\num{0.69}>\num{0.66}$&$<10^{-S}$\\
	Sleep time&\cellcolor{white}{07h40m}&$\num{0.53}>\num{0.51}$&$10^{\num{-4.27}}$&\cellcolor{white}{09h20m}&$\num{0.62}>\num{0.60}$&$10^{\num{-2.46}}$&\cellcolor{white}{08h10m}&$\num{0.50}>\num{0.46}$&$10^{\num{-3.99}}$&\cellcolor{white}{09h50m}&$\num{0.62}>\num{0.59}$&$<10^{-S}$\\
\\
	\bottomrule
	\end{tabular}

  \caption{KS statistics for the sleep/wake cycle after a permutation test of size $M=\num{.e7}$ compared seasonal deviations $\Delta K_{ks}\propto (P_s-P_w)$ to permuted deviations $\Delta K_{ks}^j$. Every analysis displays the time at $\max|\Delta K_{ks}|$, the cumulative distributions $P_s$ and $P_w$ at that index and a binary relational operator which put forward the result of the test: $=$ if the null hypothesis $H_0:P_s-P_w=0$ sustains; $>$ if the hypothesis $H_0:P_s-P_w\leq0$ does not sustain; $<$ if the hypothesis $H_0:P_s-P_w\geq0$ does not sustain; and $\neq$ if none of the null hypotheses sustains.  Next to them the $p$-value that support the result: the fraction of permutations that sustained the alternate $H_1$. Significance is taken at the standard level $\alpha=\SI{5}{\percent}=\num{.e-1.30}$. The sensitivity of the $p$-value is $M^{-1}=10^{-S}$ with $S=\num{7}$.}
  \label{tab:sleepKS}
\end{table*}

\begin{table*}[tb]
  \centering
\footnotesize\sf
\setlength{\tabcolsep}{9pt}
	\begin{tabular}{lcccccc}
	\toprule
	KS statistics&\multicolumn{3}{c}{\begin{tabular}[t]{@{}c@{}}Group 1\\Mon-Fri\\employees\end{tabular}}&\multicolumn{3}{c}{\begin{tabular}[t]{@{}c@{}}Group 3\\Sat-Sun\\non-employees\end{tabular}}\\
	&$t_m$&$P_s\leftrightarrow P_w$&$p$&$t_m$&$P_s\leftrightarrow P_w$&$p$\\
	\midrule
	\multicolumn{7}{l}{\textbf{Great Britain}}\\
	Labor onset&\cellcolor{white}{07:40}&$\num{0.25}=\num{0.23}$&$10^{\num{-0.29}}$&\cellcolor{white}{10:00}&$\num{0.67}>\num{0.60}$&$10^{\num{-1.83}}$\\
	Center of gravity&\cellcolor{white}{13:50}&$\num{0.77}=\num{0.79}$&$10^{\num{-0.13}}$&\cellcolor{white}{11:00}&$\num{0.30}=\num{0.26}$&$10^{\num{-0.33}}$\\
	Labor offset&\cellcolor{white}{16:20}&$\num{0.34}=\num{0.36}$&$10^{\num{-0.33}}$&\cellcolor{white}{18:40}&$\num{0.65}=\num{0.69}$&$10^{\num{-0.25}}$\\
	Onset to offset&\cellcolor{white}{06h30m}&$\num{0.20}=\num{0.22}$&$10^{\num{-0.19}}$&\cellcolor{white}{04h50m}&$\num{0.32}<\num{0.41}$&$10^{\num{-2.57}}$\\
	Labor time&\cellcolor{white}{07h40m}&$\num{0.48}=\num{0.52}$&$10^{\num{-1.09}}$&\cellcolor{white}{05h00m}&$\num{0.38}<\num{0.49}$&$10^{\num{-3.53}}$\\
\\
	\multicolumn{7}{l}{\textbf{France}}\\
	Labor onset&\cellcolor{white}{08:20}&$\num{0.50}=\num{0.52}$&$10^{\num{-0.42}}$&\cellcolor{white}{08:00}&$\num{0.28}>\num{0.21}$&$10^{\num{-2.57}}$\\
	Center of gravity&\cellcolor{white}{10:20}&$\num{0.10}=\num{0.09}$&$10^{\num{-0.11}}$&\cellcolor{white}{14:10}&$\num{0.47}=\num{0.44}$&$10^{\num{-0.23}}$\\
	Labor offset&\cellcolor{white}{12:00}&$\num{0.07}=\num{0.05}$&$10^{\num{-0.55}}$&\cellcolor{white}{22:30}&$\num{0.83}=\num{0.87}$&$10^{\num{-0.60}}$\\
	Onset to offset&\cellcolor{white}{08h50m}&$\num{0.51}=\num{0.49}$&$10^{\num{-0.38}}$&\cellcolor{white}{05h50m}&$\num{0.45}<\num{0.54}$&$10^{\num{-3.37}}$\\
	Labor time&\cellcolor{white}{07h30m}&$\num{0.52}=\num{0.50}$&$10^{\num{-0.25}}$&\cellcolor{white}{05h40m}&$\num{0.61}<\num{0.69}$&$10^{\num{-2.78}}$\\
\\
	\multicolumn{7}{l}{\textbf{Italy}}\\
	Labor onset&\cellcolor{white}{07:50}&$\num{0.38}>\num{0.34}$&$10^{\num{-2.23}}$&\cellcolor{white}{08:10}&$\num{0.53}=\num{0.49}$&$10^{\num{-1.10}}$\\
	Center of gravity&\cellcolor{white}{14:50}&$\num{0.81}=\num{0.83}$&$10^{\num{-0.14}}$&\cellcolor{white}{10:30}&$\num{0.25}=\num{0.22}$&$10^{\num{-0.47}}$\\
	Labor offset&\cellcolor{white}{14:50}&$\num{0.21}=\num{0.20}$&$10^{\num{-0.14}}$&\cellcolor{white}{15:00}&$\num{0.37}=\num{0.34}$&$10^{\num{-0.28}}$\\
	Onset to offset&\cellcolor{white}{09h40m}&$\num{0.56}=\num{0.58}$&$10^{\num{-0.47}}$&\cellcolor{white}{11h40m}&$\num{0.81}=\num{0.84}$&$10^{\num{-0.53}}$\\
	Labor time&\cellcolor{white}{08h40m}&$\num{0.72}=\num{0.74}$&$10^{\num{-0.57}}$&\cellcolor{white}{03h30m}&$\num{0.15}=\num{0.18}$&$10^{\num{-0.40}}$\\
\\
	\multicolumn{7}{l}{\textbf{Spain}}\\
	Labor onset&\cellcolor{white}{07:30}&$\num{0.23}=\num{0.20}$&$10^{\num{-0.79}}$&\cellcolor{white}{06:30}&$\num{0.18}=\num{0.12}$&$10^{\num{-0.53}}$\\
	Center of gravity&\cellcolor{white}{11:40}&$\num{0.34}=\num{0.31}$&$10^{\num{-0.56}}$&\cellcolor{white}{11:00}&$\num{0.26}=\num{0.22}$&$10^{\num{-0.11}}$\\
	Labor offset&\cellcolor{white}{20:10}&$\num{0.73}=\num{0.76}$&$10^{\num{-0.62}}$&\cellcolor{white}{20:30}&$\num{0.64}=\num{0.71}$&$10^{\num{-0.78}}$\\
	Onset to offset&\cellcolor{white}{11h40m}&$\num{0.81}=\num{0.84}$&$10^{\num{-0.79}}$&\cellcolor{white}{11h40m}&$\num{0.73}=\num{0.79}$&$10^{\num{-0.58}}$\\
	Labor time&\cellcolor{white}{09h10m}&$\num{0.77}=\num{0.80}$&$10^{\num{-0.76}}$&\cellcolor{white}{06h00m}&$\num{0.48}=\num{0.44}$&$10^{\num{-0.19}}$\\
\\
	\multicolumn{7}{l}{\textbf{United States}}\\
	Labor onset&\cellcolor{white}{06:50}&$\num{0.28}=\num{0.27}$&$10^{\num{-0.97}}$&\cellcolor{white}{07:50}&$\num{0.34}=\num{0.33}$&$10^{\num{-0.36}}$\\
	Center of gravity&\cellcolor{white}{12:30}&$\num{0.56}=\num{0.57}$&$10^{\num{-0.84}}$&\cellcolor{white}{21:30}&$\num{0.91}=\num{0.92}$&$10^{\num{-0.05}}$\\
	Labor offset&\cellcolor{white}{21:30}&$\num{0.86}=\num{0.87}$&$10^{\num{-0.77}}$&\cellcolor{white}{18:10}&$\num{0.55}=\num{0.56}$&$10^{\num{-0.34}}$\\
	Onset to offset&\cellcolor{white}{10h50m}&$\num{0.76}=\num{0.78}$&$10^{\num{-1.02}}$&\cellcolor{white}{06h00m}&$\num{0.43}=\num{0.42}$&$10^{\num{-0.22}}$\\
	Labor time&\cellcolor{white}{09h40m}&$\num{0.78}=\num{0.79}$&$10^{\num{-0.63}}$&\cellcolor{white}{02h10m}&$\num{0.30}=\num{0.30}$&$10^{\num{-0.09}}$\\
\\
	\bottomrule
	\end{tabular}

  \caption{Same as Table~\ref{tab:sleepKS} but for the labour cycle.}
  \label{tab:laborKS}
\end{table*}

\clearpage

\begin{table*}[t]
  \centering
\footnotesize\sf
\setlength{\tabcolsep}{3pt}
	\begin{tabular}{lcccccccccccccccccccccccccccccccccccccccccccccccc}
	\toprule
	Average values&\multicolumn{2}{c}{\begin{tabular}[t]{@{}c@{}}Group 1\\Mon-Fri\\employees\end{tabular}}&\multicolumn{2}{c}{\begin{tabular}[t]{@{}c@{}}Group 2\\Mon-Fri\\non-employees\end{tabular}}&\multicolumn{2}{c}{\begin{tabular}[t]{@{}c@{}}Group 3\\Sat-Sun\\employees\end{tabular}}&\multicolumn{2}{c}{\begin{tabular}[t]{@{}c@{}}Group 4\\Sat-Sun\\non-employees\end{tabular}}\\
	&$E_s\leftrightarrow E_w$&$p$&$E_s\leftrightarrow E_w$&$p$&$E_s\leftrightarrow E_w$&$p$&$E_s\leftrightarrow E_w$&$p$\\
	\midrule
	\multicolumn{5}{l}{\textbf{Great Britain}}\\
	Sleep offset&\textcolor{black}{06:44$=$}\textcolor{black}{06:48}&$\num{.e-0.89}$&\textcolor{black}{07:38$=$}\textcolor{black}{07:34}&$\num{.e-0.93}$&\textcolor{black}{07:08$=$}\textcolor{black}{07:17}&$\num{.e-1.04}$&\textcolor{black}{08:09$<$}\textcolor{black}{08:17}&$\num{.e-2.96}$\\
	Center of gravity&\textcolor{black}{15:06$=$}\textcolor{black}{15:03}&$\num{.e-0.66}$&\textcolor{black}{15:31$=$}\textcolor{black}{15:30}&$\num{.e-0.27}$&\textcolor{black}{15:38$=$}\textcolor{black}{15:47}&$\num{.e-1.11}$&\textcolor{black}{15:51$<$}\textcolor{black}{15:57}&$\num{.e-3.80}$\\
	Sleep onset&\textcolor{black}{23:16$=$}\textcolor{black}{23:13}&$\num{.e-0.48}$&\textcolor{black}{23:20$=$}\textcolor{black}{23:20}&$\num{.e-0.06}$&\textcolor{black}{23:43$=$}\textcolor{black}{23:49}&$\num{.e-0.47}$&\textcolor{black}{23:24$=$}\textcolor{black}{23:28}&$\num{.e-1.08}$\\
	Offset to onset&\textcolor{black}{07h28m$=$}\textcolor{black}{07h34m}&$\num{.e-1.09}$&\textcolor{black}{08h19m$=$}\textcolor{black}{08h13m}&$\num{.e-0.63}$&\textcolor{black}{07h25m$=$}\textcolor{black}{07h29m}&$\num{.e-0.20}$&\textcolor{black}{08h45m$=$}\textcolor{black}{08h49m}&$\num{.e-0.57}$\\
	Sleep time&\textcolor{black}{07h40m$=$}\textcolor{black}{07h43m}&$\num{.e-0.62}$&\textcolor{black}{08h40m$>$}\textcolor{black}{08h33m}&$\num{.e-1.69}$&\textcolor{black}{07h56m$=$}\textcolor{black}{08h02m}&$\num{.e-0.48}$&\textcolor{black}{09h03m$=$}\textcolor{black}{09h08m}&$\num{.e-0.97}$\\
	$|E_s-E_w|$ min;ave;max&\multicolumn{1}{c}{(3;4;6)\si{\minute}}&&\multicolumn{1}{c}{(1;4;7)\si{\minute}}&&\multicolumn{1}{c}{(4;7;10)\si{\minute}}&&\multicolumn{1}{c}{(4;5;8)\si{\minute}}&\\
	SEM (average)&\multicolumn{1}{c}{\SI{2}{\minute}}&&\multicolumn{1}{c}{\SI{2}{\minute}}&&\multicolumn{1}{c}{\SI{4}{\minute}}&&\multicolumn{1}{c}{\SI{2}{\minute}}&\\
	\midrule
	\multicolumn{5}{l}{\textbf{France}}\\
	Sleep offset&\textcolor{black}{06:41$>$}\textcolor{black}{06:37}&$\num{.e-3.62}$&\textcolor{black}{07:31$>$}\textcolor{black}{07:28}&$<10^{-S}$&\textcolor{black}{07:15$=$}\textcolor{black}{07:28}&$\num{.e-0.76}$&\textcolor{black}{07:55$<$}\textcolor{black}{08:00}&$\num{.e-4.40}$\\
	Center of gravity&\textcolor{black}{15:00$>$}\textcolor{black}{14:55}&$\num{.e-3.00}$&\textcolor{black}{15:25$=$}\textcolor{black}{15:21}&$\num{.e-0.37}$&\textcolor{black}{15:41$=$}\textcolor{black}{15:39}&$\num{.e-0.30}$&\textcolor{black}{15:50$=$}\textcolor{black}{15:48}&$\num{.e-0.47}$\\
	Sleep onset&\textcolor{black}{23:17$=$}\textcolor{black}{23:12}&$\num{.e-0.54}$&\textcolor{black}{23:17$>$}\textcolor{black}{23:13}&$<10^{-S}$&\textcolor{black}{23:56$>$}\textcolor{black}{23:46}&$\num{.e-5.33}$&\textcolor{black}{23:38$>$}\textcolor{black}{23:33}&$<10^{-S}$\\
	Offset to onset&\textcolor{black}{07h24m$=$}\textcolor{black}{07h24m}&$\num{.e-0.05}$&\textcolor{black}{08h14m$=$}\textcolor{black}{08h15m}&$\num{.e-1.01}$&\textcolor{black}{07h19m$<$}\textcolor{black}{07h42m}&$\num{.e-1.97}$&\textcolor{black}{08h17m$<$}\textcolor{black}{08h27m}&$<10^{-S}$\\
	Sleep time&\textcolor{black}{07h35m$=$}\textcolor{black}{07h35m}&$\num{.e-0.08}$&\textcolor{black}{08h29m$=$}\textcolor{black}{08h31m}&$\num{.e-0.40}$&\textcolor{black}{07h52m$<$}\textcolor{black}{08h06m}&$\num{.e-2.27}$&\textcolor{black}{08h42m$<$}\textcolor{black}{08h49m}&$\num{.e-2.55}$\\
	$|E_s-E_w|$ min;ave;max&\multicolumn{1}{c}{(0;3;5)\si{\minute}}&&\multicolumn{1}{c}{(2;3;4)\si{\minute}}&&\multicolumn{1}{c}{(2;12;24)\si{\minute}}&&\multicolumn{1}{c}{(2;6;10)\si{\minute}}&\\
	SEM (average)&\multicolumn{1}{c}{\SI{1}{\minute}}&&\multicolumn{1}{c}{\SI{2}{\minute}}&&\multicolumn{1}{c}{\SI{3}{\minute}}&&\multicolumn{1}{c}{\SI{2}{\minute}}&\\
	\midrule
	\multicolumn{5}{l}{\textbf{Italy}}\\
	Sleep offset&\textcolor{black}{06:45$=$}\textcolor{black}{06:47}&$\num{.e-0.74}$&\textcolor{black}{07:30$=$}\textcolor{black}{07:30}&$\num{.e-0.09}$&\textcolor{black}{06:58$=$}\textcolor{black}{07:00}&$\num{.e-0.37}$&\textcolor{black}{08:00$<$}\textcolor{black}{08:05}&$\num{.e-3.24}$\\
	Center of gravity&\textcolor{black}{15:00$=$}\textcolor{black}{14:58}&$\num{.e-0.57}$&\textcolor{black}{15:21$>$}\textcolor{black}{15:17}&$\num{.e-2.59}$&\textcolor{black}{15:20$=$}\textcolor{black}{15:21}&$\num{.e-0.17}$&\textcolor{black}{15:45$=$}\textcolor{black}{15:44}&$\num{.e-0.27}$\\
	Sleep onset&\textcolor{black}{23:10$>$}\textcolor{black}{23:06}&$\num{.e-1.80}$&\textcolor{black}{23:08$>$}\textcolor{black}{22:58}&$\num{.e-6.52}$&\textcolor{black}{23:37$=$}\textcolor{black}{23:34}&$\num{.e-0.60}$&\textcolor{black}{23:20$>$}\textcolor{black}{23:14}&$\num{.e-5.54}$\\
	Offset to onset&\textcolor{black}{07h35m$<$}\textcolor{black}{07h41m}&$\num{.e-2.52}$&\textcolor{black}{08h22m$<$}\textcolor{black}{08h32m}&$\num{.e-4.32}$&\textcolor{black}{07h21m$=$}\textcolor{black}{07h27m}&$\num{.e-0.89}$&\textcolor{black}{08h40m$<$}\textcolor{black}{08h51m}&$<10^{-S}$\\
	Sleep time&\textcolor{black}{07h39m$=$}\textcolor{black}{07h42m}&$\num{.e-0.74}$&\textcolor{black}{08h34m$<$}\textcolor{black}{08h42m}&$\num{.e-2.70}$&\textcolor{black}{07h38m$=$}\textcolor{black}{07h44m}&$\num{.e-0.97}$&\textcolor{black}{08h53m$<$}\textcolor{black}{09h03m}&$<10^{-S}$\\
	$|E_s-E_w|$ min;ave;max&\multicolumn{1}{c}{(2;4;7)\si{\minute}}&&\multicolumn{1}{c}{(0;6;10)\si{\minute}}&&\multicolumn{1}{c}{(1;4;6)\si{\minute}}&&\multicolumn{1}{c}{(1;6;11)\si{\minute}}&\\
	SEM (average)&\multicolumn{1}{c}{\SI{1}{\minute}}&&\multicolumn{1}{c}{\SI{1}{\minute}}&&\multicolumn{1}{c}{\SI{2}{\minute}}&&\multicolumn{1}{c}{\SI{1}{\minute}}&\\
	\midrule
	\multicolumn{5}{l}{\textbf{Spain}}\\
	Sleep offset&\textcolor{black}{06:59$<$}\textcolor{black}{07:05}&$\num{.e-1.61}$&\textcolor{black}{08:14$=$}\textcolor{black}{08:16}&$\num{.e-0.21}$&\textcolor{black}{07:26$<$}\textcolor{black}{07:34}&$<10^{-S}$&\textcolor{black}{08:44$<$}\textcolor{black}{08:53}&$<10^{-S}$\\
	Center of gravity&\textcolor{black}{15:31$=$}\textcolor{black}{15:33}&$\num{.e-0.60}$&\textcolor{black}{16:17$>$}\textcolor{black}{16:10}&$\num{.e-5.20}$&\textcolor{black}{16:11$=$}\textcolor{black}{16:11}&$\num{.e-0.01}$&\textcolor{black}{16:40$=$}\textcolor{black}{16:39}&$\num{.e-0.26}$\\
	Sleep onset&\textcolor{black}{23:55$>$}\textcolor{black}{23:49}&$\num{.e-2.46}$&\textcolor{black}{00:02$>$}\textcolor{black}{23:51}&$<10^{-S}$&\textcolor{black}{00:31$>$}\textcolor{black}{00:18}&$<10^{-S}$&\textcolor{black}{00:11$>$}\textcolor{black}{00:00}&$<10^{-S}$\\
	Offset to onset&\textcolor{black}{07h05m$<$}\textcolor{black}{07h16m}&$\num{.e-2.67}$&\textcolor{black}{08h12m$<$}\textcolor{black}{08h25m}&$\num{.e-3.36}$&\textcolor{black}{06h54m$<$}\textcolor{black}{07h16m}&$<10^{-S}$&\textcolor{black}{08h34m$<$}\textcolor{black}{08h53m}&$<10^{-S}$\\
	Sleep time&\textcolor{black}{07h29m$=$}\textcolor{black}{07h31m}&$\num{.e-0.33}$&\textcolor{black}{08h42m$<$}\textcolor{black}{08h48m}&$\num{.e-1.63}$&\textcolor{black}{07h56m$=$}\textcolor{black}{07h58m}&$\num{.e-0.13}$&\textcolor{black}{09h15m$<$}\textcolor{black}{09h24m}&$\num{.e-2.48}$\\
	$|E_s-E_w|$ min;ave;max&\multicolumn{1}{c}{(2;5;11)\si{\minute}}&&\multicolumn{1}{c}{(2;8;13)\si{\minute}}&&\multicolumn{1}{c}{(0;9;21)\si{\minute}}&&\multicolumn{1}{c}{(2;10;20)\si{\minute}}&\\
	SEM (average)&\multicolumn{1}{c}{\SI{2}{\minute}}&&\multicolumn{1}{c}{\SI{2}{\minute}}&&\multicolumn{1}{c}{\SI{5}{\minute}}&&\multicolumn{1}{c}{\SI{2}{\minute}}&\\
	\midrule
	\multicolumn{5}{l}{\textbf{United States}}\\
	Sleep offset&\textcolor{black}{06:08$>$}\textcolor{black}{06:06}&$\num{.e-2.14}$&\textcolor{black}{07:16$=$}\textcolor{black}{07:14}&$\num{.e-1.21}$&\textcolor{black}{06:55$=$}\textcolor{black}{06:58}&$\num{.e-0.81}$&\textcolor{black}{07:43$<$}\textcolor{black}{07:47}&$\num{.e-4.42}$\\
	Center of gravity&\textcolor{black}{14:30$>$}\textcolor{black}{14:26}&$\num{.e-4.95}$&\textcolor{black}{15:05$>$}\textcolor{black}{15:01}&$\num{.e-3.76}$&\textcolor{black}{15:12$=$}\textcolor{black}{15:11}&$\num{.e-0.27}$&\textcolor{black}{15:16$=$}\textcolor{black}{15:16}&$\num{.e-0.07}$\\
	Sleep onset&\textcolor{black}{22:53$>$}\textcolor{black}{22:49}&$\num{.e-3.65}$&\textcolor{black}{22:49$>$}\textcolor{black}{22:45}&$\num{.e-3.19}$&\textcolor{black}{23:21$>$}\textcolor{black}{23:16}&$\num{.e-2.13}$&\textcolor{black}{22:46$>$}\textcolor{black}{22:43}&$\num{.e-2.99}$\\
	Offset to onset&\textcolor{black}{07h15m$=$}\textcolor{black}{07h17m}&$\num{.e-0.87}$&\textcolor{black}{08h27m$=$}\textcolor{black}{08h29m}&$\num{.e-0.57}$&\textcolor{black}{07h34m$<$}\textcolor{black}{07h42m}&$\num{.e-2.73}$&\textcolor{black}{08h57m$<$}\textcolor{black}{09h04m}&$\num{.e-7.00}$\\
	Sleep time&\textcolor{black}{07h39m$<$}\textcolor{black}{07h41m}&$\num{.e-1.73}$&\textcolor{black}{08h58m$<$}\textcolor{black}{09h03m}&$\num{.e-2.70}$&\textcolor{black}{08h16m$<$}\textcolor{black}{08h23m}&$\num{.e-3.50}$&\textcolor{black}{09h22m$<$}\textcolor{black}{09h30m}&$<10^{-S}$\\
	$|E_s-E_w|$ min;ave;max&\multicolumn{1}{c}{(2;3;4)\si{\minute}}&&\multicolumn{1}{c}{(2;4;5)\si{\minute}}&&\multicolumn{1}{c}{(1;5;8)\si{\minute}}&&\multicolumn{1}{c}{(0;4;8)\si{\minute}}&\\
	SEM (average)&\multicolumn{1}{c}{\SI{1}{\minute}}&&\multicolumn{1}{c}{\SI{1}{\minute}}&&\multicolumn{1}{c}{\SI{1}{\minute}}&&\multicolumn{1}{c}{\SI{1}{\minute}}&\\
	\bottomrule
	\end{tabular}

  \caption{Seasonal sample average values $E_s$ and $E_w$ for the stochastic variables related to the sleep/wake cycle. For every survey and group the minimum, average and maximum value of absolute differences $|E_s-E_w|$ and the average of the standard deviation of the averages (SEM) is listed. Between every $E_s$ and $E_w$ a binary relational operator puts forward the result of the permutation test with the $p$-value in the next column. The size of the permutation test was $M=10^7$ ($S=7$).}
  \label{tab:vigil}
\end{table*}

\begin{table*}
  \centering
\footnotesize\sf
\setlength{\tabcolsep}{9pt}
	\begin{tabular}{lcccccccc}
	\toprule
	Average values&\multicolumn{2}{c}{\begin{tabular}[t]{@{}c@{}}Group 1\\Mon-Fri\\employees\end{tabular}}&\multicolumn{2}{c}{\begin{tabular}[t]{@{}c@{}}Group 3\\Sat-Sun\\employees\end{tabular}}\\
	&$E_s\leftrightarrow E_w$&$p$&$E_s\leftrightarrow E_w$&$p$\\
	\midrule
	\multicolumn{5}{l}{\textbf{Great Britain}}\\
	Labor onset&\textcolor{black}{08:50$=$}\textcolor{black}{08:57}&$\num{.e-0.83}$&\textcolor{black}{09:59$<$}\textcolor{black}{10:26}&$\num{.e-1.60}$\\
	Center of gravity&\textcolor{black}{13:13$=$}\textcolor{black}{13:10}&$\num{.e-0.25}$&\textcolor{black}{13:54$=$}\textcolor{black}{13:50}&$\num{.e-0.13}$\\
	Labor offset&\textcolor{black}{17:24$=$}\textcolor{black}{17:19}&$\num{.e-0.36}$&\textcolor{black}{17:19$=$}\textcolor{black}{16:56}&$\num{.e-0.83}$\\
	Onset of offset&\textcolor{black}{08h34m$=$}\textcolor{black}{08h22m}&$\num{.e-1.10}$&\textcolor{black}{07h20m$>$}\textcolor{black}{06h29m}&$\num{.e-3.22}$\\
	Labor time&\textcolor{black}{07h35m$>$}\textcolor{black}{07h25m}&$\num{.e-1.67}$&\textcolor{black}{06h08m$>$}\textcolor{black}{05h32m}&$\num{.e-3.25}$\\
	$|E_s-E_w|$ min;ave;max&\multicolumn{1}{c}{(3;7;12)\si{\minute}}&&\multicolumn{1}{c}{(5;28;51)\si{\minute}}&\\
	SEM (average)&\multicolumn{1}{c}{\SI{4}{\minute}}&&\multicolumn{1}{c}{\SI{10}{\minute}}&\\
	\midrule
	\multicolumn{5}{l}{\textbf{France}}\\
	Labor onset&\textcolor{black}{09:00$=$}\textcolor{black}{08:58}&$\num{.e-0.11}$&\textcolor{black}{11:08$<$}\textcolor{black}{11:41}&$\num{.e-2.32}$\\
	Center of gravity&\textcolor{black}{13:29$=$}\textcolor{black}{13:31}&$\num{.e-0.10}$&\textcolor{black}{14:52$=$}\textcolor{black}{14:49}&$\num{.e-0.12}$\\
	Labor offset&\textcolor{black}{17:43$=$}\textcolor{black}{17:47}&$\num{.e-0.19}$&\textcolor{black}{18:04$=$}\textcolor{black}{17:43}&$\num{.e-1.00}$\\
	Onset of offset&\textcolor{black}{08h43m$=$}\textcolor{black}{08h48m}&$\num{.e-0.21}$&\textcolor{black}{06h56m$>$}\textcolor{black}{06h02m}&$\num{.e-3.83}$\\
	Labor time&\textcolor{black}{06h58m$=$}\textcolor{black}{07h01m}&$\num{.e-0.41}$&\textcolor{black}{04h44m$>$}\textcolor{black}{04h20m}&$\num{.e-2.47}$\\
	$|E_s-E_w|$ min;ave;max&\multicolumn{1}{c}{(2;4;6)\si{\minute}}&&\multicolumn{1}{c}{(4;27;54)\si{\minute}}&\\
	SEM (average)&\multicolumn{1}{c}{\SI{3}{\minute}}&&\multicolumn{1}{c}{\SI{8}{\minute}}&\\
	\midrule
	\multicolumn{5}{l}{\textbf{Italy}}\\
	Labor onset&\textcolor{black}{08:24$<$}\textcolor{black}{08:32}&$\num{.e-2.02}$&\textcolor{black}{09:10$=$}\textcolor{black}{09:20}&$\num{.e-0.77}$\\
	Center of gravity&\textcolor{black}{12:57$=$}\textcolor{black}{12:56}&$\num{.e-0.09}$&\textcolor{black}{13:11$=$}\textcolor{black}{13:20}&$\num{.e-0.62}$\\
	Labor offset&\textcolor{black}{17:31$=$}\textcolor{black}{17:31}&$\num{.e-0.02}$&\textcolor{black}{17:13$=$}\textcolor{black}{17:16}&$\num{.e-0.13}$\\
	Onset of offset&\textcolor{black}{09h07m$=$}\textcolor{black}{08h59m}&$\num{.e-0.98}$&\textcolor{black}{08h03m$=$}\textcolor{black}{07h56m}&$\num{.e-0.37}$\\
	Labor time&\textcolor{black}{07h41m$=$}\textcolor{black}{07h36m}&$\num{.e-0.79}$&\textcolor{black}{06h40m$=$}\textcolor{black}{06h35m}&$\num{.e-0.43}$\\
	$|E_s-E_w|$ min;ave;max&\multicolumn{1}{c}{(0;4;8)\si{\minute}}&&\multicolumn{1}{c}{(3;7;10)\si{\minute}}&\\
	SEM (average)&\multicolumn{1}{c}{\SI{3}{\minute}}&&\multicolumn{1}{c}{\SI{5}{\minute}}&\\
	\midrule
	\multicolumn{5}{l}{\textbf{Spain}}\\
	Labor onset&\textcolor{black}{08:48$=$}\textcolor{black}{08:56}&$\num{.e-0.85}$&\textcolor{black}{09:44$=$}\textcolor{black}{09:58}&$\num{.e-0.20}$\\
	Center of gravity&\textcolor{black}{13:18$=$}\textcolor{black}{13:25}&$\num{.e-0.81}$&\textcolor{black}{14:06$=$}\textcolor{black}{14:06}&$10^{-0.00}$\\
	Labor offset&\textcolor{black}{18:03$=$}\textcolor{black}{18:03}&$\num{.e-0.01}$&\textcolor{black}{17:53$=$}\textcolor{black}{17:55}&$\num{.e-0.05}$\\
	Onset of offset&\textcolor{black}{09h15m$=$}\textcolor{black}{09h07m}&$\num{.e-0.52}$&\textcolor{black}{08h08m$=$}\textcolor{black}{07h57m}&$\num{.e-0.13}$\\
	Labor time&\textcolor{black}{07h46m$=$}\textcolor{black}{07h42m}&$\num{.e-0.44}$&\textcolor{black}{06h28m$=$}\textcolor{black}{06h22m}&$\num{.e-0.17}$\\
	$|E_s-E_w|$ min;ave;max&\multicolumn{1}{c}{(0;6;8)\si{\minute}}&&\multicolumn{1}{c}{(0;7;14)\si{\minute}}&\\
	SEM (average)&\multicolumn{1}{c}{\SI{4}{\minute}}&&\multicolumn{1}{c}{\SI{11}{\minute}}&\\
	\midrule
	\multicolumn{5}{l}{\textbf{United States}}\\
	Labor onset&\textcolor{black}{08:14$=$}\textcolor{black}{08:14}&$\num{.e-0.05}$&\textcolor{black}{10:54$=$}\textcolor{black}{10:52}&$\num{.e-0.11}$\\
	Center of gravity&\textcolor{black}{12:53$=$}\textcolor{black}{12:49}&$\num{.e-1.23}$&\textcolor{black}{14:26$=$}\textcolor{black}{14:24}&$\num{.e-0.14}$\\
	Labor offset&\textcolor{black}{17:37$=$}\textcolor{black}{17:33}&$\num{.e-1.15}$&\textcolor{black}{17:44$=$}\textcolor{black}{17:42}&$\num{.e-0.19}$\\
	Onset of offset&\textcolor{black}{09h23m$=$}\textcolor{black}{09h19m}&$\num{.e-0.91}$&\textcolor{black}{06h50m$=$}\textcolor{black}{06h49m}&$\num{.e-0.06}$\\
	Labor time&\textcolor{black}{08h01m$=$}\textcolor{black}{07h59m}&$\num{.e-0.72}$&\textcolor{black}{05h27m$=$}\textcolor{black}{05h23m}&$\num{.e-0.64}$\\
	$|E_s-E_w|$ min;ave;max&\multicolumn{1}{c}{(0;3;4)\si{\minute}}&&\multicolumn{1}{c}{(1;2;5)\si{\minute}}&\\
	SEM (average)&\multicolumn{1}{c}{\SI{1}{\minute}}&&\multicolumn{1}{c}{\SI{3}{\minute}}&\\
	\bottomrule
	\end{tabular}

  \caption{Same as Table~\ref{tab:vigil} but for the labour variables.}
  \label{tab:laborWelch}
\end{table*}

\clearpage

\begin{table*}
  \centering
\footnotesize\sf
\setlength{\tabcolsep}{4pt}
	\begin{tabular}{lccccccccccccc}
	\toprule
	Quartile differences&\multicolumn{3}{c}{\begin{tabular}[t]{@{}c@{}}Group 1\\Mon-Fri\\employees\end{tabular}}&\multicolumn{3}{c}{\begin{tabular}[t]{@{}c@{}}Group 2\\Mon-Fri\\non-employees\end{tabular}}&\multicolumn{3}{c}{\begin{tabular}[t]{@{}c@{}}Group 3\\Sat-Sun\\employees\end{tabular}}&\multicolumn{3}{c}{\begin{tabular}[t]{@{}c@{}}Group 4\\Sat-Sun\\non-employees\end{tabular}}\\
	&$\Delta Q_1$&$\Delta Q_2$&$\Delta Q_3$&$\Delta Q_1$&$\Delta Q_2$&$\Delta Q_3$&$\Delta Q_1$&$\Delta Q_2$&$\Delta Q_3$&$\Delta Q_1$&$\Delta Q_2$&$\Delta Q_3$\\
	\midrule
	\multicolumn{13}{l}{\textbf{Great Britain}}\\
	Sleep offset, $t_1$&&&&&&\textcolor{black}{$+2$}&\textcolor{black}{$-1$}&&\textcolor{black}{$-2$}&\textcolor{black}{$-1$}&\textcolor{black}{$-1$}&\textcolor{black}{$-1$}\\
	Center of gravity, $t_2$&&&&&&&\textcolor{black}{$-1$}&\textcolor{black}{$-1$}&&&&\\
	Sleep onset, $t_3$&&\textcolor{black}{$+1$}&&&&\textcolor{black}{$-1$}&\textcolor{black}{$-1$}&\textcolor{black}{$-1$}&\textcolor{black}{$-1$}&&&\textcolor{black}{$-1$}\\
	Onset to offset, $d_2$&\textcolor{black}{$-1$}&&&\textcolor{black}{$+1$}&&\textcolor{black}{$+1$}&&&\textcolor{black}{$-1$}&\textcolor{black}{$-1$}&\textcolor{black}{$-1$}&\textcolor{black}{$-1$}\\
	Sleep time, $d_1$&&\textcolor{black}{$-1$}&&&\textcolor{black}{$+1$}&\textcolor{black}{$+2$}&&&\textcolor{black}{$-1$}&\textcolor{black}{$-1$}&&\\
	\midrule
	\multicolumn{13}{l}{\textbf{France}}\\
	Sleep offset, $t_1$&&&&\textcolor{black}{$-1$}&&\textcolor{black}{$+1$}&\textcolor{black}{$-3$}&\textcolor{black}{$-1$}&&&&\\
	Center of gravity, $t_2$&\textcolor{black}{$+1$}&\textcolor{black}{$+1$}&\textcolor{black}{$+1$}&&&&&&\textcolor{black}{$+1$}&\textcolor{black}{$+1$}&&\textcolor{black}{$+1$}\\
	Sleep onset, $t_3$&&&&&&&\textcolor{black}{$+1$}&\textcolor{black}{$+2$}&\textcolor{black}{$+3$}&\textcolor{black}{$+1$}&\textcolor{black}{$+1$}&\textcolor{black}{$+1$}\\
	Onset to offset, $d_2$&\textcolor{black}{$-1$}&\textcolor{black}{$-1$}&&&&&\textcolor{black}{$-3$}&\textcolor{black}{$-1$}&\textcolor{black}{$-1$}&\textcolor{black}{$-1$}&\textcolor{black}{$-1$}&\textcolor{black}{$-1$}\\
	Sleep time, $d_1$&&&\textcolor{black}{$-1$}&&&&\textcolor{black}{$-1$}&\textcolor{black}{$-1$}&\textcolor{black}{$-1$}&\textcolor{black}{$-1$}&\textcolor{black}{$-1$}&\textcolor{black}{$-1$}\\
	\midrule
	\multicolumn{13}{l}{\textbf{Italy}}\\
	Sleep offset, $t_1$&&&&&&&&&&&&\\
	Center of gravity, $t_2$&&&\textcolor{black}{$+1$}&&\textcolor{black}{$+1$}&&&&&&&\\
	Sleep onset, $t_3$&&&\textcolor{black}{$+1$}&\textcolor{black}{$+2$}&\textcolor{black}{$+1$}&\textcolor{black}{$+1$}&&&\textcolor{black}{$+1$}&\textcolor{black}{$+1$}&\textcolor{black}{$+1$}&\\
	Onset to offset, $d_2$&&\textcolor{black}{$-1$}&\textcolor{black}{$-1$}&\textcolor{black}{$-1$}&\textcolor{black}{$-2$}&\textcolor{black}{$-1$}&\textcolor{black}{$-1$}&\textcolor{black}{$-1$}&\textcolor{black}{$-1$}&\textcolor{black}{$-1$}&\textcolor{black}{$-2$}&\textcolor{black}{$-2$}\\
	Sleep time, $d_1$&&\textcolor{black}{$-1$}&\textcolor{black}{$-1$}&\textcolor{black}{$-2$}&&\textcolor{black}{$-1$}&\textcolor{black}{$-1$}&\textcolor{black}{$-1$}&\textcolor{black}{$-1$}&\textcolor{black}{$-1$}&\textcolor{black}{$-1$}&\\
	\midrule
	\multicolumn{13}{l}{\textbf{Spain}}\\
	Sleep offset, $t_1$&\textcolor{black}{$-1$}&&&&\textcolor{black}{$-1$}&&\textcolor{black}{$-3$}&&&&&\textcolor{black}{$-1$}\\
	Center of gravity, $t_2$&&\textcolor{black}{$-1$}&&&\textcolor{black}{$+1$}&\textcolor{black}{$+1$}&&&&&\textcolor{black}{$+1$}&\textcolor{black}{$+1$}\\
	Sleep onset, $t_3$&\textcolor{black}{$+1$}&\textcolor{black}{$+1$}&&\textcolor{black}{$+2$}&&\textcolor{black}{$+1$}&&\textcolor{black}{$+2$}&\textcolor{black}{$+3$}&\textcolor{black}{$+1$}&&\textcolor{black}{$+1$}\\
	Onset to offset, $d_2$&&\textcolor{black}{$-1$}&&\textcolor{black}{$-1$}&\textcolor{black}{$-1$}&\textcolor{black}{$-2$}&\textcolor{black}{$-2$}&\textcolor{black}{$-2$}&&\textcolor{black}{$-2$}&\textcolor{black}{$-3$}&\textcolor{black}{$-2$}\\
	Sleep time, $d_1$&&&\textcolor{black}{$-1$}&&&&\textcolor{black}{$-2$}&&\textcolor{black}{$+2$}&\textcolor{black}{$-1$}&\textcolor{black}{$-1$}&\textcolor{black}{$-1$}\\
	\midrule
	\multicolumn{13}{l}{\textbf{United States}}\\
	Sleep offset, $t_1$&&&&&&&&&&&&\\
	Center of gravity, $t_2$&\textcolor{black}{$+1$}&\textcolor{black}{$+1$}&&&&\textcolor{black}{$+1$}&\textcolor{black}{$+1$}&&&&&\\
	Sleep onset, $t_3$&\textcolor{black}{$+1$}&\textcolor{black}{$+1$}&\textcolor{black}{$+1$}&\textcolor{black}{$+1$}&&\textcolor{black}{$+1$}&&&&\textcolor{black}{$+1$}&\textcolor{black}{$+1$}&\\
	Onset to offset, $d_2$&&&\textcolor{black}{$-1$}&&&\textcolor{black}{$-1$}&\textcolor{black}{$-2$}&&\textcolor{black}{$-2$}&\textcolor{black}{$-1$}&&\textcolor{black}{$-2$}\\
	Sleep time, $d_1$&\textcolor{black}{$-1$}&\textcolor{black}{$-1$}&&\textcolor{black}{$-1$}&\textcolor{black}{$-1$}&\textcolor{black}{$-1$}&&\textcolor{black}{$-2$}&\textcolor{black}{$-1$}&&\textcolor{black}{$-2$}&\textcolor{black}{$-2$}\\
	\bottomrule
	\end{tabular}

  \caption{Seasonal differences in quartiles $\Delta Q=Q_s-Q_w$ of the sleep/wake cycle. Values display units of ten minutes ---the discretization level of time use surveys---. Voids highlight null difference.}
  \label{tab:sleepQQ}
\end{table*}

\begin{table*}
  \centering
\footnotesize\sf
\setlength{\tabcolsep}{9pt}
	\begin{tabular}{lccccccc}
	\toprule
	Quartile differences&\multicolumn{3}{c}{\begin{tabular}[t]{@{}c@{}}Group 1\\Mon-Fri\\employees\end{tabular}}&\multicolumn{3}{c}{\begin{tabular}[t]{@{}c@{}}Group 2\\Sat-Sun\\employees\end{tabular}}\\
	&$\Delta Q_1$&$\Delta Q_2$&$\Delta Q_3$&$\Delta Q_1$&$\Delta Q_2$&$\Delta Q_3$\\
	\midrule
	\multicolumn{7}{l}{\textbf{Great Britain}}\\
	Labor onset, $t_1$&\textcolor{black}{$-1$}&&&\textcolor{black}{$-1$}&\textcolor{black}{$-2$}&\textcolor{black}{$-8$}\\
	Center of gravity, $t_2$&&&\textcolor{black}{$+1$}&\textcolor{black}{$-2$}&&\textcolor{black}{$+2$}\\
	Labor offset, $t_3$&&&&\textcolor{black}{$+2$}&&\textcolor{black}{$+3$}\\
	Onset to offset, $d_2$&\textcolor{black}{$+1$}&&&\textcolor{black}{$+7$}&\textcolor{black}{$+5$}&\textcolor{black}{$+2$}\\
	Labor time, $d_1$&\textcolor{black}{$+2$}&\textcolor{black}{$+1$}&\textcolor{black}{$+1$}&\textcolor{black}{$+5$}&\textcolor{black}{$+6$}&\textcolor{black}{$+1$}\\
	\midrule
	\multicolumn{7}{l}{\textbf{France}}\\
	Labor onset, $t_1$&\textcolor{black}{$-1$}&\textcolor{black}{$+1$}&&\textcolor{black}{$-3$}&\textcolor{black}{$-2$}&\textcolor{black}{$-6$}\\
	Center of gravity, $t_2$&&&&&\textcolor{black}{$-1$}&\\
	Labor offset, $t_3$&&&\textcolor{black}{$-1$}&\textcolor{black}{$+3$}&\textcolor{black}{$+1$}&\textcolor{black}{$+3$}\\
	Onset to offset, $d_2$&&\textcolor{black}{$-1$}&&&\textcolor{black}{$+10$}&\textcolor{black}{$+5$}\\
	Labor time, $d_1$&\textcolor{black}{$-1$}&&\textcolor{black}{$-1$}&&\textcolor{black}{$+3$}&\textcolor{black}{$+4$}\\
	\midrule
	\multicolumn{7}{l}{\textbf{Italy}}\\
	Labor onset, $t_1$&\textcolor{black}{$-1$}&&\textcolor{black}{$-1$}&&\textcolor{black}{$-1$}&\textcolor{black}{$-1$}\\
	Center of gravity, $t_2$&&&&\textcolor{black}{$-1$}&\textcolor{black}{$-1$}&\textcolor{black}{$-1$}\\
	Labor offset, $t_3$&&&\textcolor{black}{$-1$}&\textcolor{black}{$-1$}&&\\
	Onset to offset, $d_2$&&&\textcolor{black}{$+1$}&\textcolor{black}{$+1$}&&\textcolor{black}{$+1$}\\
	Labor time, $d_1$&&&&\textcolor{black}{$+2$}&\textcolor{black}{$-1$}&\\
	\midrule
	\multicolumn{7}{l}{\textbf{Spain}}\\
	Labor onset, $t_1$&&&\textcolor{black}{$-1$}&\textcolor{black}{$-1$}&&\textcolor{black}{$-1$}\\
	Center of gravity, $t_2$&&&\textcolor{black}{$-2$}&\textcolor{black}{$-1$}&&\textcolor{black}{$+2$}\\
	Labor offset, $t_3$&&&\textcolor{black}{$+1$}&&\textcolor{black}{$-3$}&\textcolor{black}{$+3$}\\
	Onset to offset, $d_2$&&&\textcolor{black}{$+2$}&&\textcolor{black}{$-1$}&\textcolor{black}{$+5$}\\
	Labor time, $d_1$&&&&\textcolor{black}{$+1$}&\textcolor{black}{$-1$}&\\
	\midrule
	\multicolumn{7}{l}{\textbf{United States}}\\
	Labor onset, $t_1$&&&&&\textcolor{black}{$-1$}&\\
	Center of gravity, $t_2$&&&\textcolor{black}{$+1$}&&&\textcolor{black}{$+1$}\\
	Labor offset, $t_3$&&&\textcolor{black}{$+1$}&&\textcolor{black}{$+1$}&\\
	Onset to offset, $d_2$&\textcolor{black}{$+1$}&\textcolor{black}{$+1$}&\textcolor{black}{$+1$}&&\textcolor{black}{$-1$}&\\
	Labor time, $d_1$&&&\textcolor{black}{$+1$}&\textcolor{black}{$+1$}&\textcolor{black}{$+1$}&\\
	\bottomrule
	\end{tabular}

  \caption{Same as Table~\ref{tab:sleepQQ} but for the labour statistics.}
  \label{tab:laborQQ}
\end{table*}



\begin{thebibliography}{39}%
\makeatletter
\providecommand \@ifxundefined [1]{%
 \@ifx{#1\undefined}
}%
\providecommand \@ifnum [1]{%
 \ifnum #1\expandafter \@firstoftwo
 \else \expandafter \@secondoftwo
 \fi
}%
\providecommand \@ifx [1]{%
 \ifx #1\expandafter \@firstoftwo
 \else \expandafter \@secondoftwo
 \fi
}%
\providecommand \natexlab [1]{#1}%
\providecommand \enquote  [1]{``#1''}%
\providecommand \bibnamefont  [1]{#1}%
\providecommand \bibfnamefont [1]{#1}%
\providecommand \citenamefont [1]{#1}%
\providecommand \href@noop [0]{\@secondoftwo}%
\providecommand \href [0]{\begingroup \@sanitize@url \@href}%
\providecommand \@href[1]{\@@startlink{#1}\@@href}%
\providecommand \@@href[1]{\endgroup#1\@@endlink}%
\providecommand \@sanitize@url [0]{\catcode `\\12\catcode `\$12\catcode
  `\&12\catcode `\#12\catcode `\^12\catcode `\_12\catcode `\%12\relax}%
\providecommand \@@startlink[1]{}%
\providecommand \@@endlink[0]{}%
\providecommand \url  [0]{\begingroup\@sanitize@url \@url }%
\providecommand \@url [1]{\endgroup\@href {#1}{\urlprefix }}%
\providecommand \urlprefix  [0]{URL }%
\providecommand \Eprint [0]{\href }%
\providecommand \doibase [0]{http://dx.doi.org/}%
\providecommand \selectlanguage [0]{\@gobble}%
\providecommand \bibinfo  [0]{\@secondoftwo}%
\providecommand \bibfield  [0]{\@secondoftwo}%
\providecommand \translation [1]{[#1]}%
\providecommand \BibitemOpen [0]{}%
\providecommand \bibitemStop [0]{}%
\providecommand \bibitemNoStop [0]{.\EOS\space}%
\providecommand \EOS [0]{\spacefactor3000\relax}%
\providecommand \BibitemShut  [1]{\csname bibitem#1\endcsname}%
\let\auto@bib@innerbib\@empty
\bibitem [{\citenamefont {Anglmayer}(2017)}]{EPRS2017}%
  \BibitemOpen
  \bibfield  {author} {\bibinfo {author} {\bibnamefont {Anglmayer},
  \bibfnamefont {Irmgard}}} (\bibinfo {year} {2017}),\ \href {\doibase
  10.2861/380995} {\emph {\bibinfo {title} {{EU summer-time arrangements under
  Directive 2000/84/EC Study}}}},\ \bibinfo {type} {Tech. Rep.}\ (\bibinfo
  {institution} {EPRS})\BibitemShut {NoStop}%
\bibitem [{\citenamefont {Berk}\ \emph {et~al.}(2008)\citenamefont {Berk},
  \citenamefont {Dodd}, \citenamefont {Hallam}, \citenamefont {Berk},
  \citenamefont {Gleeson},\ and\ \citenamefont {Henry}}]{Berk2008}%
  \BibitemOpen
  \bibfield  {author} {\bibinfo {author} {\bibnamefont {Berk}, \bibfnamefont
  {Michael}}, \bibinfo {author} {\bibfnamefont {Seetal}\ \bibnamefont {Dodd}},
  \bibinfo {author} {\bibfnamefont {Karen}\ \bibnamefont {Hallam}}, \bibinfo
  {author} {\bibfnamefont {Lesley}\ \bibnamefont {Berk}}, \bibinfo {author}
  {\bibfnamefont {John}\ \bibnamefont {Gleeson}}, \ and\ \bibinfo {author}
  {\bibfnamefont {Margaret}\ \bibnamefont {Henry}}} (\bibinfo {year} {2008}),\
  \bibfield  {title} {\enquote {\bibinfo {title} {{Small shifts in diurnal
  rhythms are associated with an increase in suicide: The effect of daylight
  saving}},}\ }\href {\doibase 10.1111/j.1479-8425.2007.00331.x} {\bibfield
  {journal} {\bibinfo  {journal} {Sleep and Biological Rhythms}\ }\textbf
  {\bibinfo {volume} {6}}~(\bibinfo {number} {1}),\ \bibinfo {pages}
  {22--25}}\BibitemShut {NoStop}%
\bibitem [{\citenamefont {Borb{\'{e}}ly}\ \emph {et~al.}(2016)\citenamefont
  {Borb{\'{e}}ly}, \citenamefont {Daan}, \citenamefont {Wirz-Justice},\ and\
  \citenamefont {Deboer}}]{Borbely2016}%
  \BibitemOpen
  \bibfield  {author} {\bibinfo {author} {\bibnamefont {Borb{\'{e}}ly},
  \bibfnamefont {Alexander~A}}, \bibinfo {author} {\bibfnamefont {Serge}\
  \bibnamefont {Daan}}, \bibinfo {author} {\bibfnamefont {Anna}\ \bibnamefont
  {Wirz-Justice}}, \ and\ \bibinfo {author} {\bibfnamefont {Tom}\ \bibnamefont
  {Deboer}}} (\bibinfo {year} {2016}),\ \bibfield  {title} {\enquote {\bibinfo
  {title} {{The two-process model of sleep regulation: a reappraisal}},}\
  }\href {\doibase 10.1111/jsr.12371} {\bibfield  {journal} {\bibinfo
  {journal} {Journal of Sleep Research}\ }\textbf {\bibinfo {volume}
  {25}}~(\bibinfo {number} {2}),\ \bibinfo {pages} {131--143}}\BibitemShut
  {NoStop}%
\bibitem [{\citenamefont {{BLS}}(2012)}]{ustus-2012}%
  \BibitemOpen
  \bibfield  {author} {\bibinfo {author} {\bibnamefont {{Bureau of Labor
  Statistics}},}} (\bibinfo {year} {2012}),\ \href
  {https://www.bls.gov/tus/datafiles_2013.htm} {\enquote {\bibinfo {title}
  {{A}merican {T}ime {U}se {S}urvey},}\ }\bibinfo {howpublished} {computer file
  (multi year data)}\BibitemShut {NoStop}%
\bibitem [{\citenamefont {Calandrillo}\ and\ \citenamefont
  {Buehler}(2008)}]{Calandrillo2008}%
  \BibitemOpen
  \bibfield  {author} {\bibinfo {author} {\bibnamefont {Calandrillo},
  \bibfnamefont {Steve~P}}, \ and\ \bibinfo {author} {\bibfnamefont {Dustin~E}\
  \bibnamefont {Buehler}}} (\bibinfo {year} {2008}),\ \bibfield  {title}
  {\enquote {\bibinfo {title} {{Time Well Spent: An Economic Analysis of
  Daylight Saving Time Legislation.}}}\ }\href
  {http://search.ebscohost.com/login.aspx?direct=true{\&}db=lft{\&}AN=502615679{\&}site=ehost-live}
  {\bibfield  {journal} {\bibinfo  {journal} {Wake Forest Law Review}\ }\textbf
  {\bibinfo {volume} {43}}~(\bibinfo {number} {1}),\ \bibinfo {pages}
  {45--91}}\BibitemShut {NoStop}%
\bibitem [{\citenamefont {Chenu}\ and\ \citenamefont
  {Lesnard}(2006)}]{chenu2006}%
  \BibitemOpen
  \bibfield  {author} {\bibinfo {author} {\bibnamefont {Chenu}, \bibfnamefont
  {Alain}}, \ and\ \bibinfo {author} {\bibfnamefont {Laurent}\ \bibnamefont
  {Lesnard}}} (\bibinfo {year} {2006}),\ \bibfield  {title} {\enquote {\bibinfo
  {title} {{Time Use Surveys: a Review of their Aims, Methods, and Results}},}\
  }\href {\doibase 10.1017/S0003975606000117} {\bibfield  {journal} {\bibinfo
  {journal} {European Journal of Sociology}\ }\textbf {\bibinfo {volume}
  {47}}~(\bibinfo {number} {03}),\ \bibinfo {pages} {335}}\BibitemShut
  {NoStop}%
\bibitem [{\citenamefont {Coate}\ and\ \citenamefont
  {Markowitz}(2004)}]{Coate2004}%
  \BibitemOpen
  \bibfield  {author} {\bibinfo {author} {\bibnamefont {Coate}, \bibfnamefont
  {Douglas}}, \ and\ \bibinfo {author} {\bibfnamefont {Sara}\ \bibnamefont
  {Markowitz}}} (\bibinfo {year} {2004}),\ \bibfield  {title} {\enquote
  {\bibinfo {title} {{The effects of daylight and daylight saving time on US
  pedestrian fatalities and motor vehicle occupant fatalities}},}\ }\href
  {\doibase 10.1016/S0001-4575(03)00015-0} {\bibfield  {journal} {\bibinfo
  {journal} {Accident Analysis and Prevention}\ }\textbf {\bibinfo {volume}
  {36}}~(\bibinfo {number} {3}),\ \bibinfo {pages} {351--357}}\BibitemShut
  {NoStop}%
\bibitem [{\citenamefont {van Egmond}\ \emph {et~al.}(2019)\citenamefont {van
  Egmond}, \citenamefont {Ekman},\ and\ \citenamefont
  {Benedict}}]{VanEgmond2019}%
  \BibitemOpen
  \bibfield  {author} {\bibinfo {author} {\bibnamefont {van Egmond},
  \bibfnamefont {Lieve}}, \bibinfo {author} {\bibfnamefont {Martin}\
  \bibnamefont {Ekman}}, \ and\ \bibinfo {author} {\bibfnamefont {Christian}\
  \bibnamefont {Benedict}}} (\bibinfo {year} {2019}),\ \bibfield  {title}
  {\enquote {\bibinfo {title} {{Bed and rise times during the Age of
  Enlightenment: A case report}},}\ }\href {\doibase 10.1111/jsr.12862}
  {\bibinfo  {journal} {Journal of Sleep Research}\ ,\ \bibinfo {pages}
  {e12862}}\BibitemShut {NoStop}%
\bibitem [{\citenamefont {{European Commission}}(2018)}]{Commission2018}%
  \BibitemOpen
\bibfield  {journal} {  }\bibfield  {author} {\bibinfo {author} {\bibnamefont
  {{European Commission}},}} (\bibinfo {year} {2018}),\ \href
  {https://eur-lex.europa.eu/legal-content/EN/TXT/?uri=SWD:2018:0406:FIN}
  {\emph {\bibinfo {title} {Commission Staff working document}}},\ \bibinfo
  {type} {Tech. Rep.}\ (\bibinfo  {institution} {European
  Commission})\BibitemShut {NoStop}%
\bibitem [{\citenamefont {Ferrazzi}\ \emph {et~al.}(2018)\citenamefont
  {Ferrazzi}, \citenamefont {Romualdi}, \citenamefont {Ocello}, \citenamefont
  {Frighetto}, \citenamefont {Turco}, \citenamefont {Vigolo}, \citenamefont
  {Fabris}, \citenamefont {Angeli}, \citenamefont {Vettore}, \citenamefont
  {Costa},\ and\ \citenamefont {Montagnese}}]{Ferrazzi2018}%
  \BibitemOpen
  \bibfield  {author} {\bibinfo {author} {\bibnamefont {Ferrazzi},
  \bibfnamefont {Elena}}, \bibinfo {author} {\bibfnamefont {Chiara}\
  \bibnamefont {Romualdi}}, \bibinfo {author} {\bibfnamefont {Michele}\
  \bibnamefont {Ocello}}, \bibinfo {author} {\bibfnamefont {Giovanni}\
  \bibnamefont {Frighetto}}, \bibinfo {author} {\bibfnamefont {Matteo}\
  \bibnamefont {Turco}}, \bibinfo {author} {\bibfnamefont {Stefania}\
  \bibnamefont {Vigolo}}, \bibinfo {author} {\bibfnamefont {Fabrizio}\
  \bibnamefont {Fabris}}, \bibinfo {author} {\bibfnamefont {Paolo}\
  \bibnamefont {Angeli}}, \bibinfo {author} {\bibfnamefont {Gianna}\
  \bibnamefont {Vettore}}, \bibinfo {author} {\bibfnamefont {Rodolfo}\
  \bibnamefont {Costa}}, \ and\ \bibinfo {author} {\bibfnamefont {Sara}\
  \bibnamefont {Montagnese}}} (\bibinfo {year} {2018}),\ \bibfield  {title}
  {\enquote {\bibinfo {title} {{Changes in Accident and Emergency Visits and
  Return Visits in Relation to the Enforcement of Daylight Saving Time and
  Photoperiod}},}\ }\href {\doibase 10.1177/0748730418791097} {\bibfield
  {journal} {\bibinfo  {journal} {Journal of Biological Rhythms}\ }\textbf
  {\bibinfo {volume} {33}}~(\bibinfo {number} {5}),\ \bibinfo {pages}
  {555--564}}\BibitemShut {NoStop}%
\bibitem [{\citenamefont {Franklin}(1784)}]{Franklin1784}%
  \BibitemOpen
  \bibfield  {author} {\bibinfo {author} {\bibnamefont {Franklin},
  \bibfnamefont {Benjamin}}} (\bibinfo {year} {1784}),\ \href
  {http://www.webexhibits.org/daylightsaving/franklin3.html} {\enquote
  {\bibinfo {title} {{An Economical Project}},}\ }\BibitemShut {NoStop}%
\bibitem [{\citenamefont {Franklin}(2005)}]{Franklin2005}%
  \BibitemOpen
  \bibfield  {author} {\bibinfo {author} {\bibnamefont {Franklin},
  \bibfnamefont {Benjamin}}} (\bibinfo {year} {2005}),\ \href@noop {} {\emph
  {\bibinfo {title} {{Benjamin Franklin: Autobiography, Poor Richard and Later
  Writings}}}},\ edited by\ \bibinfo {editor} {\bibfnamefont {J.A.~Leo}\
  \bibnamefont {Lemay}}\ (\bibinfo  {publisher} {Library of
  America})\BibitemShut {NoStop}%
\bibitem [{\citenamefont {Hill}\ \emph {et~al.}(2010)\citenamefont {Hill},
  \citenamefont {Desobry}, \citenamefont {Garnsey},\ and\ \citenamefont
  {Chong}}]{Hill2010}%
  \BibitemOpen
  \bibfield  {author} {\bibinfo {author} {\bibnamefont {Hill}, \bibfnamefont
  {S~I}}, \bibinfo {author} {\bibfnamefont {F.}~\bibnamefont {Desobry}},
  \bibinfo {author} {\bibfnamefont {E.~W.}\ \bibnamefont {Garnsey}}, \ and\
  \bibinfo {author} {\bibfnamefont {Y.~F.}\ \bibnamefont {Chong}}} (\bibinfo
  {year} {2010}),\ \bibfield  {title} {\enquote {\bibinfo {title} {{The impact
  on energy consumption of daylight saving clock changes}},}\ }\href {\doibase
  10.1016/j.enpol.2010.03.079} {\bibfield  {journal} {\bibinfo  {journal}
  {Energy Policy}\ }\textbf {\bibinfo {volume} {38}}~(\bibinfo {number} {9}),\
  \bibinfo {pages} {4955--4965}}\BibitemShut {NoStop}%
\bibitem [{\citenamefont {{INE}}(2010)}]{estus-2010e}%
  \BibitemOpen
  \bibfield  {author} {\bibinfo {author} {\bibnamefont {{Instituto Nacional de
  Estadística}},}} (\bibinfo {year} {2010}),\ \href {http://bit.ly/2uQFqa0}
  {\enquote {\bibinfo {title} {{S}panish {T}ime {U}se {S}urvey: {E}ncuesta de
  {E}mpleo del {T}iempo},}\ }\BibitemShut {NoStop}%
\bibitem [{\citenamefont {{Ipsos-RSL-ONS}}(2003)}]{uktus-2003}%
  \BibitemOpen
  \bibfield  {author} {\bibinfo {author} {\bibnamefont {{Ipsos-RSL and Office
  of National Statistics}},}} (\bibinfo {year} {2003}),\ \href@noop {}
  {\enquote {\bibinfo {title} {{U}nited {K}ingdom {T}ime {U}se {S}urvey 2000
  (computer file)},}\ }\bibinfo {howpublished} {3rd ed, {C}olchester, {E}ssex:
  {UK} {D}ata archive (distribuitor)}\BibitemShut {NoStop}%
\bibitem [{\citenamefont {Kamstra}\ \emph {et~al.}(2000)\citenamefont
  {Kamstra}, \citenamefont {Kramer},\ and\ \citenamefont {Levi}}]{Kamstra2000}%
  \BibitemOpen
  \bibfield  {author} {\bibinfo {author} {\bibnamefont {Kamstra}, \bibfnamefont
  {Mark~J}}, \bibinfo {author} {\bibfnamefont {Lisa~A}\ \bibnamefont {Kramer}},
  \ and\ \bibinfo {author} {\bibfnamefont {Maurice~D}\ \bibnamefont {Levi}}}
  (\bibinfo {year} {2000}),\ \bibfield  {title} {\enquote {\bibinfo {title}
  {{Losing Sleep at the Market: The Daylight Saving Anomaly}},}\ }\href
  {\doibase 10.1257/aer.90.4.1005} {\bibfield  {journal} {\bibinfo  {journal}
  {American Economic Review}\ }\textbf {\bibinfo {volume} {90}}~(\bibinfo
  {number} {4}),\ \bibinfo {pages} {1005--1011}}\BibitemShut {NoStop}%
\bibitem [{\citenamefont {Kantermann}\ \emph {et~al.}(2007)\citenamefont
  {Kantermann}, \citenamefont {Juda}, \citenamefont {Merrow},\ and\
  \citenamefont {Roenneberg}}]{Kantermann2007}%
  \BibitemOpen
  \bibfield  {author} {\bibinfo {author} {\bibnamefont {Kantermann},
  \bibfnamefont {Thomas}}, \bibinfo {author} {\bibfnamefont {Myriam}\
  \bibnamefont {Juda}}, \bibinfo {author} {\bibfnamefont {Martha}\ \bibnamefont
  {Merrow}}, \ and\ \bibinfo {author} {\bibfnamefont {Till}\ \bibnamefont
  {Roenneberg}}} (\bibinfo {year} {2007}),\ \bibfield  {title} {\enquote
  {\bibinfo {title} {{The Human Circadian Clock's Seasonal Adjustment Is
  Disrupted by Daylight Saving Time}},}\ }\href {\doibase
  10.1016/j.cub.2007.10.025} {\bibfield  {journal} {\bibinfo  {journal}
  {Current Biology}\ }\textbf {\bibinfo {volume} {17}}~(\bibinfo {number}
  {22}),\ \bibinfo {pages} {1996--2000}}\BibitemShut {NoStop}%
\bibitem [{\citenamefont {Kearney}\ \emph {et~al.}(2014)\citenamefont
  {Kearney}, \citenamefont {Chirico},\ and\ \citenamefont
  {Jarvis}}]{Kearney2014}%
  \BibitemOpen
  \bibfield  {author} {\bibinfo {author} {\bibnamefont {Kearney}, \bibfnamefont
  {James}}, \bibinfo {author} {\bibfnamefont {Stefania}\ \bibnamefont
  {Chirico}}, \ and\ \bibinfo {author} {\bibfnamefont {Andrew}\ \bibnamefont
  {Jarvis}}} (\bibinfo {year} {2014}),\ \href
  {https://ec.europa.eu/transport/sites/transport/files/facts-fundings/studies/doc/2014-09-19-the-application-of-summertime-in-europe.pdf}
  {\emph {\bibinfo {title} {Report to the European Commission
  Directorate-General for Mobility and Transport (DG MOVE)}}},\ \bibinfo {type}
  {Tech. Rep.}\ \bibinfo {number} {September}\ (\bibinfo  {institution} {ICF
  International})\BibitemShut {NoStop}%
\bibitem [{\citenamefont {Lindenberger}\ \emph {et~al.}(2019)\citenamefont
  {Lindenberger}, \citenamefont {Ackermann},\ and\ \citenamefont
  {Parzeller}}]{Lindenberger2019}%
  \BibitemOpen
  \bibfield  {author} {\bibinfo {author} {\bibnamefont {Lindenberger},
  \bibfnamefont {Lena~Marie}}, \bibinfo {author} {\bibfnamefont {Hanns}\
  \bibnamefont {Ackermann}}, \ and\ \bibinfo {author} {\bibfnamefont {Markus}\
  \bibnamefont {Parzeller}}} (\bibinfo {year} {2019}),\ \bibfield  {title}
  {\enquote {\bibinfo {title} {{The controversial debate about daylight saving
  time (DST)—results of a retrospective forensic autopsy study in
  Frankfurt/Main (Germany) over 10 years (2006–2015)}},}\ }\href {\doibase
  10.1007/s00414-018-1960-z} {\bibfield  {journal} {\bibinfo  {journal}
  {International Journal of Legal Medicine}\ }\textbf {\bibinfo {volume}
  {133}}~(\bibinfo {number} {4}),\ \bibinfo {pages} {1259--1265}}\BibitemShut
  {NoStop}%
\bibitem [{\citenamefont {{INSEE}}(2010)}]{frtus-2010e}%
  \BibitemOpen
  \bibfield  {author} {\bibinfo {author} {\bibnamefont {{INSEE}},}} (\bibinfo {year} {2010}),\
  \href@noop {} {\enquote {\bibinfo {title} {{F}rench {T}ime {U}se {S}urvey.
  {E}nqu\^ete emploi du {T}emps et {D}écisions dans les couples},}\
  }\BibitemShut {NoStop}%
\bibitem [{\citenamefont {{Istat}}(2009)}]{ittus-2010e}%
  \BibitemOpen
  \bibfield  {author} {\bibinfo {author} {\bibnamefont {{L'Istituto nazionale
  di statistica (Istat)}},}} (\bibinfo {year} {2009}),\ \href@noop {} {\enquote
  {\bibinfo {title} {{I}talian {T}ime {U}se {S}urvey: Uso del tempo},}\
  }\BibitemShut {NoStop}%
\bibitem [{\citenamefont {Luxan}(1810)}]{Luxan1810}%
  \BibitemOpen
  \bibfield  {author} {\bibinfo {author} {\bibnamefont {Luxan}, \bibfnamefont
  {Manuel}}} (\bibinfo {year} {1810}),\ \href
  {http://www.congreso.es/docu/blog/reglamento{\_}cortes{\_}1810.pdf} {\enquote
  {\bibinfo {title} {{Reglamento para el gobierno interior de las Cortes}},}\
  }\BibitemShut {NoStop}%
\bibitem [{\citenamefont {Manfredini}\ \emph {et~al.}(2018)\citenamefont
  {Manfredini}, \citenamefont {Fabbian}, \citenamefont {{De Giorgi}},
  \citenamefont {Zucchi}, \citenamefont {Cappadona}, \citenamefont {Signani},
  \citenamefont {Katsiki},\ and\ \citenamefont {Mikhailidis}}]{Manfredini2018}%
  \BibitemOpen
  \bibfield  {author} {\bibinfo {author} {\bibnamefont {Manfredini},
  \bibfnamefont {R}}, \bibinfo {author} {\bibfnamefont {F}~\bibnamefont
  {Fabbian}}, \bibinfo {author} {\bibfnamefont {A}~\bibnamefont {{De Giorgi}}},
  \bibinfo {author} {\bibfnamefont {B}~\bibnamefont {Zucchi}}, \bibinfo
  {author} {\bibfnamefont {R}~\bibnamefont {Cappadona}}, \bibinfo {author}
  {\bibfnamefont {F}~\bibnamefont {Signani}}, \bibinfo {author} {\bibfnamefont
  {N}~\bibnamefont {Katsiki}}, \ and\ \bibinfo {author} {\bibfnamefont {D~P}\
  \bibnamefont {Mikhailidis}}} (\bibinfo {year} {2018}),\ \bibfield  {title}
  {\enquote {\bibinfo {title} {{Daylight saving time and myocardial infarction:
  should we be worried? A review of the evidence.}}}\ }\href {\doibase
  10.26355/eurrev\_201802\_14306} {\bibfield  {journal} {\bibinfo  {journal}
  {European review for medical and pharmacological sciences}\ }\textbf
  {\bibinfo {volume} {22}}~(\bibinfo {number} {3}),\ \bibinfo {pages}
  {750--755}}\BibitemShut {NoStop}%
\bibitem [{\citenamefont {Mart{\'{i}}n-Olalla}(2018)}]{Martin-Olalla2018}%
  \BibitemOpen
  \bibfield  {author} {\bibinfo {author} {\bibnamefont {Mart{\'{i}}n-Olalla},
  \bibfnamefont {Jos{\'{e}}~Mar{\'{i}}a}}} (\bibinfo {year} {2018}),\ \bibfield
   {title} {\enquote {\bibinfo {title} {{Latitudinal trends in human primary
  activities: characterizing the winter day as a synchronizer}},}\ }\href
  {\doibase 10.1038/s41598-018-23546-5} {\bibfield  {journal} {\bibinfo
  {journal} {Scientific Reports 2018}\ }\textbf {\bibinfo {volume}
  {8}}~(\bibinfo {number} {1}),\ \bibinfo {pages} {5350}}\BibitemShut {NoStop}%
\bibitem [{\citenamefont
  {Mart{\'{i}}n-Olalla}(2019{\natexlab{a}})}]{Martin-Olalla2019a}%
  \BibitemOpen
  \bibfield  {author} {\bibinfo {author} {\bibnamefont {Mart{\'{i}}n-Olalla},
  \bibfnamefont {Jos{\'{e}}~Mar{\'{i}}a}}} (\bibinfo {year}
  {2019}{\natexlab{a}}),\ \bibfield  {title} {\enquote {\bibinfo {title}
  {{Comment to “Impact of Daylight Saving Time on circadian timing system: An
  expert statement”}},}\ }\href {\doibase 10.1016/J.EJIM.2019.02.006}
  {\bibfield  {journal} {\bibinfo  {journal} {European Journal of Internal
  Medicine}\ }\textbf {\bibinfo {volume} {62C}},\ \bibinfo {pages}
  {e18--e19}}\BibitemShut {NoStop}%
\bibitem [{\citenamefont
  {Mart{\'{i}}n-Olalla}(2019{\natexlab{b}})}]{Martin-Olalla2019d}%
  \BibitemOpen
  \bibfield  {author} {\bibinfo {author} {\bibnamefont {Mart{\'{i}}n-Olalla},
  \bibfnamefont {Jos{\'{e}}~Mar{\'{i}}a}}} (\bibinfo {year}
  {2019}{\natexlab{b}}),\ \bibfield  {title} {\enquote {\bibinfo {title}
  {{Scandinavian bed and rise times in the Age of Enlightenment and in the 21st
  century show similarity, helped by Daylight Saving Time}},}\ }\href {\doibase
  10.1111/jsr.12916} {\bibinfo  {journal} {Journal of Sleep Research}\ ,\
  \bibinfo {pages} {e12916}}\BibitemShut {NoStop}%
\bibitem [{\citenamefont
  {Mart{\'{i}}n-Olalla}(2019{\natexlab{c}})}]{Martin-Olalla2019b}%
  \BibitemOpen
\bibfield  {journal} {  }\bibfield  {author} {\bibinfo {author} {\bibnamefont
  {Mart{\'{i}}n-Olalla}, \bibfnamefont {Jos{\'{e}}~Mar{\'{i}}a}}} (\bibinfo
  {year} {2019}{\natexlab{c}}),\ \bibfield  {title} {\enquote {\bibinfo {title}
  {{Seasonal synchronization of sleep timing in industrial and pre-industrial
  societies}},}\ }\href {\doibase 10.1038/s41598-019-43220-8} {\bibfield
  {journal} {\bibinfo  {journal} {Scientific Reports}\ }\textbf {\bibinfo
  {volume} {9}}~(\bibinfo {number} {1}),\ \bibinfo {pages} {6772}}\BibitemShut
  {NoStop}%
\bibitem [{\citenamefont {{Meira e Cruz}}\ \emph {et~al.}(2019)\citenamefont
  {{Meira e Cruz}}, \citenamefont {Miyazawa}, \citenamefont {Manfredini},
  \citenamefont {Cardinali}, \citenamefont {Madrid}, \citenamefont {Reiter},
  \citenamefont {Araujo}, \citenamefont {Agostinho},\ and\ \citenamefont
  {Acu{\~{n}}a-Castroviejo}}]{MeiraeCruz2019}%
  \BibitemOpen
  \bibfield  {author} {\bibinfo {author} {\bibnamefont {{Meira e Cruz}},
  \bibfnamefont {M}}, \bibinfo {author} {\bibfnamefont {M.}~\bibnamefont
  {Miyazawa}}, \bibinfo {author} {\bibfnamefont {R.}~\bibnamefont
  {Manfredini}}, \bibinfo {author} {\bibfnamefont {D.}~\bibnamefont
  {Cardinali}}, \bibinfo {author} {\bibfnamefont {J.~A.}\ \bibnamefont
  {Madrid}}, \bibinfo {author} {\bibfnamefont {R.}~\bibnamefont {Reiter}},
  \bibinfo {author} {\bibfnamefont {J.~F.}\ \bibnamefont {Araujo}}, \bibinfo
  {author} {\bibfnamefont {R.}~\bibnamefont {Agostinho}}, \ and\ \bibinfo
  {author} {\bibfnamefont {D.}~\bibnamefont {Acu{\~{n}}a-Castroviejo}}}
  (\bibinfo {year} {2019}),\ \bibfield  {title} {\enquote {\bibinfo {title}
  {{Impact of Daylight Saving Time on circadian timing system: An expert
  statement}},}\ }\href {\doibase 10.1016/j.ejim.2019.01.001} {\bibfield
  {journal} {\bibinfo  {journal} {European Journal of Internal Medicine}\
  }\textbf {\bibinfo {volume} {60}}~(\bibinfo {number} {January}),\ \bibinfo
  {pages} {1--3}}\BibitemShut {NoStop}%
\bibitem [{\citenamefont {Monsivais}\ \emph {et~al.}(2017)\citenamefont
  {Monsivais}, \citenamefont {Bhattacharya}, \citenamefont {Ghosh},
  \citenamefont {Dunbar},\ and\ \citenamefont {Kaski}}]{Monsivais2017a}%
  \BibitemOpen
  \bibfield  {author} {\bibinfo {author} {\bibnamefont {Monsivais},
  \bibfnamefont {Daniel}}, \bibinfo {author} {\bibfnamefont {Kunal}\
  \bibnamefont {Bhattacharya}}, \bibinfo {author} {\bibfnamefont {Asim}\
  \bibnamefont {Ghosh}}, \bibinfo {author} {\bibfnamefont {Robin I.~M.}\
  \bibnamefont {Dunbar}}, \ and\ \bibinfo {author} {\bibfnamefont {Kimmo}\
  \bibnamefont {Kaski}}} (\bibinfo {year} {2017}),\ \bibfield  {title}
  {\enquote {\bibinfo {title} {{Seasonal and geographical impact on human
  resting periods}},}\ }\href {\doibase 10.1038/s41598-017-11125-z} {\bibfield
  {journal} {\bibinfo  {journal} {Scientific Reports}\ }\textbf {\bibinfo
  {volume} {7}}~(\bibinfo {number} {1}),\ \bibinfo {pages} {10717}}\BibitemShut
  {NoStop}%
\bibitem [{\citenamefont {Robb}\ and\ \citenamefont {Barnes}(2018)}]{Robb2018}%
  \BibitemOpen
  \bibfield  {author} {\bibinfo {author} {\bibnamefont {Robb}, \bibfnamefont
  {David}}, \ and\ \bibinfo {author} {\bibfnamefont {Thomas}\ \bibnamefont
  {Barnes}}} (\bibinfo {year} {2018}),\ \bibfield  {title} {\enquote {\bibinfo
  {title} {{Accident rates and the impact of daylight saving time
  transitions}},}\ }\href {\doibase 10.1016/J.AAP.2017.11.029} {\bibfield
  {journal} {\bibinfo  {journal} {Accident Analysis and Prevention}\ }\textbf
  {\bibinfo {volume} {111}},\ \bibinfo {pages} {193--201}}\BibitemShut
  {NoStop}%
\bibitem [{\citenamefont {Roenneberg}\ and\ \citenamefont
  {Merrow}(2016)}]{Roenneberg2016}%
  \BibitemOpen
  \bibfield  {author} {\bibinfo {author} {\bibnamefont {Roenneberg},
  \bibfnamefont {Till}}, \ and\ \bibinfo {author} {\bibfnamefont {Martha}\
  \bibnamefont {Merrow}}} (\bibinfo {year} {2016}),\ \bibfield  {title}
  {\enquote {\bibinfo {title} {{The Circadian Clock and Human Health.}}}\
  }\href {\doibase 10.1016/j.cub.2016.04.011} {\bibfield  {journal} {\bibinfo
  {journal} {Current biology : CB}\ }\textbf {\bibinfo {volume} {26}}~(\bibinfo
  {number} {10}),\ \bibinfo {pages} {R432--43}}\BibitemShut {NoStop}%
\bibitem [{\citenamefont {Roenneberg}\ \emph
  {et~al.}(2019{\natexlab{a}})\citenamefont {Roenneberg}, \citenamefont
  {Winnebeck},\ and\ \citenamefont {Klerman}}]{Roenneberg2019a}%
  \BibitemOpen
  \bibfield  {author} {\bibinfo {author} {\bibnamefont {Roenneberg},
  \bibfnamefont {Till}}, \bibinfo {author} {\bibfnamefont {Eva~C.}\
  \bibnamefont {Winnebeck}}, \ and\ \bibinfo {author} {\bibfnamefont
  {Elizabeth~B.}\ \bibnamefont {Klerman}}} (\bibinfo {year}
  {2019}{\natexlab{a}}),\ \bibfield  {title} {\enquote {\bibinfo {title}
  {{Daylight Saving Time and Artificial Time Zones – A Battle Between
  Biological and Social Times}},}\ }\href {\doibase 10.3389/fphys.2019.00944}
  {\bibfield  {journal} {\bibinfo  {journal} {Frontiers in Physiology}\
  }\textbf {\bibinfo {volume} {10}},\ \bibinfo {pages} {944}}\BibitemShut
  {NoStop}%
\bibitem [{\citenamefont {Roenneberg}\ \emph
  {et~al.}(2019{\natexlab{b}})\citenamefont {Roenneberg}, \citenamefont
  {Wirz-Justice}, \citenamefont {Skene}, \citenamefont {Ancoli-Israel},
  \citenamefont {Wright}, \citenamefont {Dijk}, \citenamefont {Zee},
  \citenamefont {Gorman}, \citenamefont {Winnebeck},\ and\ \citenamefont
  {Klerman}}]{Roenneberg2019}%
  \BibitemOpen
  \bibfield  {author} {\bibinfo {author} {\bibnamefont {Roenneberg},
  \bibfnamefont {Till}}, \bibinfo {author} {\bibfnamefont {Anna}\ \bibnamefont
  {Wirz-Justice}}, \bibinfo {author} {\bibfnamefont {Debra~J.}\ \bibnamefont
  {Skene}}, \bibinfo {author} {\bibfnamefont {Sonia}\ \bibnamefont
  {Ancoli-Israel}}, \bibinfo {author} {\bibfnamefont {Kenneth~P.}\ \bibnamefont
  {Wright}}, \bibinfo {author} {\bibfnamefont {Derk-Jan}\ \bibnamefont {Dijk}},
  \bibinfo {author} {\bibfnamefont {Phyllis}\ \bibnamefont {Zee}}, \bibinfo
  {author} {\bibfnamefont {Michael~R.}\ \bibnamefont {Gorman}}, \bibinfo
  {author} {\bibfnamefont {Eva~C.}\ \bibnamefont {Winnebeck}}, \ and\ \bibinfo
  {author} {\bibfnamefont {Elizabeth~B.}\ \bibnamefont {Klerman}}} (\bibinfo
  {year} {2019}{\natexlab{b}}),\ \bibfield  {title} {\enquote {\bibinfo {title}
  {{Why Should We Abolish Daylight Saving Time?}}}\ }\href {\doibase
  10.1177/0748730419854197} {\bibfield  {journal} {\bibinfo  {journal} {Journal
  of Biological Rhythms}\ }\textbf {\bibinfo {volume} {34}}~(\bibinfo {number}
  {3}),\ \bibinfo {pages} {227--230}}\BibitemShut {NoStop}%
\bibitem [{\citenamefont {Siegmund}\ \emph {et~al.}(1998)\citenamefont
  {Siegmund}, \citenamefont {Tittel},\ and\ \citenamefont
  {Schiefenh{\"{o}}vel}}]{Siegmund1998}%
  \BibitemOpen
  \bibfield  {author} {\bibinfo {author} {\bibnamefont {Siegmund},
  \bibfnamefont {R}}, \bibinfo {author} {\bibfnamefont {M.}~\bibnamefont
  {Tittel}}, \ and\ \bibinfo {author} {\bibfnamefont {W.}~\bibnamefont
  {Schiefenh{\"{o}}vel}}} (\bibinfo {year} {1998}),\ \bibfield  {title}
  {\enquote {\bibinfo {title} {{Activity Monitoring of the Inhabitants in
  Tauwema, a Traditional Melanesian Village: Rest/Activity Behaviour of
  Trobriand Islanders (Papua New Guinea)}},}\ }\href {\doibase
  10.1076/brhm.29.1.49.3045} {\bibfield  {journal} {\bibinfo  {journal}
  {Biological Rhythm Research}\ }\textbf {\bibinfo {volume} {29}}~(\bibinfo
  {number} {1}),\ \bibinfo {pages} {49--59}}\BibitemShut {NoStop}%
\bibitem [{\citenamefont {Smit}\ \emph {et~al.}(2019)\citenamefont {Smit},
  \citenamefont {Broesch}, \citenamefont {Siegel},\ and\ \citenamefont
  {Mistlberger}}]{Smit2019}%
  \BibitemOpen
  \bibfield  {author} {\bibinfo {author} {\bibnamefont {Smit}, \bibfnamefont
  {Andrea~N}}, \bibinfo {author} {\bibfnamefont {Tanya}\ \bibnamefont
  {Broesch}}, \bibinfo {author} {\bibfnamefont {Jerome~M.}\ \bibnamefont
  {Siegel}}, \ and\ \bibinfo {author} {\bibfnamefont {Ralph~E.}\ \bibnamefont
  {Mistlberger}}} (\bibinfo {year} {2019}),\ \bibfield  {title} {\enquote
  {\bibinfo {title} {{Sleep timing and duration in indigenous villages with and
  without electric lighting on Tanna Island, Vanuatu}},}\ }\href {\doibase
  10.1038/s41598-019-53635-y} {\bibfield  {journal} {\bibinfo  {journal}
  {Scientific Reports}\ }\textbf {\bibinfo {volume} {9}}~(\bibinfo {number}
  {1}),\ \bibinfo {pages} {17278}}\BibitemShut {NoStop}%
\bibitem [{\citenamefont {Watson}(2019)}]{Watson2019}%
  \BibitemOpen
  \bibfield  {author} {\bibinfo {author} {\bibnamefont {Watson}, \bibfnamefont
  {Nathaniel~F}}} (\bibinfo {year} {2019}),\ \bibfield  {title} {\enquote
  {\bibinfo {title} {{Time to Show Leadership on the Daylight Saving Time
  Debate.}}}\ }\href {\doibase 10.5664/jcsm.7822} {\bibfield  {journal}
  {\bibinfo  {journal} {Journal of clinical sleep medicine : JCSM : official
  publication of the American Academy of Sleep Medicine}\ }\textbf {\bibinfo
  {volume} {15}}~(\bibinfo {number} {6}),\ \bibinfo {pages}
  {815--817}}\BibitemShut {NoStop}%
\bibitem [{\citenamefont {Welch}(1947)}]{Welch1947}%
  \BibitemOpen
  \bibfield  {author} {\bibinfo {author} {\bibnamefont {Welch}, \bibfnamefont
  {B~L}}} (\bibinfo {year} {1947}),\ \bibfield  {title} {\enquote {\bibinfo
  {title} {{The generalization of Student's problem when several different
  population variances are involved}},}\ }\href {\doibase
  10.1093/biomet/34.1-2.28} {\bibfield  {journal} {\bibinfo  {journal}
  {Biometrika}\ }\textbf {\bibinfo {volume} {34}}~(\bibinfo {number} {1-2}),\
  \bibinfo {pages} {28--35}}\BibitemShut {NoStop}%
\bibitem [{\citenamefont {Willet}(1907)}]{Willet1907}%
  \BibitemOpen
  \bibfield  {author} {\bibinfo {author} {\bibnamefont {Willet}, \bibfnamefont
  {William}}} (\bibinfo {year} {1907}),\ \href@noop {} {\emph {\bibinfo {title}
  {{The waste of daylight}}}}\ (\bibinfo {address} {London})\BibitemShut
  {NoStop}%
\bibitem [{\citenamefont {Yetish}\ \emph {et~al.}(2015)\citenamefont {Yetish},
  \citenamefont {Kaplan}, \citenamefont {Gurven}, \citenamefont {Wood},
  \citenamefont {Pontzer}, \citenamefont {Manger}, \citenamefont {Wilson},
  \citenamefont {McGregor},\ and\ \citenamefont {Siegel}}]{Yetish2015}%
  \BibitemOpen
  \bibfield  {author} {\bibinfo {author} {\bibnamefont {Yetish}, \bibfnamefont
  {Gandhi}}, \bibinfo {author} {\bibfnamefont {Hillard}\ \bibnamefont
  {Kaplan}}, \bibinfo {author} {\bibfnamefont {Michael}\ \bibnamefont
  {Gurven}}, \bibinfo {author} {\bibfnamefont {Brian}\ \bibnamefont {Wood}},
  \bibinfo {author} {\bibfnamefont {Herman}\ \bibnamefont {Pontzer}}, \bibinfo
  {author} {\bibfnamefont {Paul~R}\ \bibnamefont {Manger}}, \bibinfo {author}
  {\bibfnamefont {Charles}\ \bibnamefont {Wilson}}, \bibinfo {author}
  {\bibfnamefont {Ronald}\ \bibnamefont {McGregor}}, \ and\ \bibinfo {author}
  {\bibfnamefont {Jerome~M}\ \bibnamefont {Siegel}}} (\bibinfo {year} {2015}),\
  \bibfield  {title} {\enquote {\bibinfo {title} {{Natural sleep and its
  seasonal variations in three pre-industrial societies.}}}\ }\href {\doibase
  10.1016/j.cub.2015.09.046} {\bibfield  {journal} {\bibinfo  {journal}
  {Current biology : CB}\ }\textbf {\bibinfo {volume} {25}}~(\bibinfo {number}
  {21}),\ \bibinfo {pages} {2862--2868}}\BibitemShut {NoStop}%
\end{thebibliography}
\end{document}